\documentstyle[12pt,aaspp4,epsf]{article}
\lefthead{Nagar et al. 1998}
\righthead{Early-Type Seyfert Sample}
\begin{document}

\title{Radio Structures of Seyfert Galaxies. VIII.
A Distance and Magnitude Limited Sample of Early-Type Galaxies}
\author{Neil M. Nagar, Andrew S. Wilson}
\affil{Department of Astronomy, University of Maryland, College Park, MD
20742; neil@astro.umd.edu, wilson@astro.umd.edu}
\author{John S. Mulchaey}
\affil{Observatories of the Carnegie Institute of Washington, 
813 Santa Barbara Street, Pasadena, CA 91101; mulchaey@ociw.edu}
\author{Jack F. Gallimore}
\affil{Max-Plank Institut f\"{u}r extraterrestriche Physik, Postfach
1603, D-85740 Garching bei M\"{u}nchen, Germany; jfg@hethp.mpe-garching.mpg.de}
\received{June 9, 1998}
\accepted{August 31, 1998}

\begin{center}
\textbf{To appear in ApJS, Vol. 120 \#2, February 1999}
\end{center}

\begin{abstract}

The VLA has been used at 3.6 and 20~cm to image a sample of about 50 early-type
Seyfert galaxies with recessional velocities less than 7,000 km s$^{-1}$
and total visual magnitude less than 14.5. 
Emission-line ([OIII] and H$\alpha$+[NII]) and
continuum (green and red) imaging of this sample has been presented in
a previous paper.  In this paper, we present the radio results,
discuss statistical relationships between the radio and other properties
and investigate these relationships within the context of unified
models of Seyferts.
The mean radio luminosities of early-type Seyfert 1's (i.e. Seyfert 1.0's,
1.2's and 1.5's) and Seyfert 2.0's are found to be similar (consistent with
the unified scheme) and the radio luminosity is independent of morphological
type within this sample. The fraction of resolved radio sources is larger in
the Seyfert 2.0's (93\%) than in the Seyfert 1's (64\%). However, the
mean radio extents of Seyfert 2.0's and Seyfert 1's are not
significantly different, though this result is limited by the small number
of resolved Seyfert 1's. 

The nuclear radio structures of Seyfert 2.0's in the early-type sample
tend to be aligned with the [OIII] and H$\alpha$+[NII] structures even 
though the radio extents are smaller than the [OIII] and H$\alpha$+[NII]
extents by a factor of $\sim$~2~--~$>$5. 
This alignment, previously known for individual Seyferts with
`linear' radio sources, is here shown to be characteristic of early-type
Seyfert galaxies as a class. 
Seyfert 2.0's in the early-type sample also show a significant  
alignment between the emission-line ([OIII] and H$\alpha$+[NII]) axes
and the major axis of the host galaxy.
These alignments are consistent with a picture in which the ionized gas
represents ambient gas predominantly co-planar with the galaxy stellar disk.
This ambient gas is ionized by nuclear radiation that may escape preferentially
along and around the radio axis, and is compressed in shocks driven by
the radio ejecta.
We use this alignment to constrain the product of the velocity of the 
radio ejecta and the period of any large angle precession 
of the inner accretion disk and jet : V$_{ejecta}~{\times}~P~\geq$~2~kpc.

An investigation of a larger sample of Seyferts reveals the
unexpected result that the Seyfert 1's with the largest radio  
extent ($\geq$~1.5~kpc) are all of type Seyfert 1.2.
It appears that classification as this type of intermediate 
Seyfert depends on some factor other than the relative
orientation of the nuclear obscuring torus to the line of sight.
Among all the other Seyferts, the distribution of radio extent with Seyfert
intermediate type is consistent with the expectations of the unified scheme.

\end{abstract}


\section{Introduction}

Centimeter wave emission from active galactic nuclei (AGN) is unaffected
by the extinction that plagues some other wavebands and can be imaged at  
high ($\lesssim$1{\arcsec}) spatial resolution with instruments 
like the Very Large Array (VLA).
These two factors can be exploited to provide important clues towards
an understanding of the workings of their central engines.

A number of samples of Seyfert galaxies have been observed at resolutions 
$\lesssim$1{\arcsec} with the VLA. A radio flux-limited sample of Markarian
Seyferts was studied by Ulvestad \& Wilson (1984a, hereafter Paper~V). A 
distance-limited sample, comprising the 57 Seyfert galaxies known 
as of mid-1983 with recession velocity less than 4,600 km s$^{-1}$ 
and declination $>-$45{\arcdeg}, was mapped by Ulvestad \& Wilson 
(1984b, hereafter Paper~VI; 1989, hereafter Paper~VII).
More recently,  Kukula et al. (1995) have imaged
46 of the 48 Seyferts in a magnitude-limited (m$_{pg}<$ 14.5) sample
(the CfA Seyfert sample).
Together, these samples represent less than 15\% of all known Seyferts.
Lower spatial resolution radio surveys have also been completed e.g.
the CfA (Edelson 1987)  and the 12{\micron} Seyfert galaxy samples 
(Rush, Malkan \& Edelson 1996).  
However, lower resolution surveys are contaminated by  
disk radio emission and thus do not provide direct information on
nuclear properties.

These surveys have largely supported the unified model (see
Antonucci 1993, Urry \& Padovani 1995 for reviews) of AGN.
Ulvestad \& Wilson (Paper~VII) concluded that the  trend for Seyfert 2 
galaxies to have stronger and larger radio sources (found in Papers V and
VI) is only marginally 
statistically significant.  Giuricin et al. (1990) also found no clear 
evidence for a significant difference
in the major radio properties (radio luminosity, radio-to-optical luminosity,
radio spectral index, and radio size) of Seyfert 1's and Seyfert 2's, while 
Rush et al. (1996) showed that there is no significant difference between the 
average 6--20~cm spectral indices of Seyfert 1's and Seyfert 2's.
Interestingly, Edelson (1987) found that while
there is no significant difference between the radio luminosities
of Seyfert 1 and Seyfert 2 galaxies, the normalized radio to total optical
luminosity of Seyfert 2 galaxies is greater than that of Seyfert 1 galaxies
by a factor of $\simeq$2, at a 99\% confidence level. 

The radio properties of Seyferts are also relevant 
to the excitation and properties of the narrow line region.
The radio luminosities are known to be correlated with both the narrow
emission-line luminosities (de Bruyn \& Wilson 1978; Whittle 1985, 1992c)
and the narrow emission-line widths (Wilson \& Willis 1980; 
Whittle 1985, 1992b).
Recent high resolution imaging (using HST and the VLA or MERLIN) shows a close 
spatial association between the  radio and the high excitation gas 
structures in almost all Seyfert 2's studied (Bower et al. 1994, 1995;
\cite{capet96}; Falcke, Wilson \& Simpson 1998). 
Capetti et al. (1996) and Falcke et al. (1998) find strong evidence
that the line-emitting gas is compressed by
the shocks created by the passage of the radio-emitting outflow; the
interaction increases the ambient density, causing increased line emission.
They also find that radio lobes produce emission-line gas in shell 
or bow-shock like morphologies 
formed by the sweeping up of material by the ejected and expanding lobes,
while radio jets produce linear emission-line structures, 
perhaps formed by the lateral expansion of hot gas around the jet axis.
The radio and `ionization cone' axes are also closely correlated 
(Wilson \& Tsvetanov 1994) implying that photons from the central engine
escape preferentially along the radio axis, a model originally 
indicated by the fact that 
high excitation ionized gas is detected well beyond the inner radio
lobes (Unger et al. 1987b).

Mulchaey, Wilson \& Tsvetanov (1996a, hereafter MWZ) have reported an optical
imaging survey of a 
sample of 57 early-type Seyfert galaxies; the galaxies were imaged in
[OIII]$\lambda$5007, H$\alpha$ + [NII]${\lambda}{\lambda}$6548,~6583 and
the nearby continua (green and red). 
The motivation for choosing early-type Seyferts
came from Haniff, Ward \& Wilson (1991) who found that most clearly defined
extended, high excitation nebulosities are
preferentially found in Seyfert galaxies of type S0.  
Almost all of the galaxies studied by MWZ showed extended emission in [OIII]
and H$\alpha$+[NII]. 

In the present paper, we present a high resolution radio survey of the 
early-type Seyfert galaxy sample of MWZ. We use the results to 
investigate unified models of Seyfert galaxies and the
relationships between radio and emission-line properties and between 
emission-line structures and the host galaxies. A separate paper (Nagar 
\& Wilson 1998) discusses the orientations of the radio ejecta of Seyfert
nuclei with respect to their host galaxy disks. 
For consistency with MWZ and with earlier papers in this series,
all distance-dependent quantities  have been calculated
from recessional velocities assuming H$_0$ = 50 km s$^{-1}$ Mpc$^{-1}$.
We use [OIII] to denote the [OIII]~$\lambda$5007 line, and 
H$\alpha$+[NII] to denote H{$\alpha$}~+~[NII] $\lambda\lambda$6548,~6583. 
We use `Seyfert~1' to denote Seyfert 1.0 through Seyfert 1.5, and
treat Seyfert 1.8's  and Seyfert 1.9's independently of 
Seyfert 2.0's (`Seyfert 2.0' emphasizes that this
class does not include Seyfert 1.8's and Seyfert 1.9's), as Maiolino
\& Rieke (1995) present evidence that Seyfert 1.8's and 1.9's are
type 1 nuclei seen through a 100 pc-scale  torus coplanar with the 
galaxy disk and not type 1 nuclei partially obscured by an inner pc-scale
torus. 

We have made extensive use of the following catalogs and databases -
``Third Reference Catalogue of Bright Galaxies'' 
(de Vaucouleurs et al. 1991, hereafter RC3), 
``Uppsala General Catalogue of Galaxies'' (Nilson 1973, hereafter UGC),
``ESO/Uppsala Survey of the ESO(B) Atlas'' (Lauberts 1982, hereafter ESO), 
``Extended Southern Galactic Catalog'' (Corwin et al. 1998, hereafter ESGC), 
``The NASA/IPAC Extragalactic Database'' 
(see e.g. Helou et al. 1991, hereafter NED), ``The Lyon-Meudon Extragalactic
Database'' (see e.g. Paturel et al. 1997, hereafter LEDA), and
the STScI Digitized Sky Survey (hereafter DSS).

\section{Sample and Observations}
A detailed description of the early-type Seyfert sample selection and a 
discussion of its completeness can be
found in MWZ and Mulchaey, Wilson \& Tsvetanov (1996b). 
Briefly, Seyfert galaxies were selected from
the catalogs of Hewitt \& Burbidge (1991), Huchra (1989), \&
Ve\'ron-Cetty and Ve\'ron (1991).
All Seyferts with total magnitude m$_V$~$\leq$14.5, recessional
velocity cz~$<$~7,000 km s$^{-1}$ and morphological type E, S0 or S0/a
were included in the MWZ early-type ``Sample I.''
Our radio sample consists of all galaxies in this sample
that are north of declination $\delta$=$-$41{\arcdeg}
and an additional 3 objects (NGC~7743, Mkn~335 and Mkn~612)
later found to satisfy the selection criterion. The final early-type
sample for which statistical studies are reported in Section~5 therefore 
consists of 43 Seyfert galaxies; 14 Seyfert 1's, 
2 Seyfert 1.9's and 27 Seyfert 2.0's.

New radio observations of 47 galaxies were made in the A- and A/B-hybrid
configurations of the VLA between December 1992 and February 1993
(see e.g. \cite{tet80} for a description of the VLA configurations).
These 47 galaxies comprise the early-type Seyfert sample of 43, minus
two galaxies (MCG--5-23-16 and Mrk~1239) which were not re-observed as
high resolution VLA maps exist in the literature 
(\cite{uw6} and Ulvestad, Antonucci \& Goodrich 1995), 
plus four misclassified Seyferts
(Mrk~938, Mrk~565, Mrk~577 and NGC~5077), and two Seyferts not included in
the sample studied in Section~5 (Mrk~10, host galaxy of morphological
type Sb and NGC~6251, a radio-loud object).
Each source was observed consecutively at L (20~cm)  and X (3.6~cm) band 
(central frequencies 1465 MHz  and 8440 MHz, respectively).
Two galaxies, Mrk~10 and NGC~513, were observed at only L band.
The time spent on each source was typically 15 to 20 min in
each band. Source observations were sandwiched between two 
1 min observations of a nearby calibrator. 
The two sets of VLA intermediate frequency (IF) channels used
gave a total bandwidth of 100 MHz at each band. Data
from the two sets of IF's were calibrated separately and then 
combined and mapped.
Observations of 0134+329 and 1328+307 were used to set the flux density
scale to that of Baars et al. (1977). 
Most of the maps were made at the highest resolution possible 
(uniform weighting with Brigg's robust parameter set to --4
in the AIPS software). In some of the noisier maps,
it was necessary to use either uniform weighting with
a robust parameter of 0 or natural weighting to achieve better signal to
noise at the expense of resolution.
The stronger sources were iteratively self-calibrated and mapped.
The r.m.s. noise in the final maps is typically 80 to 150 $\mu$Jy for
the 20~cm maps and 40 to 90 $\mu$Jy for the 3.6~cm maps. The 
southern sources ($\delta$ $\lesssim$~--20{\arcdeg}) have
significantly larger noise values.

\section{Results}
In this section we present contour maps (Figures 1--15) for all the
observed sources and tables (Tables 1--5) of derived source properties.
Of the 47 galaxies observed, 
Mrk~577 was not detected at either 3.6 or 20~cm, which is 
not surprising as this is not a Seyfert galaxy (MWZ),
and NGC~4117 was not detected at 3.6~cm. 
No useful data were obtained on MCG--2-27-9 at either 3.6 or 20~cm as 
the pointing center was incorrect by $\sim$85{\arcsec}.
A cross marks the optical position of the galaxy on the contour map
whenever this position falls within the field of view of the radio map.
In the 3.6~cm maps, the semi-length of the arms of the cross represents 
the error in the position of the optical nucleus. In the 20~cm maps, 
we use a minimum cross size of 1{\arcsec}  in order to
keep the cross visible. Most optical positions and errors are taken from
Clements (1981; 1983) and Argyle \& Eldridge (1990); these positions
have typical internal errors of $\pm0{\farcs}1-0{\farcs}2$.
Radio positions and flux densities of most unresolved or slightly resolved
components have been determined by Gaussian model fitting.  In the case
of more extended components, the position of the flux maximum is given. 
For such extended structure,
or for slightly resolved sources for which Gaussian model fitting is 
inappropriate, the flux has been summed within a box containing all
significant emission.

Table~1 summarizes the properties of all the 47 galaxies which have been
newly observed with the VLA plus the two additional galaxies in the 
early-type radio sample that have been previously observed with the VLA
(Section~2).
The leftmost columns in Table~1 are as follows :
(1)~galaxy name; (2)~other common name; 
(3) and (4)~radio position (B1950). 
When more than one radio component is present, the position listed was
determined as follows. If a component appears on both the 3.6 and
20~cm maps, we give its position.
If more than one component satisfies this criterion,
we have listed the position of the strongest one. If multiple components have
similar strengths, the component closest to the optical position is listed.
The other columns of Table~1 are :
(5)~host galaxy type from the RC3 or UGC catalog; (6)~Seyfert type.
If broad permitted lines are present we follow, whenever possible, the 
Seyfert classification scheme adopted by Whittle (1992a) which
is based on the flux ratio R $=$ F$_{[OIII]}$ / F$_{H\beta}$,
where F$_{[OIII]}$ is the [OIII] flux and F$_{H\beta}$ the total
(broad plus narrow) H{$\beta$} flux.  These types are (with R values) 
Sey 1.0 (R $\leq$ 0.3),
Sey 1.2 (0.3 $<$ R $\leq$ 1), Sey 1.5 (1 $<$ R $\leq$ 4),
Sey 1.8 (R $>$ 4) and Sey 1.9 (assigned if only broad
H$\alpha$ is seen). As in Whittle (1992a), some 
Seyferts with highly variable H$\beta$ flux are classified as Sey 1.5.
We have used emission-line fluxes listed in Whittle (1992a) and Winkler (1992); 
for objects for which we could find no measurement of R, we use
the classification given in NED.
The source of the Seyfert classification for galaxies with broad permitted
lines can be found in the comments column (column 13);
(7)~radio structure, where, as in 
Paper~V: L = linear; D = diffuse; A = ambiguous; S = slightly resolved; and
U = unresolved. Single component sources which, after Gaussian deconvolution,
are larger than one half the beam size are considered `S'. These sources
are $>$ 1.12 times the beam size before deconvolution. Sources which,
after Gaussian deconvolution, were smaller than one half the beam size,
but showed clear signs of an extension in either the 3.6~cm or 20~cm contour
maps are tentatively considered to be slightly resolved and are listed
as `(S)'. Similarly, multi-component sources with apparently `linear'
structure in which the reality of
the weaker components is not completely certain, are listed as `(L)';
(8)~total flux from all components at 3.6~cm;
(9)~total flux from all components at 20~cm;
(10)~largest linear extent (in kpc) of the radio emission in 
the 3.6 and 20~cm maps. 
For unresolved sources, two numbers are listed, representing an upper limit
of half the beam size at 3.6 and 20~cm, respectively.  For
slightly resolved sources, we use the source size after Gaussian deconvolution.
For sources with multiple unresolved components, we use the peak to peak
distance between the most widely separated components. The extents for
``(S)'' and ``(L)'' sources have been measured assuming that the extensions 
are real and these values are listed in brackets;
(11)~host galaxy morphological type (Hubble T parameter), taken from the RC3 or 
UGC catalogs unless otherwise mentioned. The uncertainty
in T, if listed in the catalog, is given in brackets;
(12)~heliocentric recessional velocity of the galaxy taken from NED;
(13)~comments according to the key at the bottom of the table. 

Table~2 lists the positions and fluxes of each component for all 
multiple component sources.
Components have been named according to apparent morphology
e.g. `Core', `Ext' (Extended), `Center', `Jet', `Arm', and/or by direction.
In several cases, multiple components in the 3.6~cm map cannot be spatially
separated in the 20~cm map. The total 20~cm flux of these components
is then listed opposite the first of these components and the 20~cm flux 
of the other component(s) is left blank. 

Table~3 lists relevant major axis position angles (P.A.'s), in degrees,
of various structures for all galaxies in the early-type radio sample 
that are resolved in the radio. All P.A.'s are measured north through east,
with 0{\arcdeg} $\leq$ P.A. $\leq$ 180{\arcdeg}.
In order to provide a measure of their reliability, 
each of the radio P.A.'s and the galaxy major axis P.A.'s 
are assigned a quality flag - `a' (highest quality) through
`d' (lowest quality). An explanation of the flags is given in Table~4;
a more detailed discussion on the reasons for choosing the specific limits 
for each quality flag can be found in Nagar \& Wilson (1998).
The columns of Table~3 are :
(1)~name; (2)~Seyfert type; 
(3)~P.A. in 3.6~cm map published in this paper and its quality flag;
(4)~P.A. in 20~cm  map published in this paper and its quality flag; 
(5)~adopted radio P.A., derived from column (3) and (4) or from the 
literature, and its quality flag. When curved structure is present,
the P.A. of the smallest scale extension is listed;
(6)~and (7)~P.A. of the [OIII] emission and the H${\alpha}$+[NII] emission
at a surface brightness of 2 x 10$^{-16}$ ergs cm$^{-2}$ s$^{-1}$ 
(arcsec)$^{-2}$, from
MWZ. This is the level of the lowest contour in the images of
MWZ. Typical errors are $\pm$10{\arcdeg} and values in brackets
have higher uncertainties.
(8)~P.A. of the green continuum emission at a surface
brightness of 2 x 10$^{-18}$ ergs cm$^{-2}$ s$^{-1}$ {\AA}$^{-1}$
(arcsec)$^{-2}$, from MWZ. 
This is the level of the lowest contour in the images of MWZ.
Typical errors are $\pm$5{\arcdeg} and values in brackets have larger 
uncertainties. 
(9)~P.A. of the major axis of the host galaxy, obtained as indicated
in column 10, and our adopted quality flag. It is notable that, when
the major axis of the galaxy is determined photometrically, it 
is measured at an extent which is typically $\geq$~2 times that
at which the green continuum P.A. is measured;
(10)~comments according to the key at the bottom of the table.
The r.m.s. noise and the contour levels
(in multiples of the r.m.s. noise) of each map (Figures 1--15)
are listed in Table~5. 

\notetoeditor{Please place Figures 1--15 here on consecutive pages}

\begin{figure}
\figurenum{1}
\plotfiddle{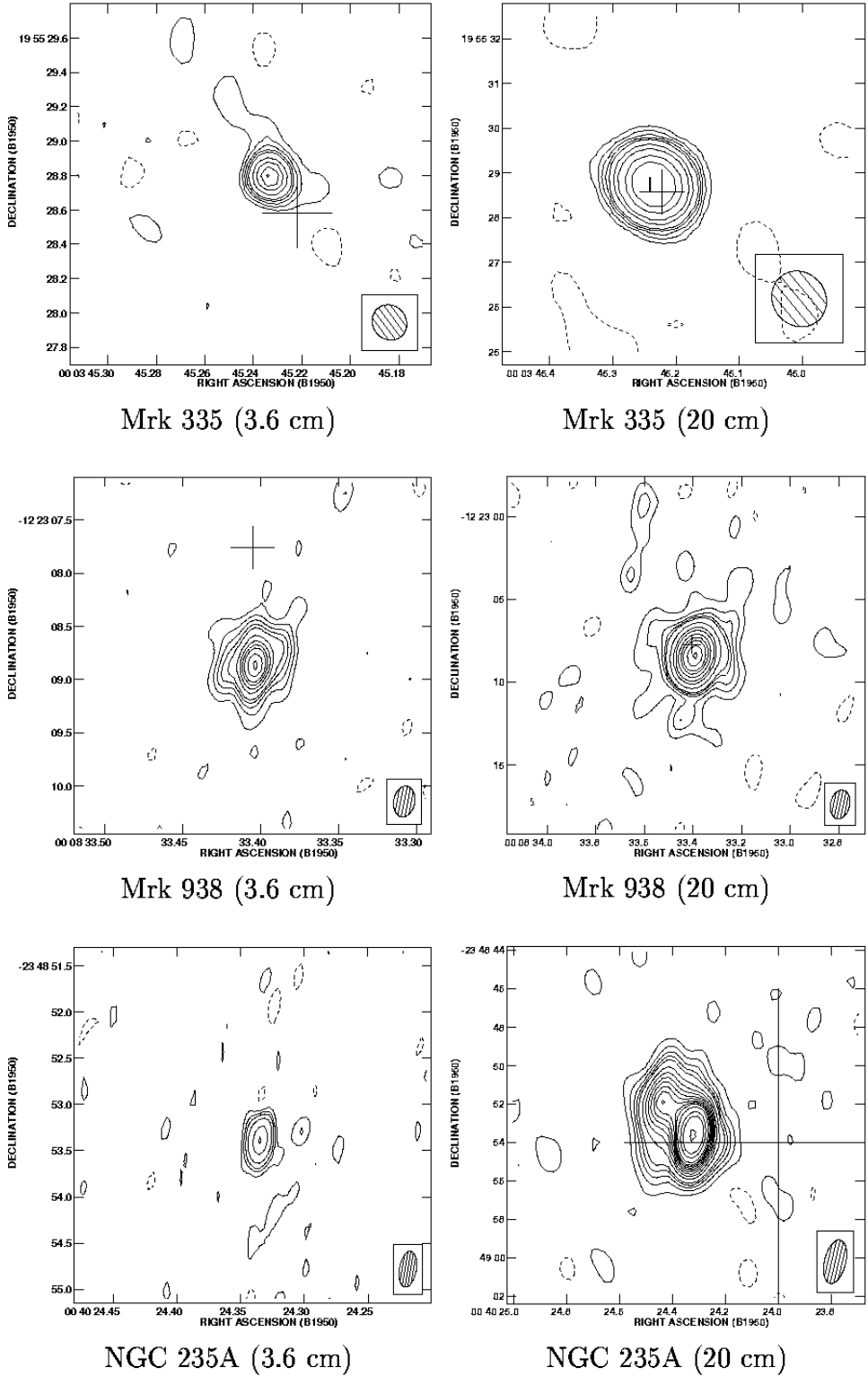}{8.5in}{0}{92}{92}{-290}{-60}
\caption{3.6~cm and 20~cm VLA maps. See Table 5 for contour levels.}
\end{figure}

\begin{figure}
\figurenum{2}
\plotfiddle{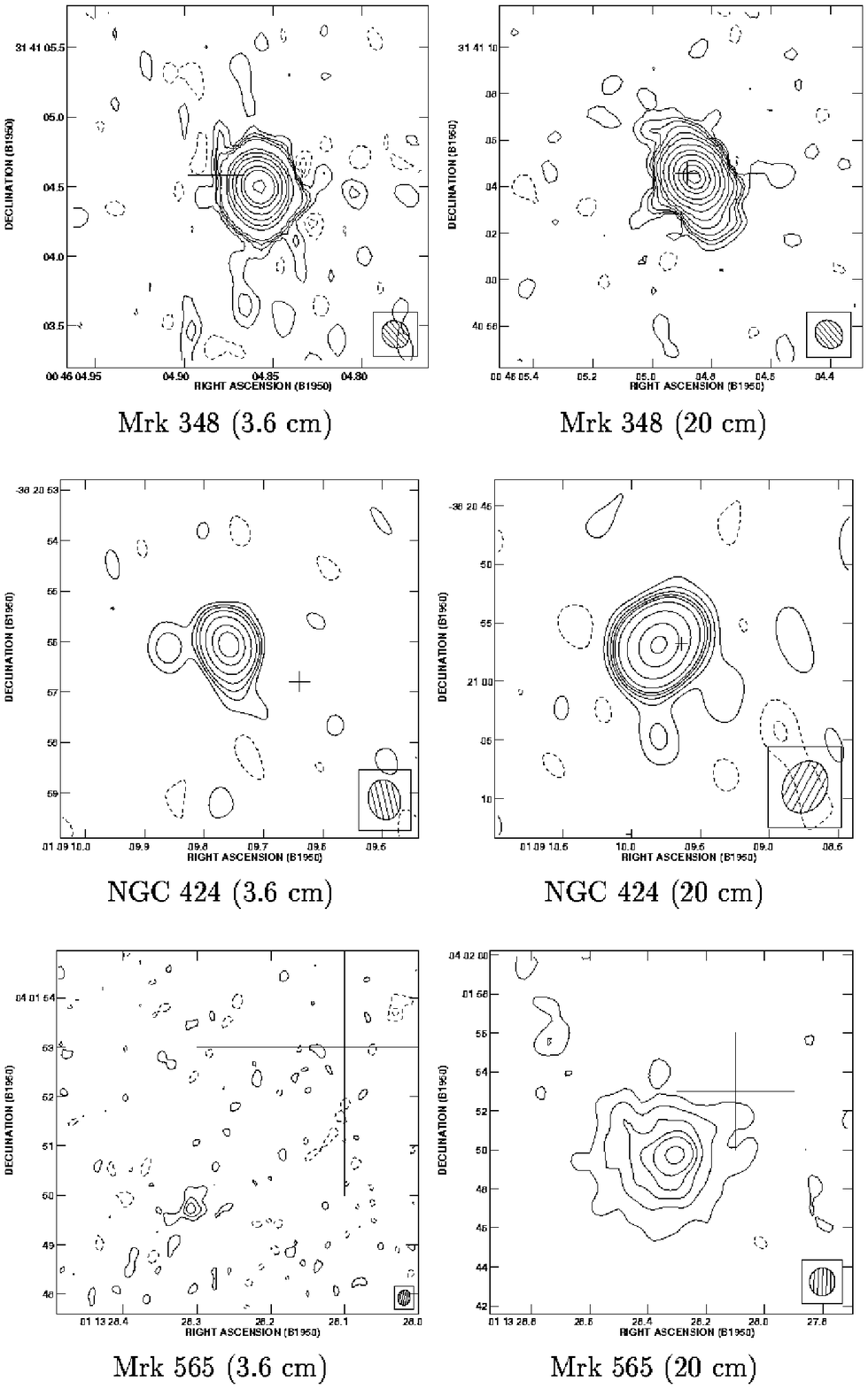}{8.5in}{0}{92}{92}{-290}{-60}
\caption{3.6~cm and 20~cm VLA maps. See Table 5 for contour levels.}
\end{figure}

\begin{figure}
\figurenum{3}
\plotfiddle{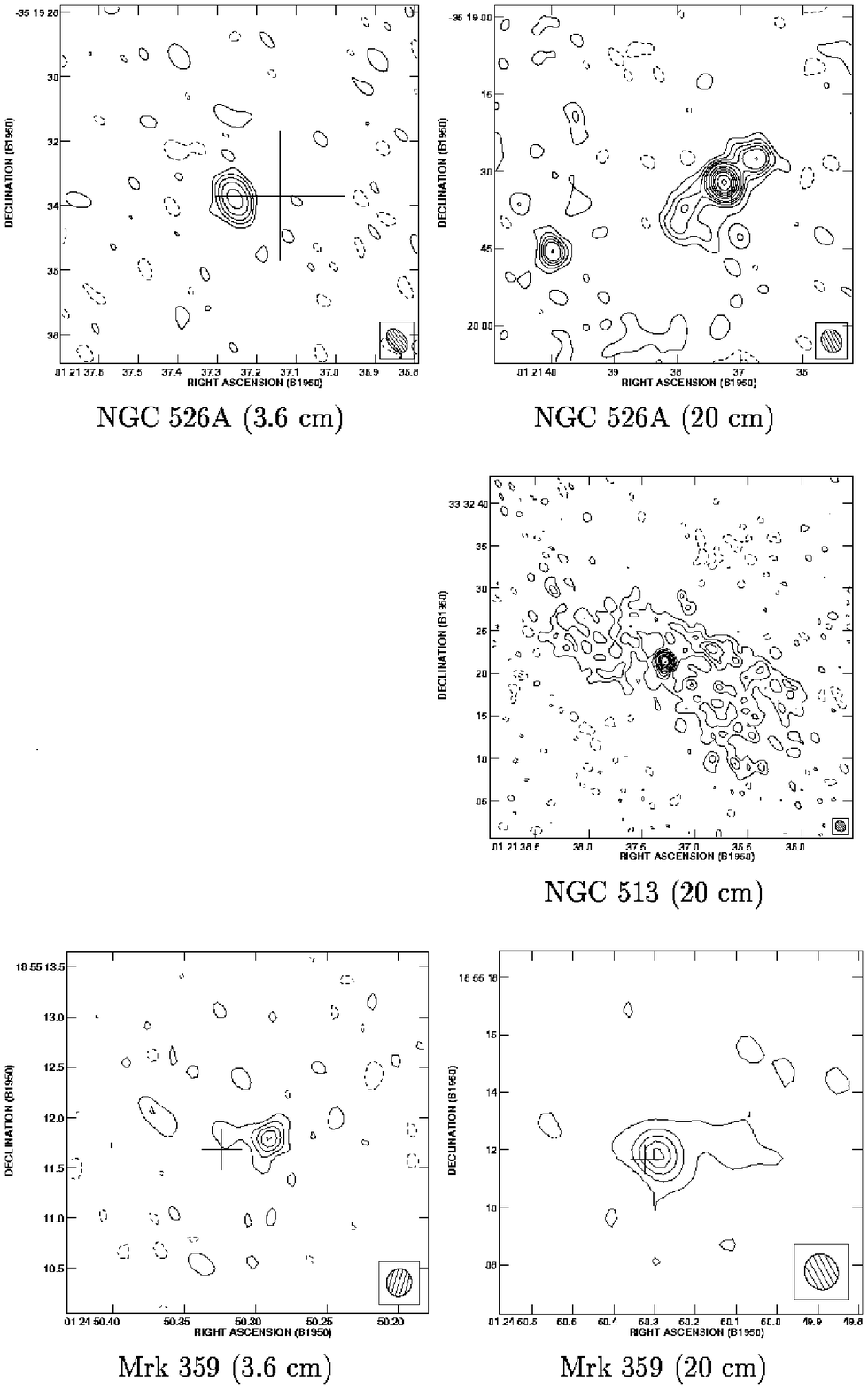}{8.5in}{0}{92}{92}{-290}{-60}
\caption{3.6~cm and 20~cm VLA maps. See Table 5 for contour levels.}
\end{figure}

\begin{figure}
\figurenum{4}
\plotfiddle{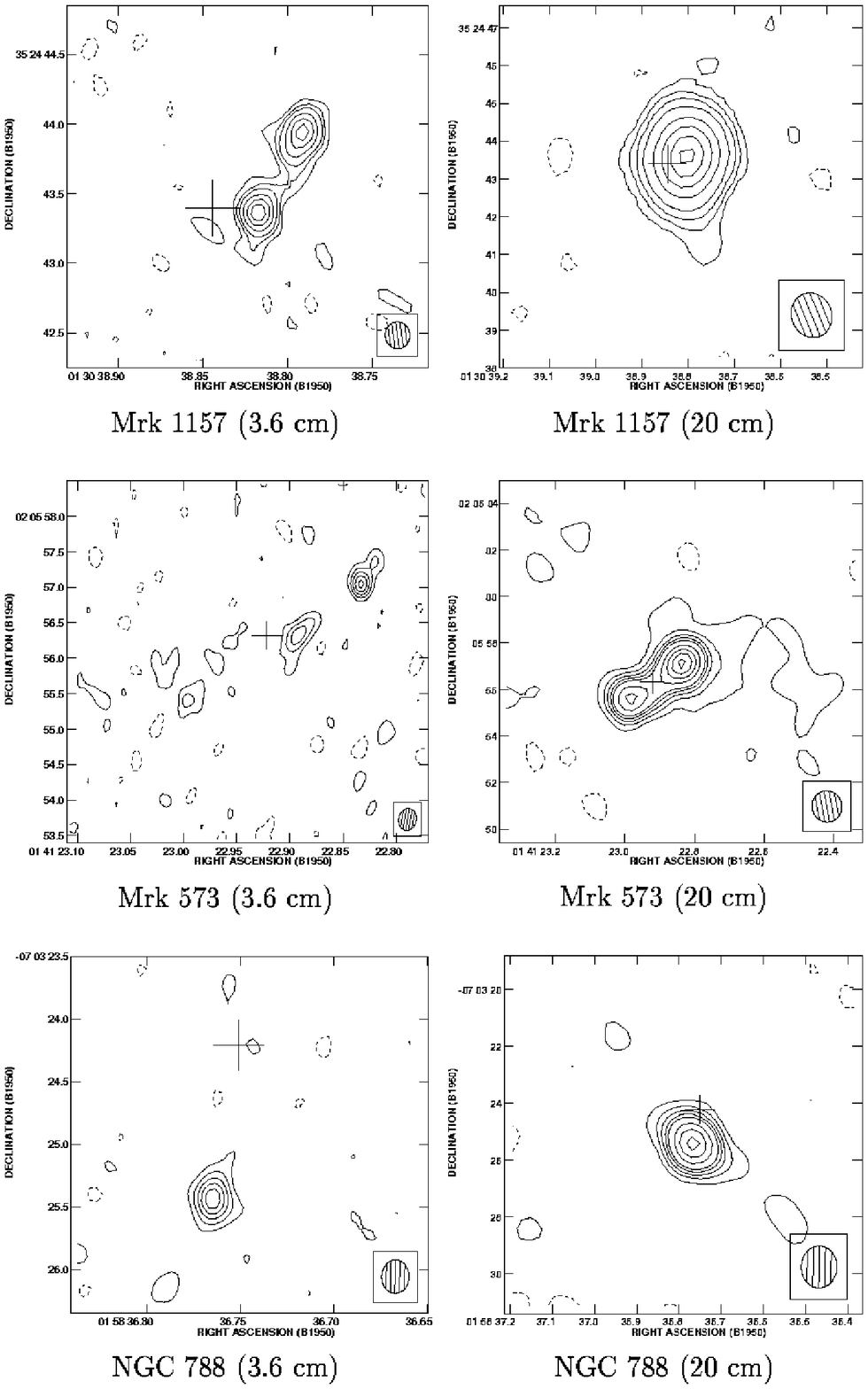}{8.5in}{0}{92}{92}{-290}{-60}
\caption{3.6~cm and 20~cm VLA maps. See Table 5 for contour levels.}
\end{figure}

\begin{figure}
\figurenum{5}
\plotfiddle{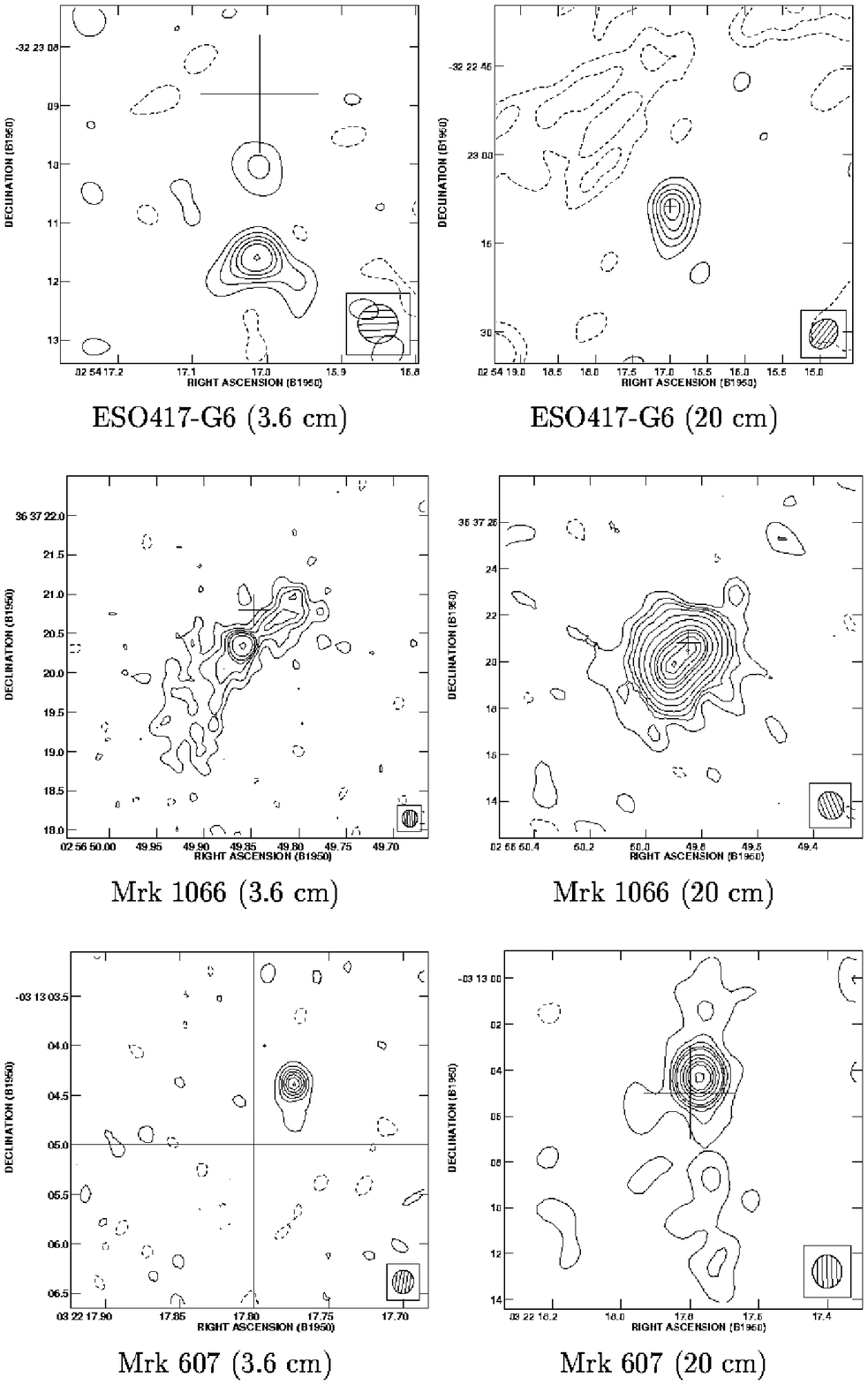}{8.5in}{0}{92}{92}{-290}{-60}
\caption{3.6~cm and 20~cm VLA maps. See Table 5 for contour levels.}
\end{figure}

\begin{figure}
\figurenum{6}
\plotfiddle{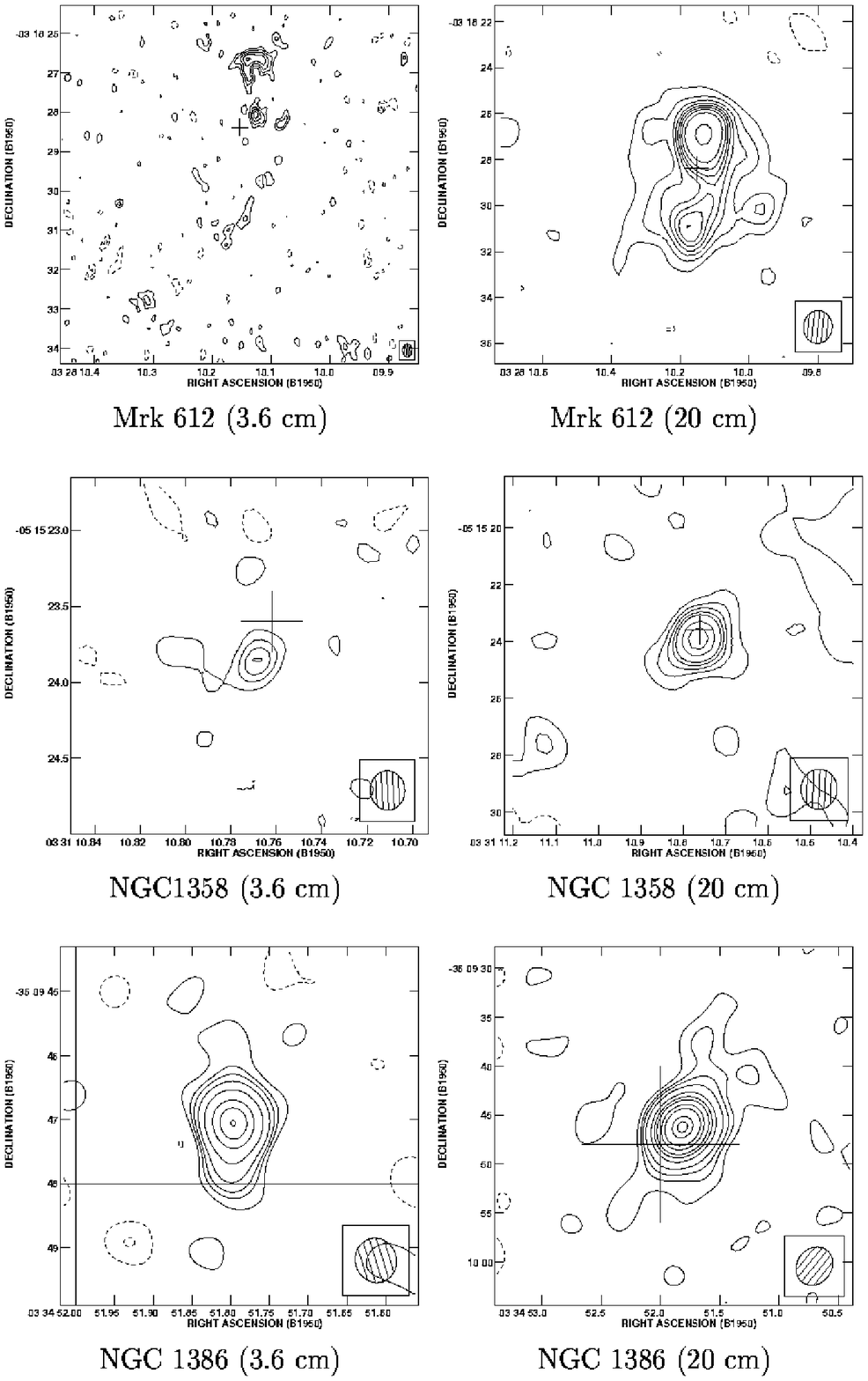}{8.5in}{0}{92}{92}{-290}{-60}
\caption{3.6~cm and 20~cm VLA maps. See Table 5 for contour levels.}
\end{figure}

\begin{figure}
\figurenum{7}
\plotfiddle{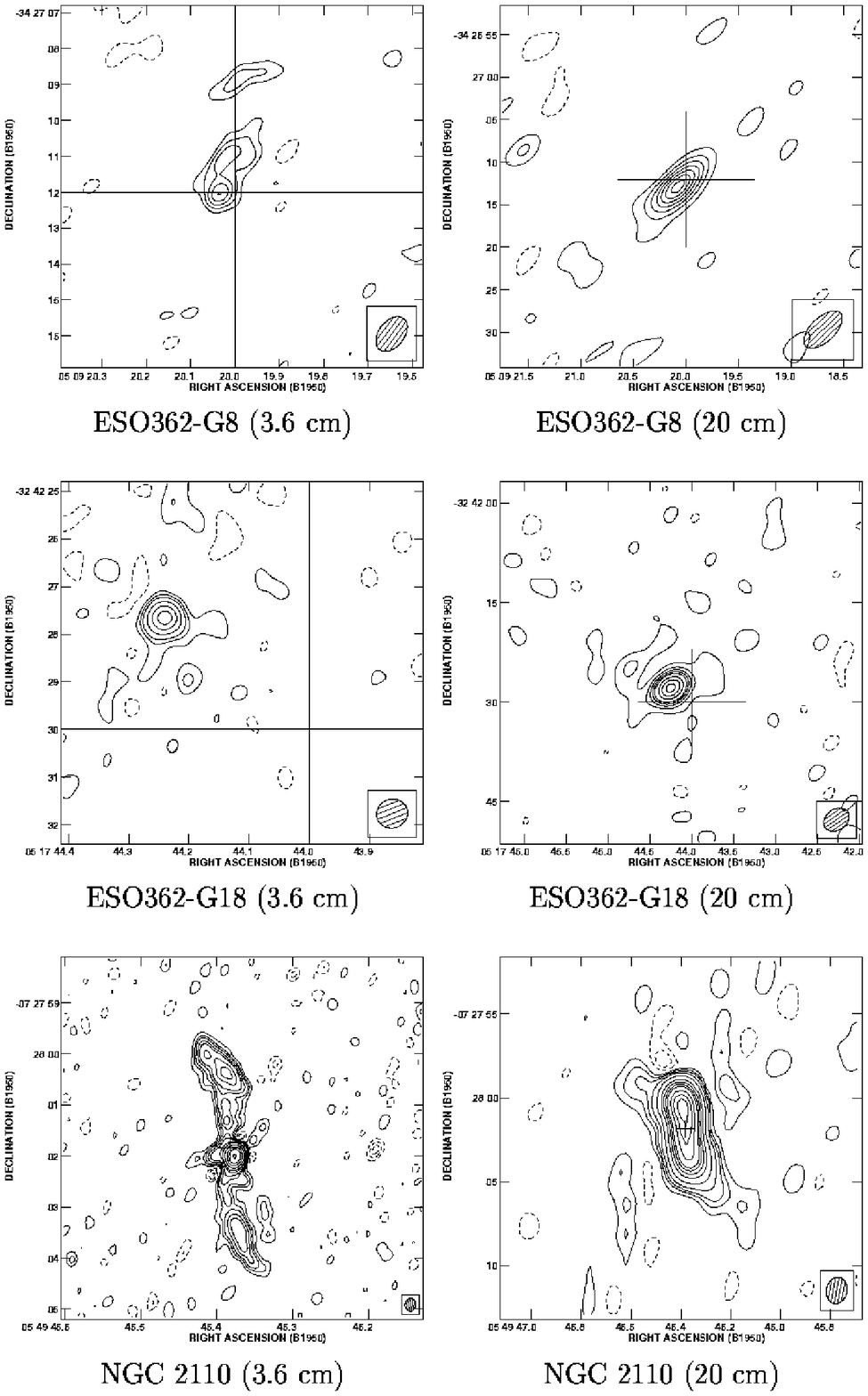}{8.5in}{0}{92}{92}{-290}{-60}
\caption{3.6~cm and 20~cm VLA maps. See Table 5 for contour levels.}
\end{figure}

\begin{figure}
\figurenum{8}
\plotfiddle{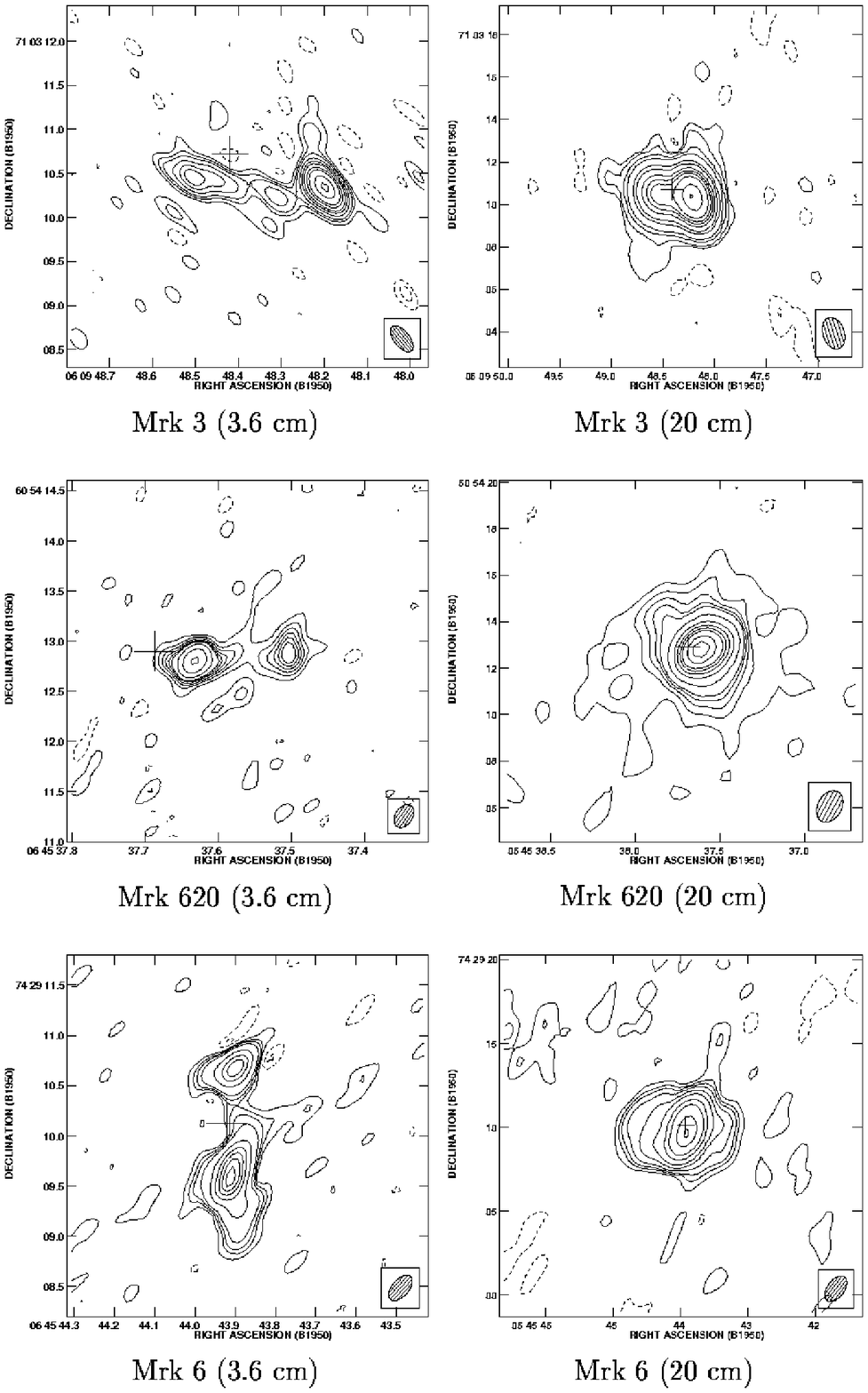}{8.5in}{0}{92}{92}{-290}{-60}
\caption{3.6~cm and 20~cm VLA maps. See Table 5 for contour levels.}
\end{figure}

\begin{figure}
\figurenum{9}
\plotfiddle{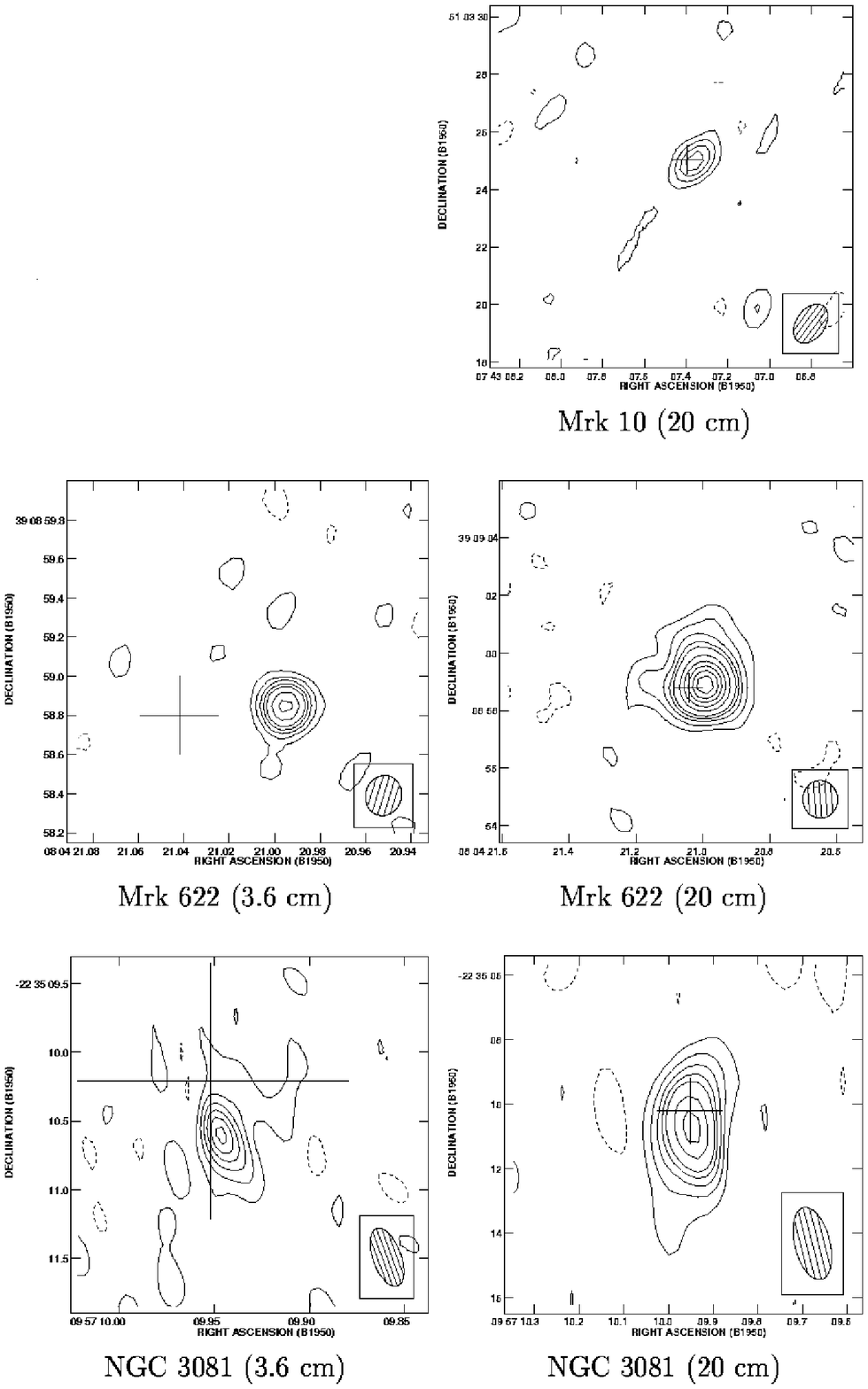}{8.5in}{0}{92}{92}{-290}{-60}
\caption{3.6~cm and 20~cm VLA maps. See Table 5 for contour levels.}
\end{figure}

\begin{figure}
\figurenum{10}
\plotfiddle{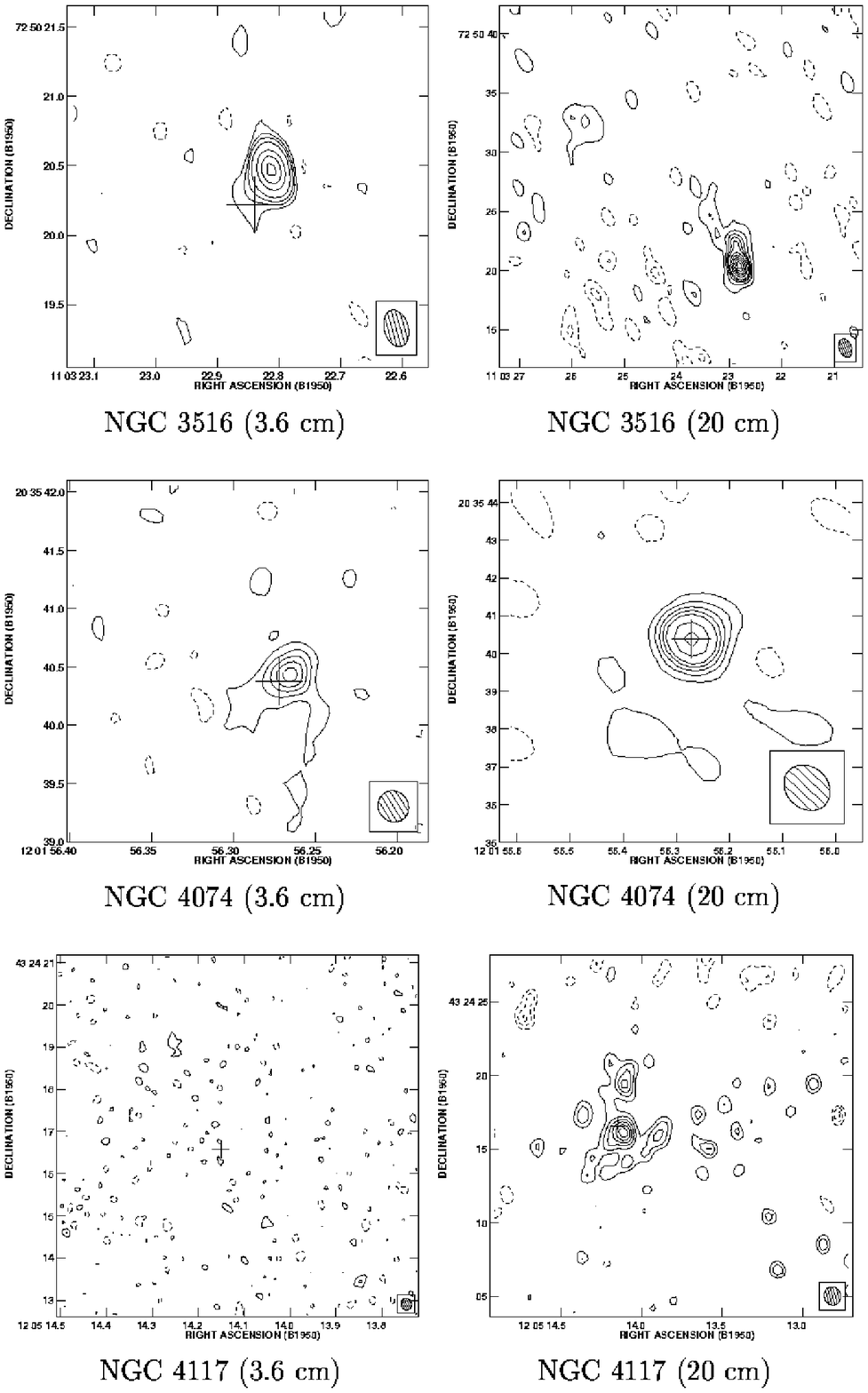}{8.5in}{0}{92}{92}{-290}{-60}
\caption{3.6~cm and 20~cm VLA maps. See Table 5 for contour levels.}
\end{figure}

\begin{figure}
\figurenum{11}
\plotfiddle{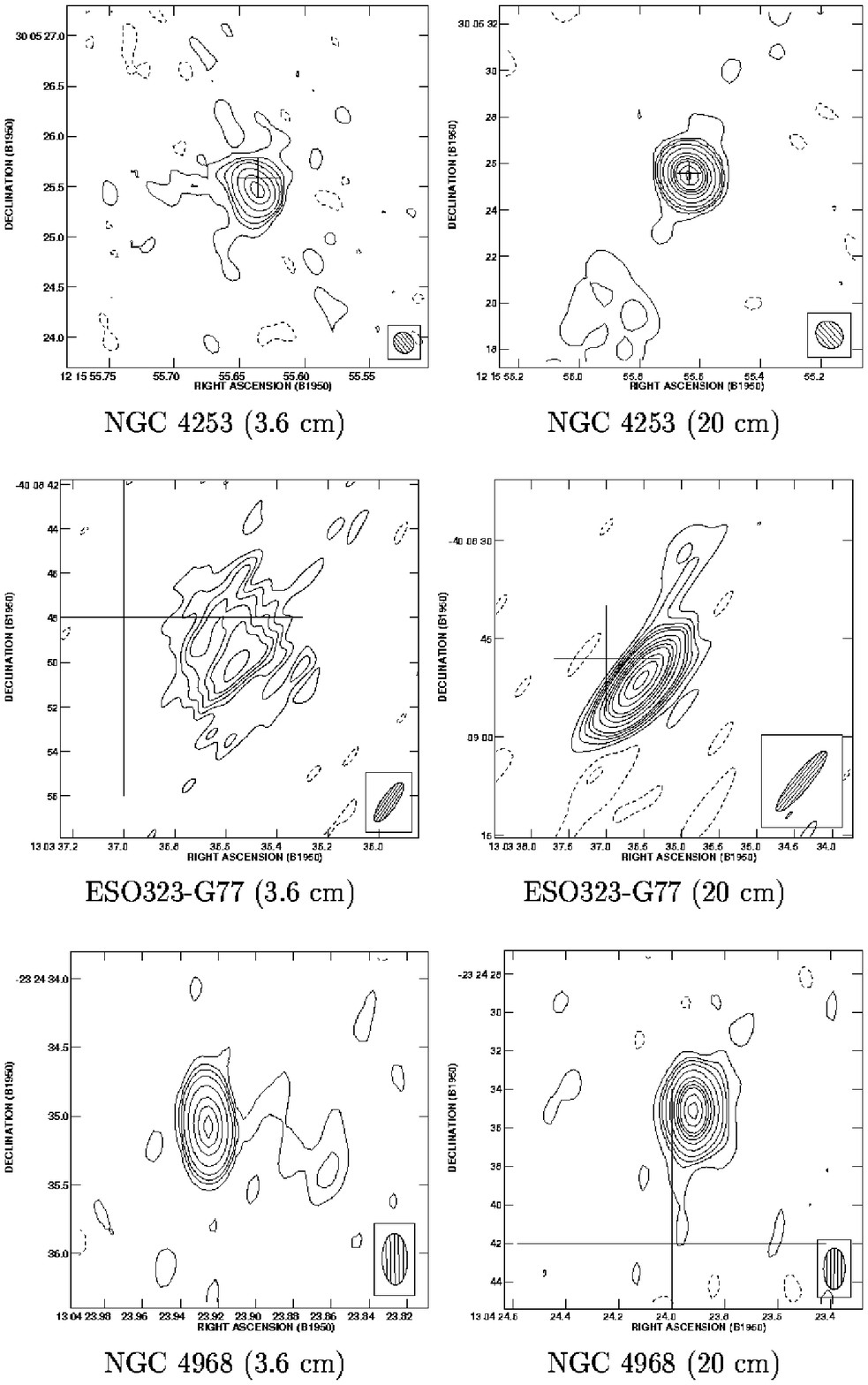}{8.5in}{0}{92}{92}{-290}{-60}
\caption{3.6~cm and 20~cm VLA maps. See Table 5 for contour levels.}
\end{figure}

\begin{figure}
\figurenum{12}
\plotfiddle{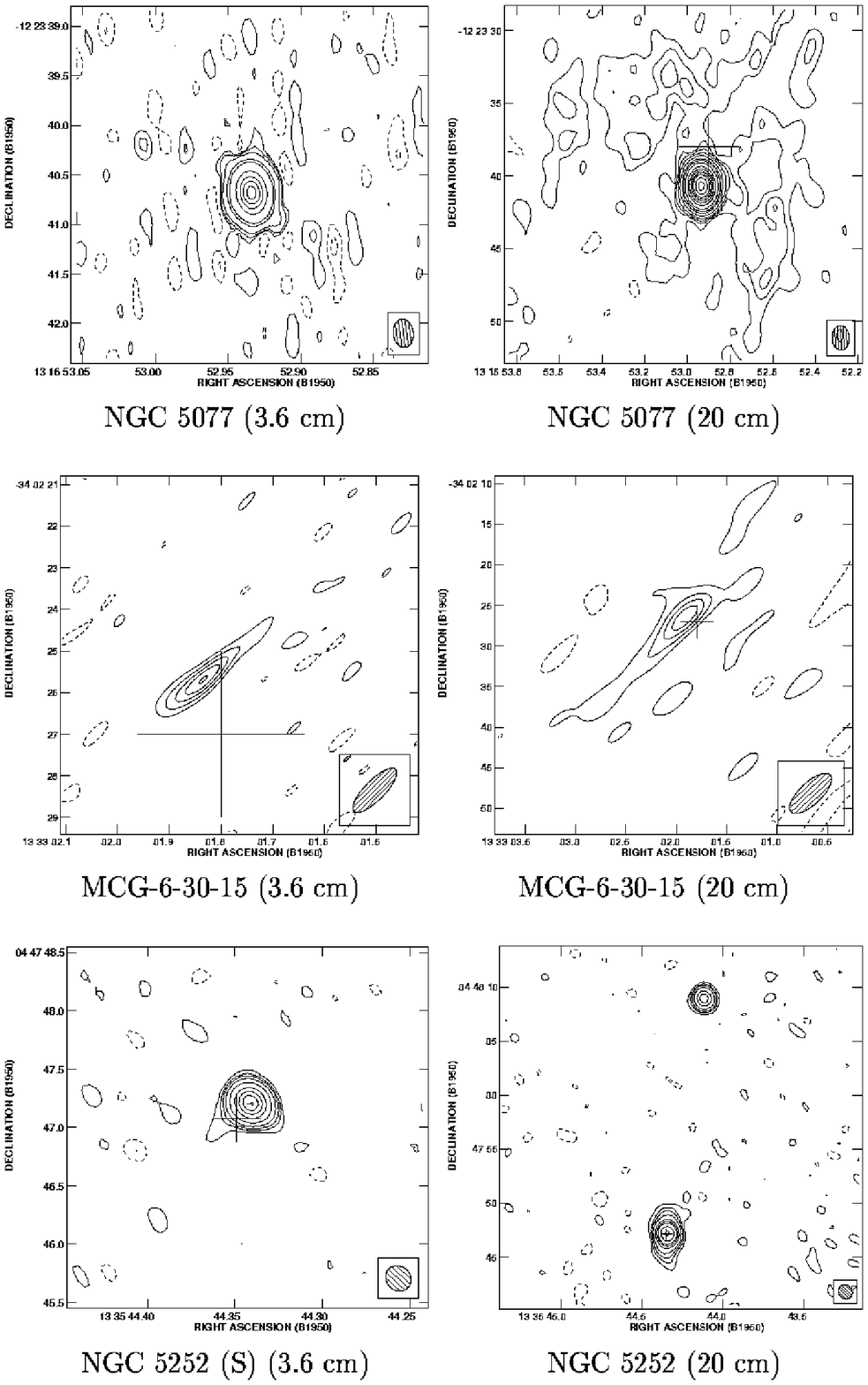}{8.5in}{0}{92}{92}{-290}{-60}
\caption{3.6~cm and 20~cm VLA maps. See Table 5 for contour levels.}
\end{figure}

\begin{figure}
\figurenum{13}
\plotfiddle{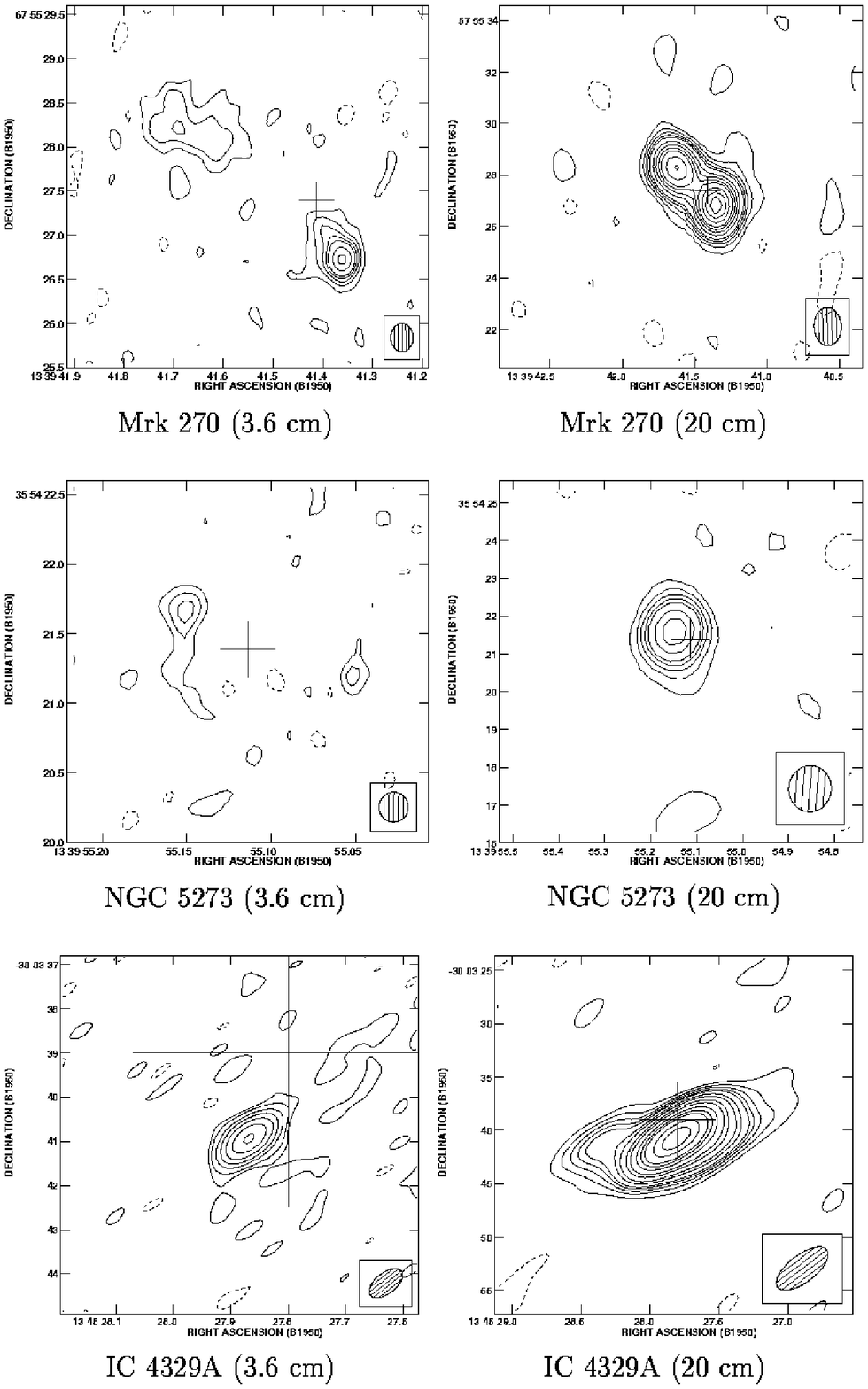}{8.5in}{0}{92}{92}{-290}{-60}
\caption{3.6~cm and 20~cm VLA maps. See Table 5 for contour levels.}
\end{figure}

\begin{figure}
\figurenum{14}
\plotfiddle{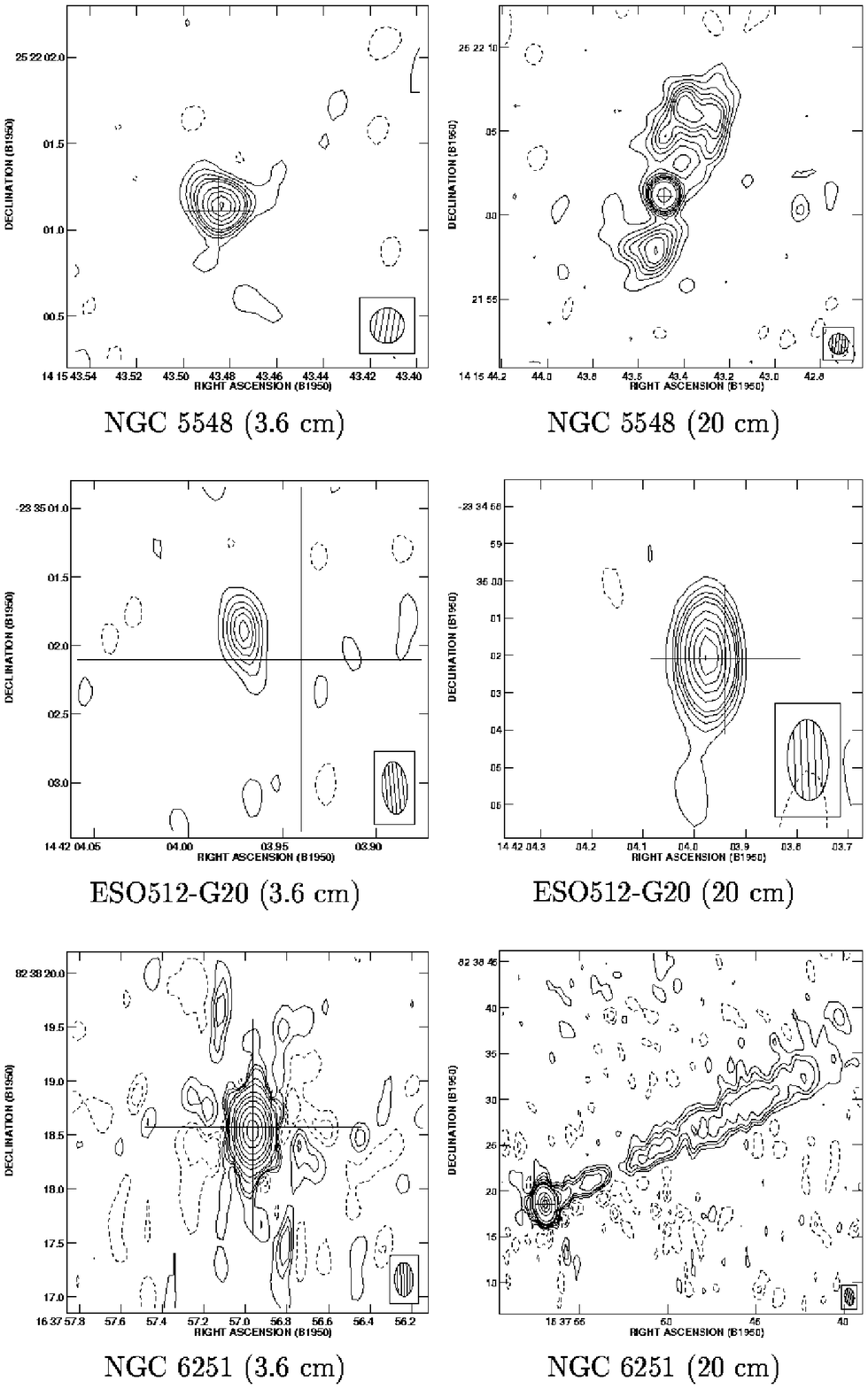}{8.5in}{0}{92}{92}{-290}{-60}
\caption{3.6~cm and 20~cm VLA maps. See Table 5 for contour levels.}
\end{figure}

\begin{figure}
\figurenum{15}
\plotfiddle{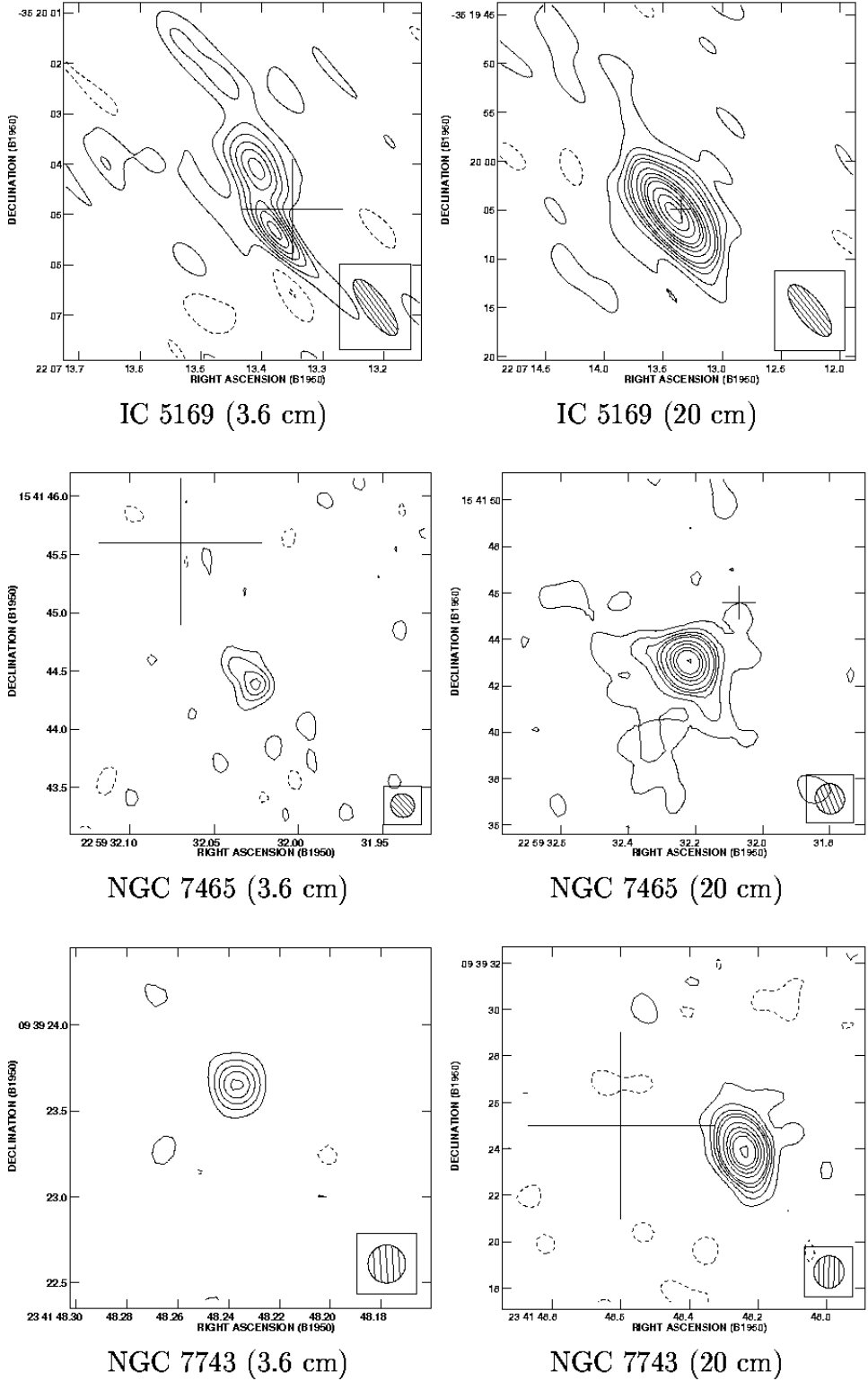}{8.5in}{0}{92}{92}{-290}{-60}
\caption{3.6~cm and 20~cm VLA maps. See Table 5 for contour levels.}
\end{figure}

\section{Notes on individual objects}
Comments on many individual galaxies are listed below. Further
comments on the optical properties of most of the sources can be found in MWZ.
A number of radio sources apparently unrelated to the Seyfert galaxy
were detected in many of the maps.  Most of these extraneous sources 
are far enough from the antenna pointing
center that they are appreciably bandwidth smeared.
We refer to the stronger of these sources as `confusing'; they are
hard to subtract and the resulting maps are noisier than usual.
A few extraneous sources appear within the $\sim$1{\farcm}2~$\times$~1{\farcm}2
field of view of the 20~cm maps centered on the Seyfert galaxy.
These extraneous sources are close enough to the antenna pointing center
that they are not bandwidth smeared in the 20~cm maps. They
have been deconvolved and mapped and  their positions and fluxes are
given in the galaxy comments.
In this section, radio P.A.'s are listed such that
0{\arcdeg}$\leq$P.A.$\leq$360{\arcdeg}. 

\paragraph{Mrk~335 (type 1.0), Fig. 1.} -
Our flux is in good agreement with  the value of 2.05 mJy at 8.4 GHz obtained
by Kukula et al. (1995). There is a confusing source $\sim$7{\arcmin} away.
Whittle (1992a) lists the host galaxy as compact and does not
list a Hubble T parameter. We have used the NED classification of 
S0/a for the host galaxy (T=0).  

\paragraph{Mrk~938 (misclassified Seyfert), Fig. 1.} - 
The 3.6~cm source is extended in P.A. 125{\arcdeg}.
Gaussian deconvolution of the 20~cm source suggests that it 
is unresolved, though the contours show a faint extension 
in a direction consistent with the 3.6~cm extension. 
The summed flux at 3.6~cm is 7.5 mJy.
There is a source (2.5  mJy at 20~cm) $\sim$1{\arcmin} south at position 
$\alpha$=00$^h$08$^m$32$^s$.69, 
$\delta$$=-$12{\arcdeg}24{\arcmin}02{\farcs}1 (B1950.0).
MWZ claim that this is not a Seyfert galaxy. They find the [OIII]
emission weak,
while the H$\alpha$+[NII] is much stronger over the entire galaxy, implying
that the ionization of the gas is not related to any Seyfert activity.
The printed edition of RC3 (1991) lists a P.A. of 30{\arcdeg} for the
galaxy major axis while the updated online version at NED does not list a P.A.
ESGC lists a P.A. of 30{\arcdeg}
and an external diameter of 2{\farcm}19 x 1{\farcm}29 in B, though
this measurement is listed as uncertain.
Garnier et al. (1996) measures a major axis P.A. of 36{\arcdeg} at an
external diameter of  1{\farcm}62 x 0{\farcm}71 in B.

\paragraph{NGC~235A (type 1?), Fig. 1.} - 
A double Gaussian fit to the 20~cm map gives fluxes of 25.5 mJy 
and 12.5  mJy. The summed flux is 38.5 mJy, consistent with
the result of the double Gaussian fit. 
The NE component is not seen in the 3.6~cm image.
There is a confusing source $\sim$2{\arcmin} away.
The Seyfert type of this galaxy is uncertain.
MWZ, the catalog of V\'eron-Cetty and V\'eron (1996) and NED list this galaxy 
as a Seyfert 1.
Keel (1996) and the catalog of Huchra (1989)
list it as a Seyfert 2. 
Maia et al. (1987) have published its spectrum, and find that it exhibits 
broad H$\alpha$ emission. They note that their classification is
preliminary because the spectrum was not corrected for
underlying stellar absorption, interstellar reddening, and blending of lines.
We therefore list this galaxy as a Seyfert~1?, and use a classification 
of Seyfert~1 for all further analysis.
The H$\alpha$+[NII] and excitation maps of MWZ show an extension in P.A.
similar to the P.A. of the radio extension. 
RC3 lists a galaxy major axis P.A. of 117{\arcdeg} for NGC~235A,
and no P.A. for NGC~235B.

\paragraph{Mrk~348 (type 2), Fig. 2.} - 
Gaussian deconvolution suggests that the bright core of the source is
unresolved in both the 3.6 and 20~cm maps. 
There appear to be faint extensions to the NNE and SSW at 20~cm.
Our measured fluxes of 238 mJy and 302 mJy at 3.6~cm and 20~cm, respectively,  
are low compared with some previous measurements;
earlier observations of this source (Paper~V) gave
fluxes of 480 and 565 mJy at 6 and 2~cm, respectively. 
VLBI observations of the source (\cite{nb83}) reveal a triple structure 
in P.A. 170{\arcdeg}. The radio source structure is therefore classified 
as `L'. The P.A. of the major axis of the galaxy is not listed in 
RC3 or UGC. Heckman et al. (1982) mapped the HI in this galaxy and
derived P.A.$\sim$0{\arcdeg} for the outer HI ring.
A deeper HI map (\cite{simpet87}) reveals a complex kinematic
structure (their Figure~3), in which the kinematic major axis appears to 
be along P.A. 170{\arcdeg}.
Falcke et al. (1998) and Capetti et al. (1996) have used high
resolution radio and optical images to investigate the detailed
correlation of radio structure and ionized gas in this galaxy.

\paragraph{NGC~424 (type 2), Fig. 2.}
The 20~cm map clearly shows an extension in a direction different
from the beam extension. 
The deconvolved source size is less than half the beam size; however
the extension, in P.A. 96{\arcdeg}, is probably real. 
The 20~cm map in Paper~VII which is at a higher resolution than 
our 20~cm map, also clearly shows an extension to the E. 
The central source in the 
3.6~cm image is unresolved. The weaker source to the east appeared during
the self-calibration process.  The P.A. joining the two sources is 
consistent with the direction of the extension in the 20~cm map.
Our 3.6~cm map is
similar to the 6~cm map of the same source in Paper~VII (where it
is listed as TOL 0109-383), in which the
weak component to the E is also seen.
We classify
this source as `S', and adopt a radio P.A. of 96{\arcdeg}.
Paper~VII commented on the flat spectrum of this source -
${\alpha}$=0.17${\pm}$0.07 
(S$_{\nu}$ ${\propto}$ ${\nu}^{-{\alpha}}$) between
6 and 20~cm. Our measured flux of 12.2 mJy at 3.6~cm confirms that the 
spectrum of the core source is flat to even shorter wavelengths.

\paragraph{Mrk~565 (misclassified Seyfert), Fig. 2.}
The 20~cm emission is diffuse. 
A Gaussian model fit  to the central peak  suggests a slight extension
in P.A. 136{\arcdeg}, in agreement with the contour map.
There is a confusing source 5{\farcm}5 away and another
very strong confusing source 25{\arcmin} away.
MWZ find that this is a misclassified Seyfert with very weak [OIII] emission.
The 20~cm morphology is similar to that seen in the [OIII] image (MWZ).

\paragraph{NGC~526A (type 1.9), Fig. 3.} 
At 20~cm NGC~526A is a triple source.
The NW component is itself slightly resolved in P.A. 113{\arcdeg}
$\pm$~5{\arcdeg}.
The distance between the peaks of the SE and NW components is 17{\arcsec}, and 
the SE component extends another few arcseconds to the SE.
Only the central source in NGC~526A is detected at 3.6~cm. 
NGC~526B, which lies some 35{\arcsec} to the SE of NGC~526A,
has also been detected at both 3.6~cm and 20~cm.
There is a confusing source $\sim$3{\arcmin} away. 
The directions of the extensions in the  [OIII], H$\alpha$+[NII], and 
excitation maps of MWZ agree with that of the triple radio source.
MWZ, the catalog of Huchra (1989) and Whittle (1992a)
list NGC~526A as a Seyfert 2. The catalog of Ve\'ron-Cetty and V\'eron (1996)
and NED list it as a Seyfert 1.5.
Winkler (1992) discusses the presence of broad H$\alpha$ and the
absence of broad H$\beta$ in their and other observations of this source. 
We therefore classify this object as a Seyfert 1.9.
The P.A. of the major axis of NGC~526 is listed in RC3 
as 112{\arcdeg} which is the P.A. between NGC~526A and NGC~526B.
We have measured a major axis P.A. of 60{\arcdeg} for NGC~526A
from the outer isophotes of a DSS image.

\paragraph{NGC~513 (type 2), Fig. 3.}
This galaxy was not observed at 3.6~cm.
The extended emission in the 20~cm map has a maximum extent of 32{\arcsec} 
in P.A. $\simeq$ 58{\arcdeg}.
A Gaussian deconvolution of  the central source indicates an extent
of 1{\farcs}1  in P.A. 
167{\arcdeg}. We use this value for the radio P.A. as we are interested
in the nuclear structure rather than the diffuse extended emission.
The morphology of the 20~cm extended emission
is similar to that seen in the [OIII] and H$\alpha$+[NII]
images (MWZ). All these images align well with the galaxy disk suggesting
an origin related to stellar processes rather than the nuclear activity.
The morphological type is listed as .S?.... in RC3, so we
do not list a  T parameter value for this galaxy in Table~1. 

\paragraph{Mrk~359 (type 1.5), Fig. 3.}
The 20~cm nuclear source sits on a plateau of faint, possibly real emission.
Gaussian deconvolution suggests the nuclear source is extended in P.A.
$\simeq$69{\arcdeg}, but the source size is less than half the beam size.
Gaussian deconvolution of the 3.6~cm source suggests that it is unresolved;
however, the contours show a faint extension in P.A.$\sim$80{\arcdeg},
more or less consistent with the P.A. in the 20~cm map. We therefore list the 
source as `(S)' and adopt an overall radio P.A. of 75{\arcdeg}.
There is a strong (800 mJy at 20~cm) double-lobed confusing source - IC~0115 -
some 10{\arcmin} away which is difficult to clean out.
Mrk~359 is a narrow line Seyfert 1 (\cite{veil91}).

\paragraph{Mrk~1157 (type 2), Fig. 4.}
The position of the 20~cm peak is listed in Table~1 and the elongated
20~cm source
includes both the 3.6~cm components.
The NW component is slightly resolved at 3.6~cm in P.A. 153{\arcdeg}.
The 20~cm flux measured here agrees with an earlier measurement
of 24.5 mJy (Paper~VII).
Combination of the present 3.6~cm fluxes with those at 6~cm reported
in Paper~VII gives spectral indices of $\simeq$ 0.9 and 0.7
for the NW and SE components, respectively.
The axis of the radio double is well aligned with the extension in the 
[OIII] image of MWZ; higher
resolution optical maps are required for a detailed comparison.

\paragraph{Mrk~573 (type 2), Fig. 4.}
The 3.6~cm map shows both 20~cm components and a third 
close to the optical nucleus. The NW and SE components are in P.A. 
309{\arcdeg} and 120{\arcdeg} relative to the central source, respectively. 
An overall radio P.A. of 125{\arcdeg} is adopted, in excellent agreement with
earlier maps (\cite{uw5}; \cite{kuket95}; \cite{fws98}).
Our measured fluxes at 3.6~cm are similar to previously 
measured 3.6~cm fluxes of 0.78, 0.63 and 0.53 mJy  
for the NW, central and SE components, respectively (Kukula et al. 1995). 
Falcke et al. (1998) use higher resolution and sensitivity radio and HST
images to investigate the detailed correlations between the 
radio structure and the ionized gas in this galaxy (see also \cite{capet96}).
There is a confusing source $\sim$9{\arcmin} away.
The P.A. of the major axis of the host galaxy has not been listed in the RC3
or UGC catalogs, as both catalogs find the galaxy to be almost circular. 
We have confirmed this using a deeper second generation DSS image.

\paragraph{Mrk~577 (misclassified Seyfert).}
This galaxy has not been detected at either 3.6~cm or 20~cm.
There is a confusing double source $\sim$7{\arcmin} away. 
The [OIII] emission is very weak in this galaxy (MWZ), so that the
radio non-detection is not unexpected.

\paragraph{NGC~788 (type 2), Fig. 4.}
A Gaussian fit to the 20~cm source indicates an 
extension in P.A. 62{\arcdeg}. Since the deconvolved source size is half
the beam size we classify this source as `(S)'.
Paper~VII has measured fluxes of 2.2 mJy and
1.2 mJy at 20~cm and 6~cm, respectively.
There is a double source (17.6 mJy at 20~cm) $\sim$80{\arcsec} from
the pointing center, at $\alpha$=01$^h$58$^m$39$^s$.67,
$\delta$=--07{\arcdeg}04{\arcmin}30{\farcs}8 (B1950.0)
and a confusing source $\sim$8{\arcmin} away. 
The P.A. of the major axis of the host galaxy is not given in RC3
(which lists log~R$_{25}$ = 0.12).  ESGC lists a major axis P.A. of 75{\arcdeg}
at an external diameter of 4{\farcm}07 x 2{\farcm}57  in B.
We measure a major axis P.A. of 125{\arcdeg}, at an extent of 
4{\farcm}1~$\times$~3{\farcm}3,  from a DSS image.

\paragraph{ESO417-G6 (type 2), Fig. 5.}
A Gaussian fit to the 20~cm source 
suggests a P.A. of 6{\arcdeg}. The closer inspection of the contour map 
shows that while the extension starts at this P.A. the lower
contours to the south show a P.A. closer to 170{\arcdeg}.
The reality of the northern source in the
3.6~cm map is doubtful, but it is consistent with the N-S elongation
seen at 20~cm. We adopt a radio classification of `S' and
an overall radio P.A. of 0{\arcdeg}.
There is a strong confusing source $\sim$11{\farcm}5 away.
ESO lists the galaxy major axis P.A. as 130{\arcdeg} but considers it 
uncertain.
RC3 also lists this P.A. as 130{\arcdeg}. DSS images show that this 
P.A. is probably affected by a companion object; 
the galaxy itself is quite circular.

\paragraph{Mrk~1066 (type 2), Fig. 5.}
A single Gaussian model fit to the 20~cm source indicates a P.A. of 
137{\arcdeg}.
In the 3.6~cm image, the central source has a flux of 4.8 mJy.  The  NW
extension is at P.A. $\simeq$ 305{\arcdeg} w.r.t. the core and has a
 summed flux of 5.1 mJy. The SE
extension is at P.A. $\simeq$ 140{\arcdeg} w.r.t to the core; its summed
flux of 6.5 mJy is less certain as the emission is more diffuse. 
We  adopt a P.A. of 130{\arcdeg} for the 3.6~cm image.
Our 20~cm flux agrees with the measurement of 94.3~mJy in Paper~VII.
The axis of the radio source is similar to that of the [OIII]
emission and the extent of the radio emission is similar to the
extent of the brighter parts of the [OIII] image.
Bower et al. (1995) have used the same 3.6~cm  radio data 
in a detailed comparison with HST emission-line images and ground-based
optical spectroscopic data.

\paragraph{Mrk~607 (type 2), Fig. 5.}
The extension of the lowest contour in the 3.6~cm map is probably noise.
The N-S extension in the 20~cm map is also probably not real.
There are three confusing sources, the closest being 500{\arcsec} away.

\paragraph{Mrk~612 (type 2), Fig. 6.}
The  northern source in the 20~cm map is extended 
in P.A. $\simeq$10{\arcdeg} and the peak to peak 
P.A. of the two components is 170{\arcdeg}.
The 3.6~cm image consists of a knot close to the optical nucleus,
emission to the north elongated 
perpendicular to the source axis, suggestive of a transverse shock,
and faint emission to the S.
The peak to peak P.A. of the two strongest components in the 
3.6~cm image is 10{\arcdeg}, in agreement with the extension of the 
northern component at 20~cm.
The southern component of the 20~cm image is seen very faintly in the 
3.6~cm map as a few weak knots with a total flux between 0.2 and 0.7~mJy.  
The overall radio P.A. of the object is taken as 10{\arcdeg}.
There is a confusing source $\sim$9{\arcmin} away.
The P.A. of the major axis of the host galaxy is listed in the printed
edition of RC3 (1991) as 95{\arcdeg} but is not listed in the
updated online version available at NED.
A second generation DSS image shows that the outer isophotes of 
the galaxy are ill-defined.

\paragraph{NGC~1358 (type 2), Fig. 6.}
A Gaussian deconvolution of the 20~cm source  suggests a small extension in
P.A. between 116{\arcdeg} and 138{\arcdeg}, consistent with the 
extension seen in the lower contours.
It is not certain if the faint extension seen in the 3.6~cm map is real. 
We classify the source as `(S)' with a 20~cm (and overall radio)
P.A. of  120{\arcdeg}. 
Paper~VII measured a 20~cm flux of 3.8 mJy (in agreement with the present
flux) and a 6~cm flux of 1.2 mJy.  
There is a weak confusing source 6{\arcmin} away and a stronger confusing
source 13{\farcm}5 away, both of which do not clean out very well.
ESGC lists galaxy major axis P.A. = 165{\arcdeg}
at an external diameter of  4{\farcm}07 x 3{\farcm}08 in B.
The printed edition of RC3 (1991) lists major axis P.A. = 165{\arcdeg} and
log~R$_{25}$ = 0.1. However the updated online version of RC3 
(available at NED) does not list a P.A. Second generation DSS images show
that the galaxy is very close to circular.

\paragraph{NGC~1386 (type 2), Fig. 6.}
There is a clear extension in the lower contours of the 3.6~cm map in
P.A.  $\sim$170{\arcdeg}. 
This extension is present before  the self-calibration
process and remains after multiple iterations with self-calibration.
The source is therefore listed as `S' with P.A. $\simeq$ 170{\arcdeg}.
This value is different from the P.A.
of $-$125{\arcdeg} reported in Paper~VI from a 6~cm map; 
we believe the present 3.6~cm map is more reliable.
There is a strong confusing source $\sim$16{\arcmin} away. The 20~cm
flux of 28.8 mJy measured here is marginally higher  than the value of 
23.0$\pm$2.0 mJy  published in Paper~VI.

\paragraph{ESO362-G8 (type 2), Fig. 7.}
Gaussian deconvolution of  the 20~cm source suggests that it is unresolved. 
The 3.6~cm map is noisy but shows a faint extension $\simeq$2{\arcsec}
to the north. The detached component 3{\arcsec} N of the peak in the 
3.6~cm map may not be real.
The P.A. between the two peaks separated by 1{\arcsec} in the 3.6~cm map
is 170{\arcdeg}, but since there is significant emission  to the west
of the northern peak 
we adopt a 3.6~cm P.A. of 165{\arcdeg}.
There are two confusing sources at 3{\arcmin} and $\sim$9{\arcmin}, 
and another weak source 2{\farcm}5 away.
The radio extension is in a similar direction to the extension seen
on  a similar spatial scale in the [OIII] and excitation maps of MWZ.

\paragraph{ESO362-G18 (type 1.5), Fig. 7.}
A Gaussian fit to the 20~cm source indicates 
a deconvolved size of about half the beam size and  
a P.A. in the direction of the beam; this apparent extension is
probably not real.  
There is a double-lobed confusing source $\sim$5{\arcmin} away.

\paragraph{NGC~2110 (type 2), Fig. 7.}
A single component Gaussian fit to the 20~cm source indicates a P.A. of
9{\arcdeg}. The P.A. of the ridge line of the 3.6~cm jets is also 9{\arcdeg}. 
A Gaussian deconvolution of  the central peak in the 3.6~cm map
gives a flux of
77.6 mJy. The total summed flux in the central peak is 81.2 mJy
(the difference of 3.6 mJy is probably due to the small eastern extension).
Radio maps of this  source have been previously published by
Ulvestad \& Wilson (1983) and in Paper~VI.
H$\alpha$+[NII] and [OIII] images have been published
by Mulchaey et al. (1994) and Wilson, Baldwin \& Ulvestad (1985) and
the relationship between the radio and emission-line structures has been
discussed by Mulchaey et al. (1994).
The P.A. of the major axis of the host galaxy is not listed in the RC3
or UGC catalogs. ESGC lists a major axis P.A. of 20{\arcdeg}  at
an external diameter of 2{\farcm}45 x 1{\farcm}86 in B.
Wilson \& Baldwin (1985) find that the kinematic axis
of the inner $\sim$10{\arcsec} of the galaxy is 161{\arcdeg}$\pm$2{\arcdeg}
while the photometric axis of the inner $\sim$30{\arcsec} of the
galaxy is 163{\arcdeg}$\pm$3{\arcdeg}.

\paragraph{Mrk~3 (type 2), Fig. 8.}
The 3.6~cm map shows a triple structure in a direction consistent with 
the extension in the 20~cm map. 
The 3.6~cm flux of the central component (listed as `E1' in Table~2)
has been measured by both
Gaussian deconvolution and flux summation and found to be 7.5$\pm$1 mJy.
The images are similar to those published in Paper~V; higher resolution
maps are given by Kukula et al. (1993).  Capetti et al. (1996)  provide 
a detailed discussion of the correlation of
radio and emission-line structure in this galaxy.
RC3 lists a log~R$_{25}$ of 0.06 and does not list a major axis P.A. 
Thompson \& Martin (1988) used enlarged Sky Survey prints
to measure a major axis P.A. of 
15{\arcdeg} for the brighter isophotes. A DSS image shows that the outer
isophotes of this galaxy are close to circular.

\paragraph{Mrk~620 (type 2), Fig. 8.}
After Gaussian deconvolution, the core of the 20~cm map is elongated
$\simeq$1{\farcs}1 FWHM in P.A. 95{\arcdeg}, which agrees with
the axis of the double source seen at 3.6~cm. The 20~cm contours to the NE
suggest a larger scale extension in P.A. $\simeq$40{\arcdeg},
similar to the elongation of the H$\alpha$+[NII] image of MWZ.

\paragraph{Mrk~6 (type 1.5), Fig. 8.}
The central source in the 20~cm map is extended in P.A. 177{\arcdeg},
in agreement with the P.A. found in \cite{uw5}.
The larger scale structure in the 20~cm map (listed as `Ext' in Table~2) 
has P.A.~$\simeq$~75{\arcdeg}, in agreement with the 
P.A.~$\simeq$~87{\arcdeg} observed by Baum et al. (1993).
At 3.6~cm, the source is a triple with a roughly N-S axis.
The weak southernmost extension is labelled SW in Table~2. 
Higher resolution radio maps are published in Kukula et al. (1996).

\paragraph{Mrk~10 (type 1.2), Fig. 9.}
This galaxy was observed at only 20~cm.
There is a 9 mJy double source in the same field at
$\alpha$=07$^h$43$^m$15$^s$.31, 
$\delta$=61{\arcdeg}04{\arcmin}52{\farcs}9 (B1950).
There is also a strong confusing source 4{\farcm}5 away.

\paragraph{Mrk~622 (type 2), Fig. 9.}
The 20~cm map has large noise stripes running across it due to 
confusing sources. Gaussian deconvolution of the 20~cm source 
suggests an extension in P.A. 0{\arcdeg}.
There are two nearby confusing sources  $\sim$2{\farcm}2 and 2{\farcm}5 away.
Garnier et al. (1996) measure an external diameter of 
0{\farcm}81 x 0{\farcm}35 in B and a galaxy major axis P.A. of 91{\arcdeg},
while RC3 lists log~R$_{25}$ = 0.03 and does not list a major axis P.A. 
A first generation DSS image confirms that the galaxy is nearly circular.

\paragraph{MCG$-$5-23-16 (type 2).}
Paper~VI lists this source as `S' with an extent of 75 pc but does not
list the P.A. of the extension, or publish a map of the source. 

\paragraph{Mrk~1239 (type 1.5).}
Radio maps of this source have been published in Ulvestad et al. (1995).
The host galaxy type is listed as `E-S0' in NED, while  
Whittle (1992a) lists the host galaxy type as `compact'.

\paragraph{NGC~3081 (type 2), Fig. 9.}
The 20~cm map shows an extension to the NW and a Gaussian model 
fit indicates a P.A. of 158{\arcdeg}. However, the deconvolved source size is 
smaller than the beam size and the map has noise stripes
along P.A. $\sim$158{\arcdeg}.
Gaussian deconvolution of the 3.6~cm source suggests a P.A. of 170{\arcdeg},
roughly consistent with the P.A. in the 20~cm image. Like the
20~cm map, the 3.6~cm map has noise stripes along P.A. $\sim$158{\arcdeg} 
and the deconvolved source size is slightly smaller than the beam size.
Since the 3.6 and 20~cm sources display a consistent P.A. of extension
we list this source as `(S)'. The 20~cm flux of 3.5~mJy is higher than the
value of 2.5~mJy given in Paper~VII.
There is a confusing source $\sim$10{\arcmin} away, and a very strong
confusing source more than 20{\arcmin} away. These do not clean out well and 
contribute noise to the final maps.
The optical morphology of the host galaxy is known to be complex
(Storchi-Bergmann et al. 1996).
The major axis P.A. is listed as 158{\arcdeg} in
the ESO catalog, but this determination is listed as uncertain.
RC3 also lists a P.A. of 158{\arcdeg}, while 
the ESO Surface Photometry Catalog lists a P.A. of 74{\arcdeg}.
Buta \& Purcell (1998) have measured a photometric galaxy major axis P.A. of 
71{\arcdeg}, a photometric inclination of 34{\arcdeg}, a kinematic major
axis P.A. of 97{\arcdeg} and a kinematic inclination of 39--48{\arcdeg}.
They argue that the kinematic major axis P.A. and the photometric inclination
are more reliable.
Pogge (1989) has published H$\alpha$+[NII] and [OIII] images of this object.

\paragraph{MCG--2-27-9 (type 2).}
An error caused the pointing center to be off by $\sim$85{\arcsec} so 
we do not publish these maps.
Colbert et al. (1996) have published a lower resolution radio map
of this source.

\paragraph{NGC~3516 (type 1.2), Fig. 10.}
The 20~cm map shows an extension to the north in P.A. $\sim$ 8{\arcdeg}, but
the 3.6~cm map shows only the unresolved nucleus.
There is a strong confusing source $\sim$4{\arcmin} away 
which is very difficult to clean out. 
Miyaji, Wilson \& P\'erez-Fournon (1992) present deeper radio images of
this source which
reveal a P.A. of 10{\arcdeg} for the central source and a larger scale
northern extension, the P.A. of which increases from 10{\arcdeg} to 20{\arcdeg} 
with increasing distance from the nucleus. We measure a radio extent of
0{\farcs}65 (0.17 kpc) in our 20~cm map but use an extent of 4 kpc from 
Miyaji et al. (1992).
Miyaji et al. (1992) measured  fluxes at 20~cm of 4.31 and 2.17 mJy for the 
N and S components of the central source, respectively. 
We do not completely resolve these two components but our total flux of
9.4 mJy is significantly higher than the sum of Miyaji et al.'s (1992)
fluxes.
UGC lists a host galaxy major axis P.A. of 55{\arcdeg}, a blue
size of 2{\farcm}1~$\times$~1{\farcm}8 and a red size of 
2{\farcm}3~$\times$~2{\farcm}0.
However, RC3 does not list a P.A. for the major axis.
Thompson \& Martin (1988) used enlarged Sky Survey prints to derive a P.A. 
of 55{\arcdeg} for the brighter isophotes, but list this as an unsure 
determination.
We used a POSS-E Red plate from DSS to measure a P.A. of  55{\arcdeg}.
Arribas et al. (1997) find the stellar-kinematic line of nodes to be 
53{\arcdeg}$\pm$5{\arcdeg} over the inner $\sim$10{\arcsec} of the galaxy.
Since the kinematic and photometric axes are in close agreement
we use a major axis P.A. of 55{\arcdeg} with a quality flag of `a'.
Ho et al. (1997) use the criterion of Whittle (1992a) to classify this
object as a Seyfert 1.2, which agrees with the classification in
Whittle (1992a).

\paragraph{NGC~4074 (type 2), Fig. 10.}
A Gaussian deconvolution of the 20~cm source suggests that it is unresolved,
but the contours appear to have an extension in P.A. $\approx$ $-$45{\arcdeg}.
The source is weak at 3.6~cm but a Gaussian deconvolution suggests
an elongation in P.A. 131{\arcdeg}.
The extension in the  3.6~cm source appears in
both uniformly and naturally weighted maps and is in the same direction
as the extension seen in the 20~cm map. We therefore list this source
as `S', with a radio P.A. of 131{\arcdeg}.
There is a confusing source $\sim$6{\farcm}7 away.
This galaxy is not listed in RC3, UGC or ESO.
We used second generation DSS images to measure a major axis P.A. of
127{\arcdeg} at an extent of 1{\farcm}1~$\times$~0{\farcm}9.

\paragraph{NGC~4117 (type 2), Fig. 10.}
There is a large confusing `stripe' on the 20~cm map
but no confusing source could be found within a radius of a degree.
The maps are therefore noisy. The source is quite weak and the 
structure would normally be considered unreliable,
but the 20~cm map of \cite{uw7} shows a similar N-S extension.
We therefore classify the radio morphology as `(L)'
at 20~cm in P.A. 177{\arcdeg}. 
The galaxy is not detected at 3.6~cm. Paper~VII lists 
a 20~cm flux of 2.8 mJy for this object, which is marginally higher 
than the present value, and a flux upper limit of 0.6 mJy at 6~cm, so that
our non-detection at 3.6~cm is not surprising.

\paragraph{NGC~4253 (type 1.5), Fig. 11.}
The 20~cm source appears more circular than the beam and a Gaussian
deconvolution indicates a P.A. of 169{\arcdeg}.
However, this P.A. is in the direction of a sidelobe
pattern and the source size is only
1.1 times the beam size, so the reality of the apparent extent is suspect.
The lower contours on the 
3.6~cm source exhibit extensions both in P.A. $\sim$160{\arcdeg}
(in agreement with the direction of the possible extension at 20~cm) and 
P.A. $\sim$ 55{\arcdeg}. A Gaussian deconvolution 
indicates an extension in P.A. 32{\arcdeg} for the brighter isophotes.
This source has also been observed with the VLA in `A configuration'
at 3.6  and 6~cm by Ulvestad et al. (1995).
They find the source to have an extension in P.A. 22{\arcdeg}$\pm$4{\arcdeg}
at 3.6~cm (with a weaker extension in  P.A. --30{\arcdeg})
and in P.A. 12{\arcdeg}$\pm$5{\arcdeg} at 6~cm.
Ulvestad \& Wilson (Paper~V) observed this source at 6~cm and derived
a P.A. of 16{\arcdeg}. Kukula et al. (1995) observed this source
at 3.6~cm with similar resolution to ours and find an extension in P.A. 
27{\arcdeg}.
The source therefore appears to be extended in both P.A. $\sim$27{\arcdeg}
(on a scale of 0{\farcs}25) and P.A. 160{\arcdeg} (on a scale of 0{\farcs}3). 
We classify this source as `S', and adopt radio P.A.~$\simeq$~27{\arcdeg}
as representative of the source axis on the smallest scales.
The 20~cm flux measured  here is consistent with the value of 
36.4 mJy published in Paper~VII.  UGC gives a host galaxy diameter of
0{\farcm}89 x 0{\farcm}89 in B  and 0{\farcm}89 x 0{\farcm}79 in R. 
Takase et al. (1987) find a galaxy diameter of 0{\farcm}9 x 0{\farcm}7.
MacKenty (1990) finds the galaxy to have a major axis P.A. of 69{\arcdeg}
(at a major axis extent of 41{\arcsec}), and a minor to major 
axis ratio (b/a) of 0.89. 
We used second generation DSS images to measure a P.A. of 60{\arcdeg}
at an extent of 1{\farcm}3~$\times$~1{\farcm}15.

\paragraph{ESO323-G77 (type 1.2), Fig. 11.}
Both the 20~cm and 3.6~cm maps of this southern source are noisy.
The 20~cm source is  extended in P.A.
24{\arcdeg} $\pm$ 15{\arcdeg}, in agreement with the extension  at 3.6~cm.
A naturally weighted image at 3.6~cm 
shows a double source in P.A. $\simeq$ 40{\arcdeg}.
There is a confusing source $\sim$ 3{\farcm}7 away.

\paragraph{NGC~5077 (misclassified Seyfert; Liner type 1.9), Fig. 12.}
Gaussian deconvolution shows that the galaxy is unresolved
at both 3.6 and 20~cm. The source has a flat spectrum
between 3.6 and 20~cm. 
Previous high resolution VLA observations of this source (\cite{wh84})
give fluxes of 109, 90 and 111 mJy at 
1.5, 4.9 and 15 GHz, respectively; the source was unresolved in high
resolution maps at all three frequencies.
The large scale features in the 20~cm self-calibrated map were not
seen by Wrobel \& Heeschen (1984) but may be real
considering the morphology of the extended emission seen in the
[OIII] and H$\alpha$+[NII] images of MWZ.

\paragraph{MCG--6-30-15 (type 1.2), Fig. 12.}
The source is unresolved at both 3.6 and 20~cm.
There is another source (76 mJy at 20~cm) at 
$\alpha$=13$^h$33$^m$06$^s$.07,
$\delta$=--34{\arcdeg}04{\arcmin}49{\farcs}0 (B1950.0).
The position listed in Table~1 is that of the peak at 3.6~cm.
The 20~cm source position is offset by $\sim$ 1{\farcs}4 to the SE 
and is at $\alpha$=13$^h$33$^m$01$^s$.93,
$\delta$=--34{\arcdeg}02{\arcmin}26{\farcs}6 (B1950.0), 
but is consistent with the 3.6~cm position given the weakness of the source.
The 20~cm flux measured here is more than twice the value of 
1.7$\pm$0.7~mJy published in Paper~VI.

\paragraph{NGC~5252 (type 1.9), Fig. 12.}
Gaussian deconvolution of the nuclear (southern) source in the 20~cm map
shows it to be extended in 
P.A. 170{\arcdeg}, in close agreement with the P.A. of 171{\arcdeg} between
the two components. The northern source is unresolved.
A Gaussian deconvolution of the core component in the 20~cm map indicates a 
flux of 12.7 mJy while the summed flux is 13.5 mJy. 
Our results are consistent with those of Wilson \& Tsvetanov (1994)
and confirm the relatively flat spectrum of the nucleus 
($\alpha^{20}_{3.6} = $0.32). The extent of the S (nuclear) component
in the 20~cm  map is 0.65 kpc. Wilson \& Tsvetanov (1994) suggest
the N component is associated with NGC~5252  rather than being
an unrelated background source. We therefore  adopt a radio extent of 14.79 kpc
which is the distance between the two components.
Emission line fluxes for the S (nuclear) component have been taken from
Acosta-Pulido et al. (1996), who also list earlier determinations of the flux.
Acosta-Pulido et al. (1996) find evidence for a nuclear broad H$\alpha$ line,
but no broad H$\beta$ line, which leads to
a classification of Seyfert 1.9 for this object.

\paragraph{Mrk~270 (type 2), Fig. 13.}
The 20~cm map shows two components.
At 3.6~cm, both components are detected in a 
natural weighted (robust=4) map. 
The NE source is weakly detected in a uniformly weighted (robust=0) map.
Our maps are in agreement with those of Paper~V.
The P.A. and separation of the radio double agree well with the
major axis P.A. and extent of the brighter
[OIII] and H$\alpha$+[NII] extension (MWZ); higher resolution optical images
are required for a detailed comparison.
The P.A. of the host galaxy is not listed in RC3. UGC lists the
galaxy extent as 1{\farcm}1~$\times$~1{\arcmin} in B and
1{\farcm}1~$\times$~1{\farcm}1 in R. Second generation DSS images confirm that
the galaxy is nearly circular.

\paragraph{NGC~5273 (type 1.5), Fig. 13.}
A Gaussian deconvolution suggests that the 20~cm source is extended in P.A.
179{\arcdeg}, but the deconvolved source size is less than half
the beam size. 
The southern extension to the source at 3.6~cm (in P.A.$\sim$175{\arcdeg})
does not appear in a uniformly weighted (robust=--4) map but does appear
in the uniformly weighted (robust=0) map and in naturally weighted maps.
A Gaussian deconvolution of the 3.6~cm
source suggests a P.A. of 170{\arcdeg}; again, the deconvolved source
size is less than half the beam size.
Ulvestad \& Wilson (Paper~VI) have observed this source at 6 and 20~cm 
and list the source as `S'.
Their unpublished map shows an extension in P.A. 5{\arcdeg}. 
We therefore adopt a radio source classification of `S' and
a radio P.A. of 5{\arcdeg}.
Paper~VII gives a 20~cm flux of 2.5 mJy, in agreement with our measured
value, and a 6~cm flux of 0.9 mJy. There is a confusing source 9{\farcm}5 away.
H$\alpha$+[NII] and [OIII] images have been published by Pogge (1989).
Whittle (1992a) classifies this object as a Seyfert 1.9.
Ho et al. (1997) have detected broad lines in both H{$\alpha$} and
H$\beta$. They have used the criteria of Whittle (1992a) to 
classify this object as a Seyfert 1.5.  

\paragraph{IC 4329A (type 1.2), Fig. 13.}
The extension in P.A. $\sim$75{\arcdeg}
in the 20~cm map is present before the self-calibration process.
Higher  resolution 20~cm and 6~cm maps of this source (Unger et al. 1987a) 
show  a diffuse 6{\arcsec} extension in P.A.$\sim$285{\arcdeg}.
There is another source in the same field (8 mJy at 20~cm)
at $\alpha$=13$^h$46$^m$25$^s$.94, 
$\delta$=--30{\arcdeg}02{\arcmin}25{\farcs}4 (B1950.0).
The source is suspected to be variable (Unger et al. 1987a).
The printed edition of RC3 (1991) lists an erroneous host galaxy P.A. of 
63{\arcdeg}, while the updated online version (available at NED) lists
the correct P.A. of 45{\arcdeg}.

\paragraph{NGC~5548 (type 1.2), Fig. 14.}
The northern component seen in the 20~cm map is limb-brightened,
while the nuclear component is unresolved (cf. \cite{wu82}).  
The P.A.'s of the peaks of the northern and southern extensions 
relative to the core are 348{\arcdeg} and 170{\arcdeg}, respectively.
Neither of the extensions are detected in the 3.6~cm map.
There are two other sources in the field, one at
$\alpha$=14$^h$15$^m$39$^s$.76, 
$\delta$=25{\arcdeg}21{\arcmin}56{\farcs}0 (B1950.0) and a 
double source at $\alpha$=14$^h$15$^m$35$^s$.82, 
$\delta$=25{\arcdeg}22{\arcmin}08{\farcs}3 (B1950.0). 
There is also a confusing source $\sim$4{\arcmin} away.
The radio sources in the field around NGC~5548 are discussed in detail
by Wilson \& Ulvestad (1982).
The flux ratio R$=$ F$_{[OIII]}$ / F$_{H\beta}$ listed by
Whittle (1992a) leads to a classification of Seyfert 1.2 while
Ho et al. (1997) use the criterion of Whittle (1992a) to classify this
object as a Seyfert 1.5.  We use a classification of Seyfert 1.2 as 
long term monitoring of the BLR in NGC~5548 (\cite{petet91}) consistently 
gives R~$\lesssim$~0.75. 
UGC lists a major axis P.A. of 110{\arcdeg} at an extent of 
1{\farcm}7~$\times$~1{\farcm}5.
The optical morphology of this galaxy is, however, complex, and we do not
list a major axis P.A. for it.

\paragraph{NGC~6251 (type 2), Fig. 14.}
The jet of this famous radio galaxy (see e.g. Perley, Bridle \& Willis 1984) 
is not detected in the 3.6~cm map.  
The jet extends well beyond the synthesized field of view of
our 20~cm map and we detect only the innermost part of it.
For this reason, the total 20~cm flux listed in Table~1, and
the 20~cm flux of the jet listed in Table~2, are underestimates
of the actual fluxes.

\paragraph{NGC~7465 (type 2), Fig. 15.}
In the 20~cm map, the source sits on a large, low level plateau so the flux
value is uncertain. Gaussian deconvolution of the
3.6~cm source suggests a P.A. of 32{\arcdeg}, but the source
size is less than half the beam size. There is a significant displacement
(3{\arcsec}) between the apparent positions of the 3.6~cm and 20~cm sources.
The VLA calibrator manual lists a position code of 
`Terrible' (`positional error $>$ 0{\farcs}15, and frequently much greater')
for 2247+140, the calibrator used at 20~cm.
Since the 3.6~cm calibrator, 2251+158, has a position code of `A'  
(positional error $<$0{\farcs}002) and the 3.6~cm resolution is higher 
that the 20~cm resolution, the position of the 3.6~cm source is more reliable
and is listed in Table~1.
The position angle of the major axis of the host galaxy is not
listed in the RC3 or UGC catalogs. UGC lists an
external diameter of 1{\farcm}2~$\times$~0{\farcm}7 in B.
A second generation DSS image reveals a diameter of $\sim$1{\farcm}7
but shows clear signs of interaction between NGC~7465 and two nearby galaxies.

\section{Statistical Results}

In this section, we perform a statistical study of the radio properties
of Seyfert galaxies. We include only the 43 galaxies  in
the early-type Seyfert sample (Section~2), except for Section~5.4, where
the sample is expanded to include other Seyfert galaxies.

To test for correlations in our data, we use the techniques of ``survival 
analysis'' as coded in the ASURV software package (Lavalley, Isobe \&
Feigelson 1992).
We use the 5 available univariate tests - Gehan's generalized Wilcoxon test
with permutation variance, Gehan's generalized Wilcoxon test with
hypergeometric variance, the logrank test, the Peto-Peto test
and the Peto-Prentice test - to estimate the probability that two distributions 
are not derived from the same parent population.  Note that for completely
``uncensored'' data, i.e. data which contains no upper or lower limits,
the Peto-Peto test reduces to the logrank test and the Peto-Prentice
method reduces to Gehan's Wilcoxon test.
We use the 3 available bivariate tests - Cox proportional hazard model, 
generalized Kendall's $\tau$, and Spearman's $\rho$ - to test for
correlations between two variables within a sample. 
The result of each bivariate test is a value which represents the
probability that the two variables are not correlated.
We consider a result to be statistically significant only if
all relevant ASURV statistical test results are 
less than 0.05 and ``weakly'' significant if all relevant ASURV test
results are less than 0.10, or if all except one test result
are less than 0.05.
We use the Kaplan-Meier estimator, also from the ASURV package, to calculate 
the mean of a distribution which contains censored (upper or lower limit)
data points.
The numerical results of the statistical tests are listed in Table~6 
(univariate tests) and Table~7 (bivariate tests).

All P.A. measurements are subject to uncertainty.
In an attempt to keep track of these uncertainties, we have provided 
quality flags, as outlined in Section~3 and Table~4 to each radio
P.A. and host galaxy P.A. determination. The color coding used in the 
following histograms reflects this quality control - dark shades are used for
more reliable data and lighter shades for less reliable data. The 
P.A.'s of the [OIII], H$\alpha$ and green emission from MWZ have also
been color coded according to data quality, and this coding is expained
in the relevant figure captions. 
Because of the small number of objects in our sample, data quality
flags have been ignored when doing statistical tests (Tables 6 and 7).

\notetoeditor{Please place the multiple panels of Figures 16--20 in
a double column format with panels (a), (c), (e) and (g) in the left column
and panels (b), (d), (f) and (h) in the right column.}

\subsection{Sample Properties, radio extent and luminosity}

\begin{figure}[!ht]
\figurenum{16}
\plottwo{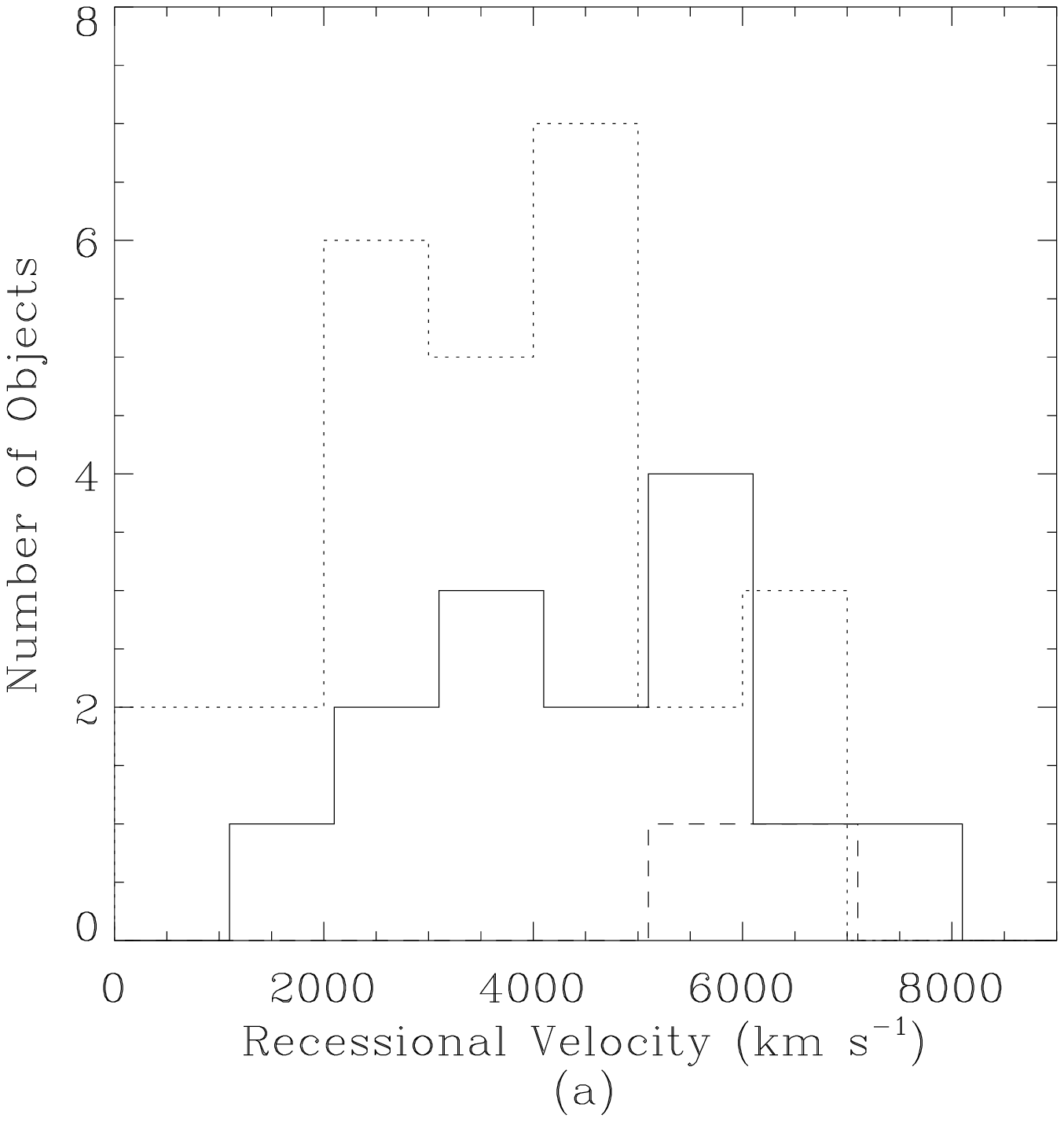}{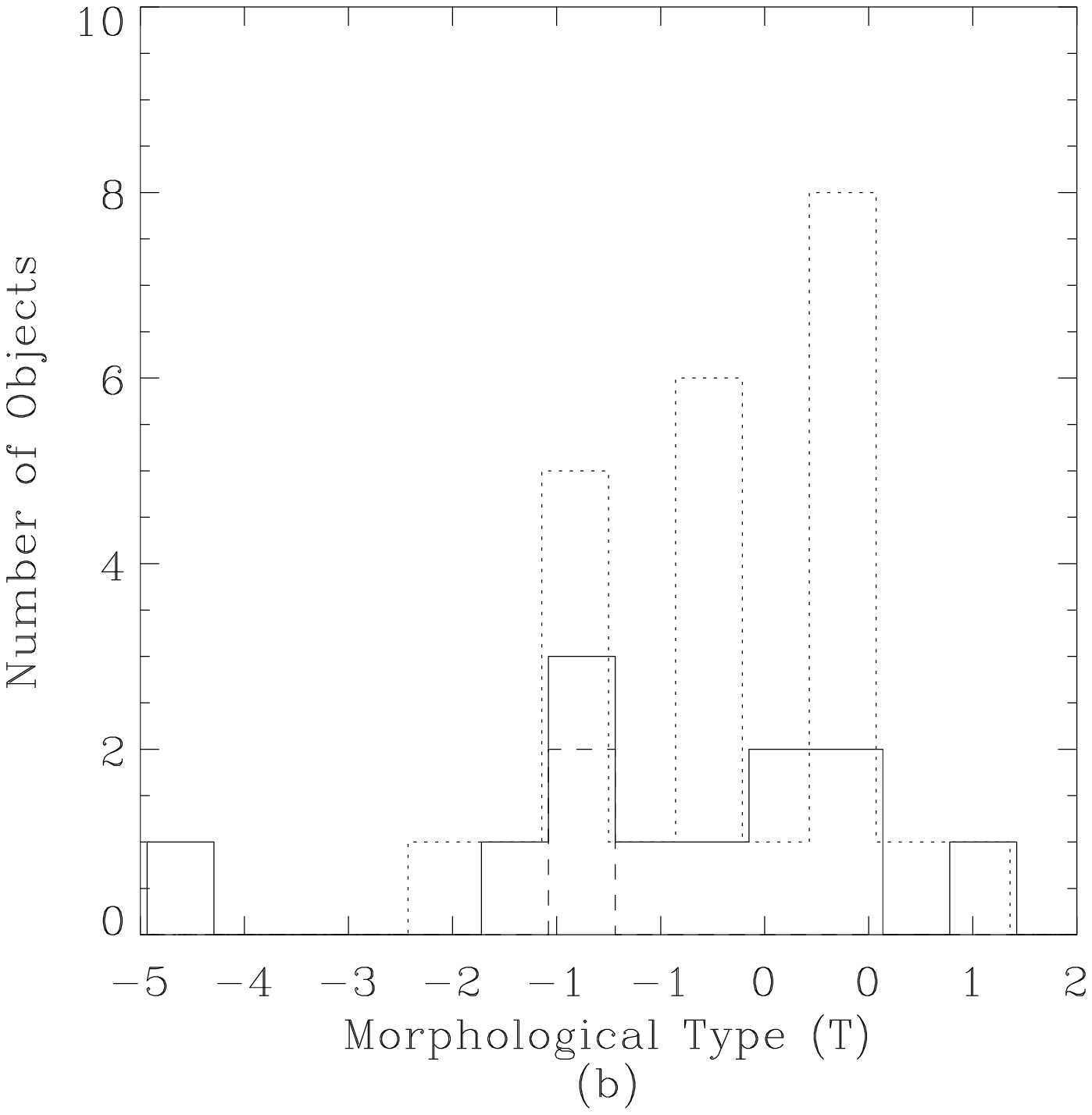}
\caption
{\small{Overall properties of the early-type Seyfert sample.
Seyfert 1's (Seyfert 1.0's through Seyfert 1.5's)
are plotted with solid lines, Seyfert 1.8's and 1.9's with dashed lines,
and Seyfert 2.0's with dotted lines.  For clarity, the histograms 
are slightly offset along the x axis.
(a)~Histogram showing the distributions of recessional velocity; 
(b)~histogram showing the distributions of morphological type.
}}
\end{figure}

Some properties of the sample are shown in Figures~16 and 17. 
The Seyfert 1's and Seyfert 2.0's in the early-type radio 
sample have similar distributions of V$_{rec}$
and T (Figure~16) though the Seyfert 1's are marginally more distant.
A histogram of 20~cm luminosities (Figure~17a) shows that the 20~cm
luminosities of Seyfert 1's and Seyfert 2.0's are similar. The
distributions of the 3.6~cm luminosities (not shown) give a similar result.
The mean 20~cm luminosity (P$_{20}$) of Seyfert 1's 
(10$^{22.72}$ Watts Hz$^{-1}$) and Seyfert 2.0's 
(10$^{22.75}$ Watts Hz$^{-1}$) are not significantly different.
The 20~cm radio luminosity of the sample
is uncorrelated with galaxy morphological type (Figure~17b) for both Seyfert
types. This is perhaps not surprising given the rather narrow range of
morphological types in the sample. As seen in Figure~17c, there is no trend
for the fraction of sources with resolved structure to
decrease with increasing recessional velocity.
The 20~cm luminosity is significantly correlated with radio structural class 
(i.e. `U', `S' and `L', see Figure~17d) for 
Seyfert 2.0's, with the `L's having the largest luminosity,
while these two variables are uncorrelated for Seyfert 1's (Table~7).
Consistent with the unified model, Seyfert 2.0's  have a higher fraction of
resolved radio sources (25/27 or 93\%) than Seyfert 1's 
(9/14 or 64\%); statistical
tests show that this difference is significant (Table~6) , but the results are
limited by the small number of Seyfert 1's.
The two Seyfert 1.9's (NGC~5252 and NGC~526A) are `L'-class
radio sources and are the most radio-extended objects in the sample 
(Figure~17e). 
The next three largest radio sources are Seyfert 1.2's (NGC~3516, IC~4329A
and NGC~5548). The existence of these large Seyfert 1's makes the mean
Seyfert~1 radio extent (1.2~kpc$\pm$0.4~kpc) greater than the mean Seyfert~2.0
radio extent (0.6~kpc$\pm$0.1~kpc) despite the fact that the Seyfert~1's
have a lower fraction of resolved radio sources (the mean sizes were
calculated using the Kaplan-Meier estimator, including upper limits).
This difference in mean extents is, however, not statistically significant
(Table~6). As Figures~17e and 17f suggest, Seyfert 2.0's show a weak trend for
more radio-extended objects to have larger radio luminosities
while Seyfert 1's do not show a correlation between these two quantities
(Table~7). 
We emphasise that studies involving the linear extent of radio sources
are limited by ambiguity in the precise definition of this quantity.

\begin{figure}
\figurenum{17}
\plotfiddle{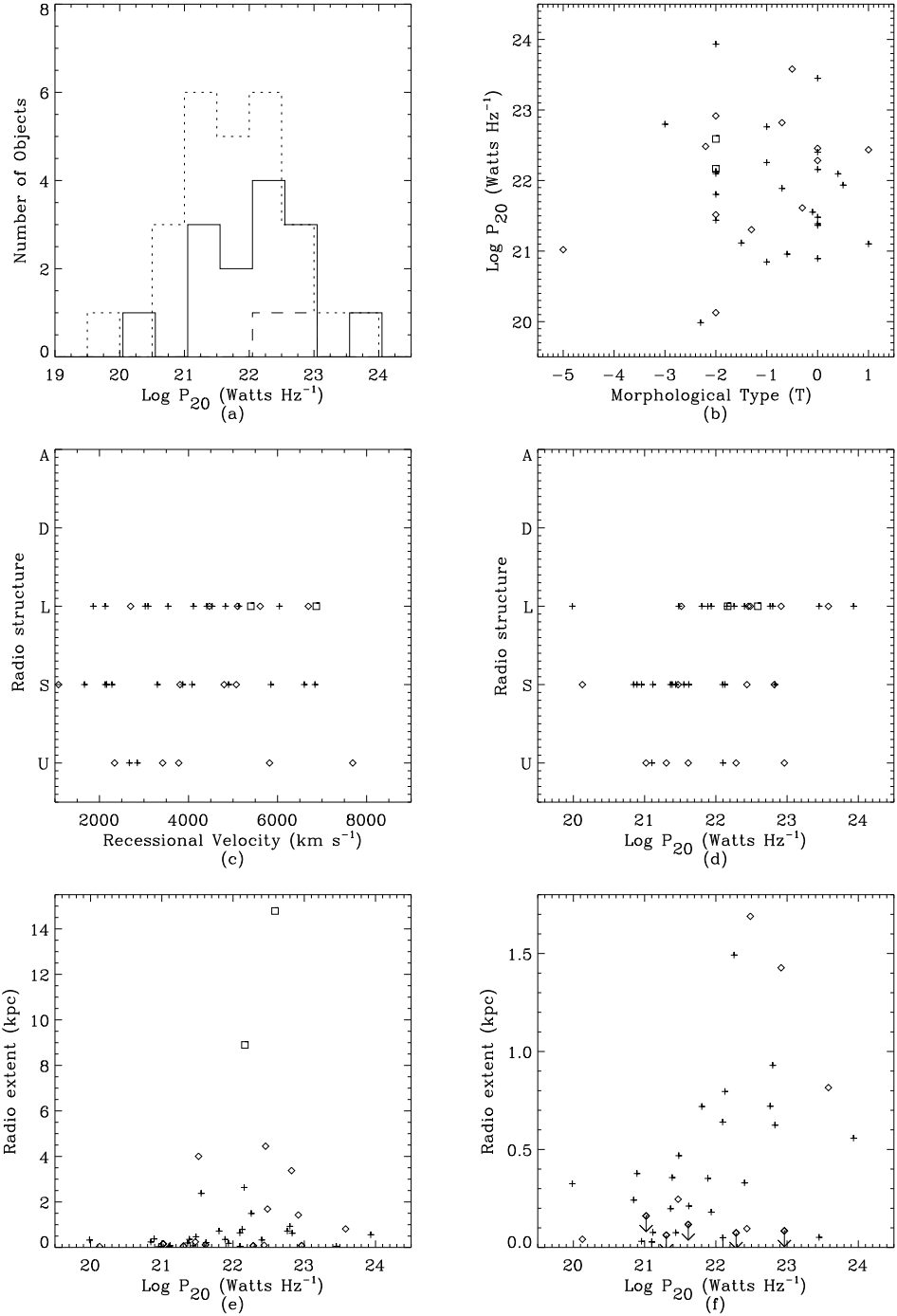}{7.0in}{0}{82}{82}{-250}{-50}
\caption
{\small{Relationships between radio luminosity, structure, extent, 
galaxy morphological
type and recessional velocity for the early-type Seyfert sample.
Seyfert 1's (Seyfert 1.0's through Seyfert 1.5's) are plotted with
solid lines and diamonds, Seyfert 1.8's and 1.9's with dashed lines and
squares, and Seyfert 2.0's with dotted lines and crosses.
(a)~Histogram showing the distribution of 20~cm radio luminosity. For 
clarity, the histograms are slightly offset along the x axis;
(b)~relationship between radio luminosity and galaxy morphological type;
(c)~relationship between radio structure and recessional velocity. 
(d)~relationship between radio structure and 20~cm radio luminosity;
(e)~relationship between radio extent and 20~cm radio luminosity.
(f)~same as (e), but with an enlarged y scale.
}}
\end{figure}

\subsection{The relationship between radio and emission-line gas} 
The radio and high excitation ionized gas structures in Seyfert 2's are
already known to be highly correlated from VLA and HST images of individual
galaxies (e.g. Bower et al. 1994, 1995; \cite{capet96}; \cite{fws98}). 
Previous work has shown that the ionized gas radiating emission lines
is ionized by photons escaping from the central source 
(e.g. Wilson and Tsvetanov 1994) and compressed through interactions with
the radio ejecta (e.g. Capetti et al. 1996).
It is therefore interesting to compare the alignments
and the total luminosities of the emission line and radio structures. 
The distribution of P.A.$_{Radio}$--P.A.$_{[OIII]}$ (Figure~18)
for Seyfert 2.0's shows a strong trend for the two axes to align,
while the same distribution for Seyfert 1's is not significantly 
different from a uniform distribution (Table~6). 
The absence of a correlation for
Seyfert 1's is, however, not surprising given that there are only 7 
galaxies in the early-type radio sample with a measurement of 
P.A.$_{Radio}$~--~P.A.$_{[OIII]}$.
In fact, the distributions of P.A.$_{Radio}$--P.A.$_{[OIII]}$
for Seyfert 1's and Seyfert 2.0's are not significantly
different from each other (Table~6).
The distribution of P.A.$_{Radio}$--P.A.$_{H\alpha}$ (Figure~18) for
Seyfert 2.0's also shows a strong tendency for the axes to align
while the same distribution for Seyfert 1's is not significantly different 
from a uniform distribution (Table~6).
Again, there are only 7 galaxies in the distribution for Seyfert 1's.
Previous studies have shown this alignment between radio and emission-line
structures for individual galaxies or a small number of galaxies,
usually selected to show extended, class `L', radio emission
(e.g. Haniff, Wilson \& Ward 1988; Whittle et al. 1988). Our
result is the first to demonstrate that this alignment is
characteristic of early-type Seyfert galaxies {\textit{as a class}} 
in a well-defined
sample. In our sample, the radio source is smaller than
the scale on which the axis of the emission-line gas is measured
(by a factor of $\sim$2~--~$>$5 for Seyfert 2.0's),
suggesting that the alignment results from the anisotropic escape of ionizing
photons preferentially along and around the radio axes, or compression of gas
by outflows aligned with the radio source but on a larger scale.

\begin{figure}[!ht]
\figurenum{18}
\plotfiddle{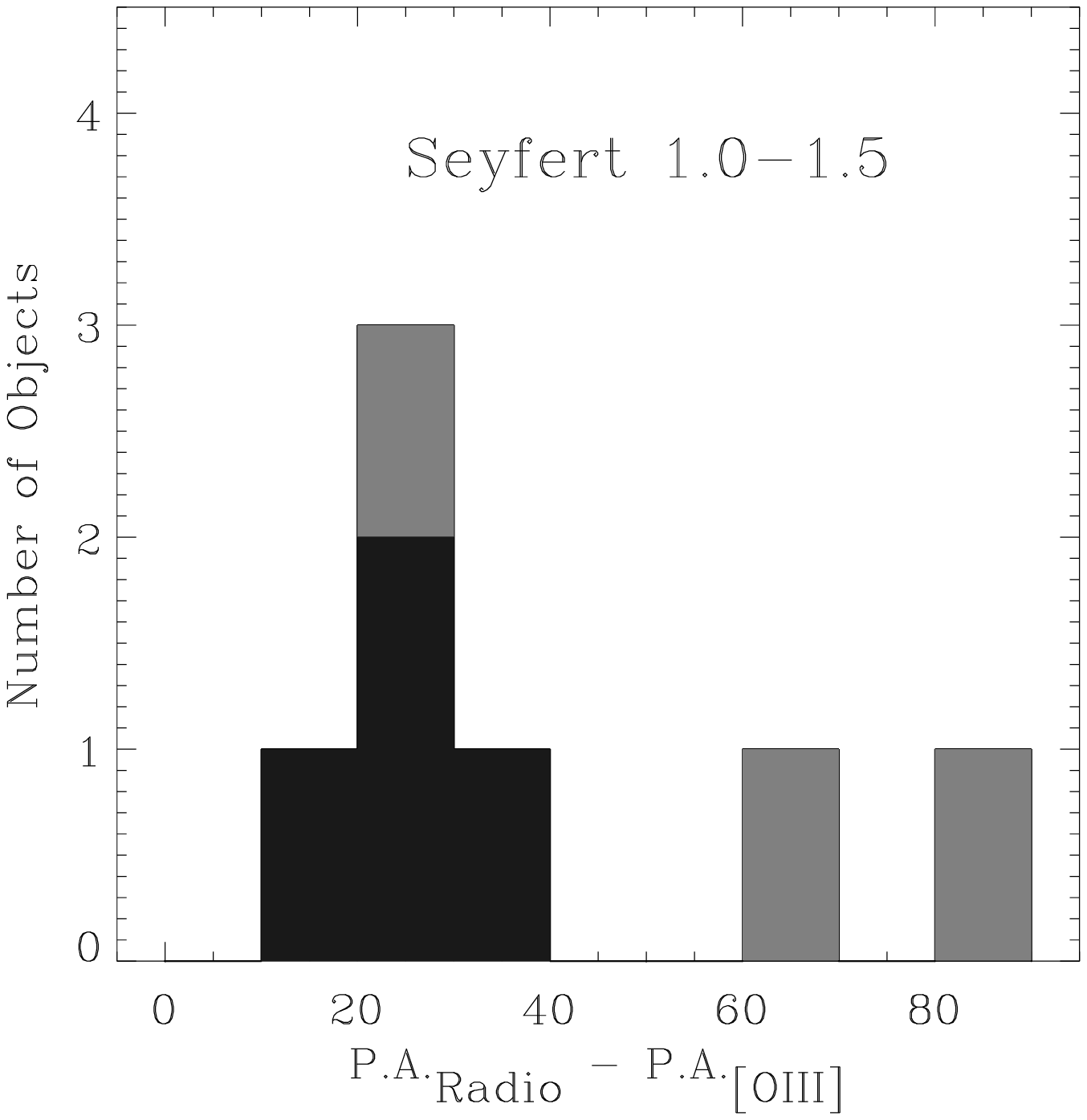}{2.0in}{0}{42}{42}{-240}{-80}
\plotfiddle{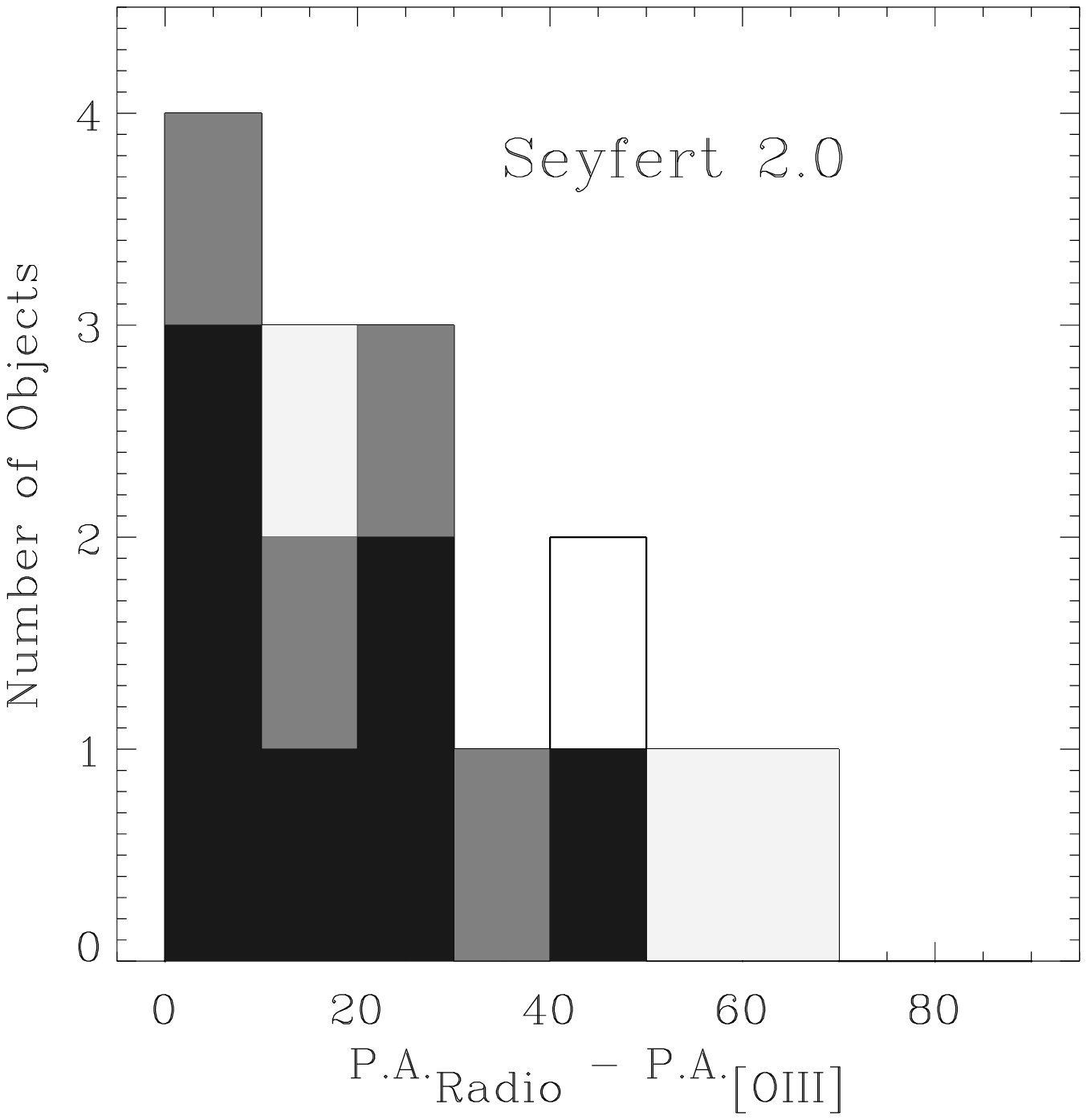}{2.0in}{0}{42}{42}{-050}{80}
\plotfiddle{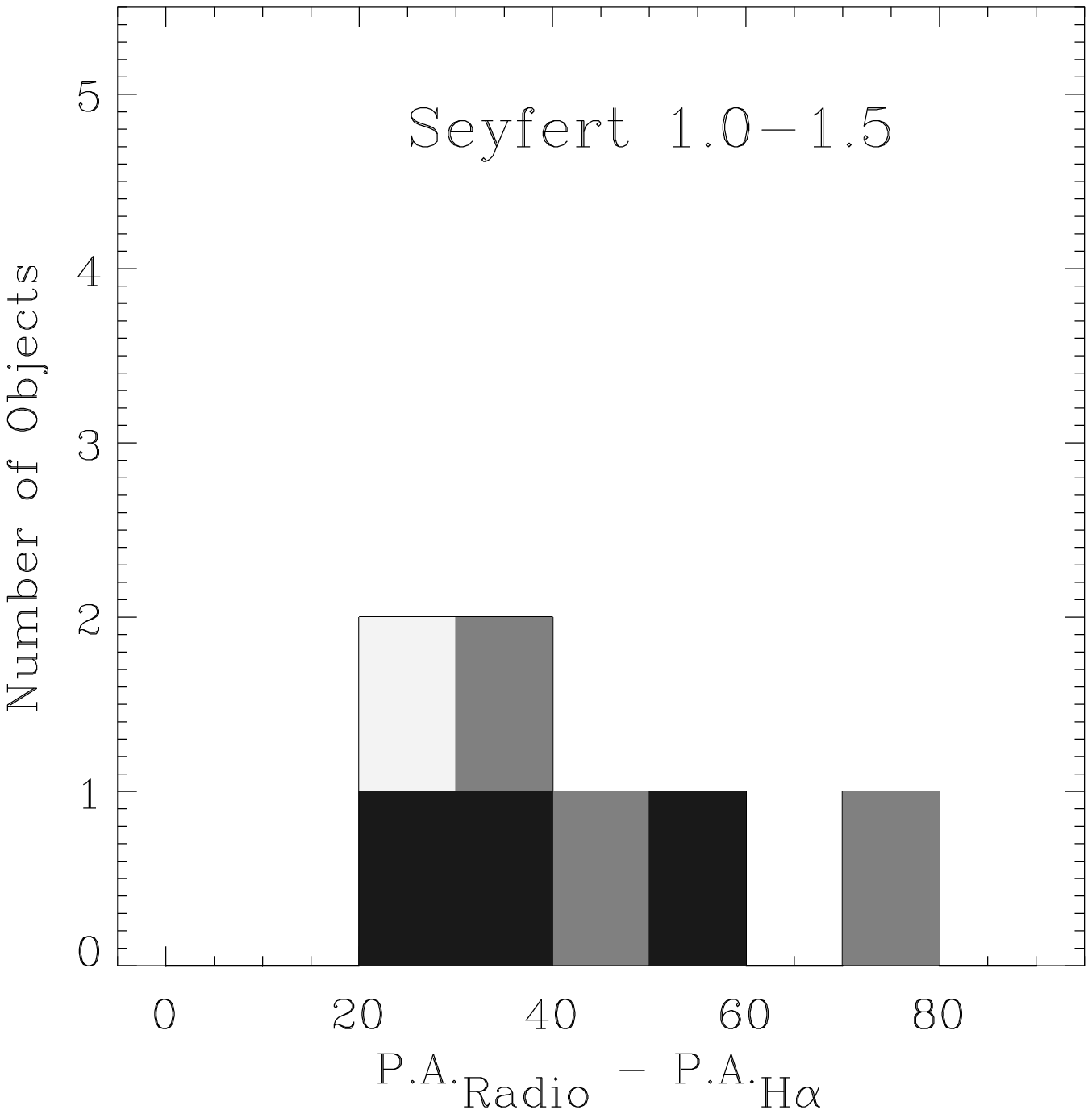}{2.0in}{0}{42}{42}{-240}{45}
\plotfiddle{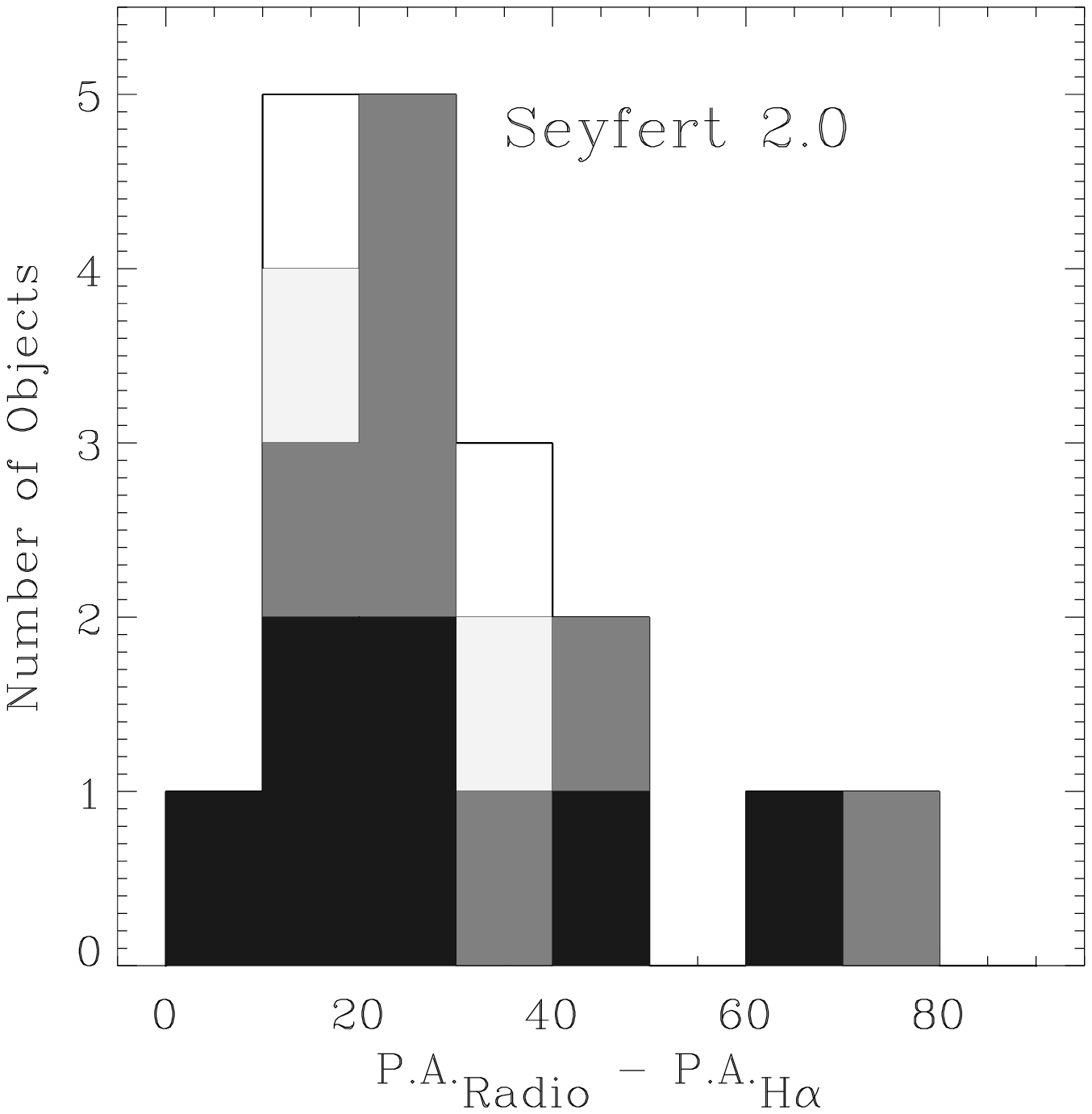}{2.0in}{0}{42}{42}{-050}{205}
\vspace{-4in}
\caption
{\small{Histogram of the difference between the radio and emission-line
P.A.'s for the early-type Seyfert sample. 
Seyfert 1's (Seyfert 1.0's through Seyfert 1.5's) are plotted on
the left; Seyfert 2.0's are plotted on the right. The two Seyfert 1.9's
in the early-type sample are not plotted. Quality flags for P.A.$_{Radio}$
are explained in Table~4. Quality flags for P.A.$_{emission-line}$
are `a', except for values in brackets in Table~3 which are assigned 
`b'.  Objects with quality flag `a'
for both P.A.$_{Radio}$ and P.A.$_{emission-line}$ 
are shaded black. 
Objects with one  flag `a' and the other `b'  for
P.A.$_{Radio}$ and P.A.$_{emission-line}$ are shaded in dark
grey. Objects with flag `b' for both P.A.$_{Radio}$ and
P.A.$_{emission-line}$ are shaded light grey. 
All others are in white.
}}
\end{figure}

The alignment between the radio and the extended ionized gas also
places a strong constraint on any large angle precession of the inner 
accretion disk, e.g., as theorized to result from radiative instabilities
(\cite{pri97}) or as inferred from observations in 4C~29.47 
(\cite{conmit84}). We consider two cases for the excitation of the
ionized gas: \newline
(i)~The extended ionized gas is photoionized by a central source and is
produced by the intersection
of the nuclear ionization cone with the disk of the galaxy. 
In this case, as demonstrated for NGC~4151 (Pedlar et al. 1993; 
Boksenberg et al. 1995) and in simulations (Mulchaey et al.
1996b), the position angles of the extended emission-line gas
and the ionization cone axis are not necessarily the same. Hence 
even if the axes of the ionization cone and radio ejecta are 
exactly aligned, there can be a significant offset 
(30{\arcdeg} in the case of NGC~4151) between the position angles of 
the radio structure and the extended ionized gas.
We find that the extended ionized gas is preferentially aligned
with the disk of the galaxy for the early-type Seyfert 2.0's (see
Section~5.3 and Figure~20), so the scatter of $\sim$30{\arcdeg}
about 0{\arcdeg} in the
distribution of P.A.$_{Radio} -$ P.A.$_{[OIII]}$ (Figure~18) for
Seyfert 2.0's might be completely explained by such a projection effect.
Consider a typical object in the sample with emission-line extent 2~kpc and
radio extent 0.5~kpc.  The difference between the travel times of the 
ionizing radiation and the radio ejecta must be significantly less 
(by at least a factor of 4, if we consider a precession angle $>$~60{\arcdeg}).
than any precession period, $P$, of the collimating disk for the radio
and ionization cone axes to be so closely aligned.
If the velocity of the radio ejecta in Seyferts is 
sub-relativistic, as suggested by the proper motion of radio knots in
NGC~4151 (\cite{ulvet98}) and by the modeling of the interaction between
radio ejecta and high-excitation gas in NGC~1068 (\cite{wilulv87})
and NGC~5929 (\cite{whiet86}), the crossing
time of the ionizing radiation is negligible ($\sim$10$^4$~yr) and we must
have V$_{ejecta}~{\times}~P~\gtrsim$ 2~kpc. Under a number of assumptions,
Pringle (1997) calculates that
the characteristic precession time of the accretion disk due to
the radiative instability is 
t$_{var}~=~2~{\times}~10^6~M_8~\alpha^{-1}$~yr, 
where M$_8$ is the mass of the central black hole in units of 
10$^8$~M$_{\sun}$ and $\alpha$ is the usual viscosity parameter.
If we use typical values of 10$^8$~M$_{\sun}$ for the mass of a Seyfert
black hole (e.g. \cite{korric95}) and $\alpha$~=~1
then V$_{ejecta}$ $\gtrsim$ 1000~km~s$^{-1}$, which is not unreasonable.
Better determinations of the velocity of radio
ejecta in Seyferts will provide more stringent constraints on P; \newline
(ii)~The extended ionized gas is caused by a large scale, self-ionized
outflow (e.g. \cite{bicet98}). The precession period of the collimating disk
disk is once more
constrained by the difference in the travel times of the radio ejecta and the
larger scale outflow. Similar considerations as described above apply,
and V$_{ejecta}$ now refers to the radio ejecta or ionized gas, whichever
has the longer crossing time.

The relationships between the radio and emission-line luminosities
are shown in Figure~19.
At a fixed nuclear radio luminosity, Seyfert 2.0's appear weaker in [OIII] 
and H$\alpha$ as compared to Seyfert 1's, in agreement with 
the findings of Whittle (1985). 
The 20~cm luminosity is uncorrelated with both the
[OIII] luminosity and the H$\alpha$ luminosity for Seyfert 2.0's (Table~7). 
For Seyfert 1's, however, the 20~cm luminosity is significantly correlated 
with both the [OIII] luminosity and the H$\alpha$ luminosity (Table~7). 
These two correlations are significant even after deleting
the single Seyfert 1 with low [OIII] and H$\alpha$ luminosity (ESO 512-G20);
in this case the probabilities that the 20~cm luminosity is not correlated
with the [OIII] luminosity and H$\alpha$ luminosity are 0.03--0.06 and 
0.015-0.06, respectively. 
Previous studies with larger samples of Seyferts have found strong 
correlations between L$_{[OIII]}$ and P$_{20}$ 
(de~Bruyn \& Wilson 1978; Whittle 1985; Whittle 1992c).
The scatter in our plot of P$_{20}$ vs [OIII], which contains
26 objects (Figure~19a), is similar to that in the plot of [OIII] vs 
P$_{20}$ in Whittle (1992c), which contains $\sim$80 objects (his
Figure~3). It is therefore likely that the reason that our results 
are not completely consistent with Whittle (1992c) is  the smaller
number of objects in our sample.

\begin{figure}[!ht]
\figurenum{19}
\plottwo{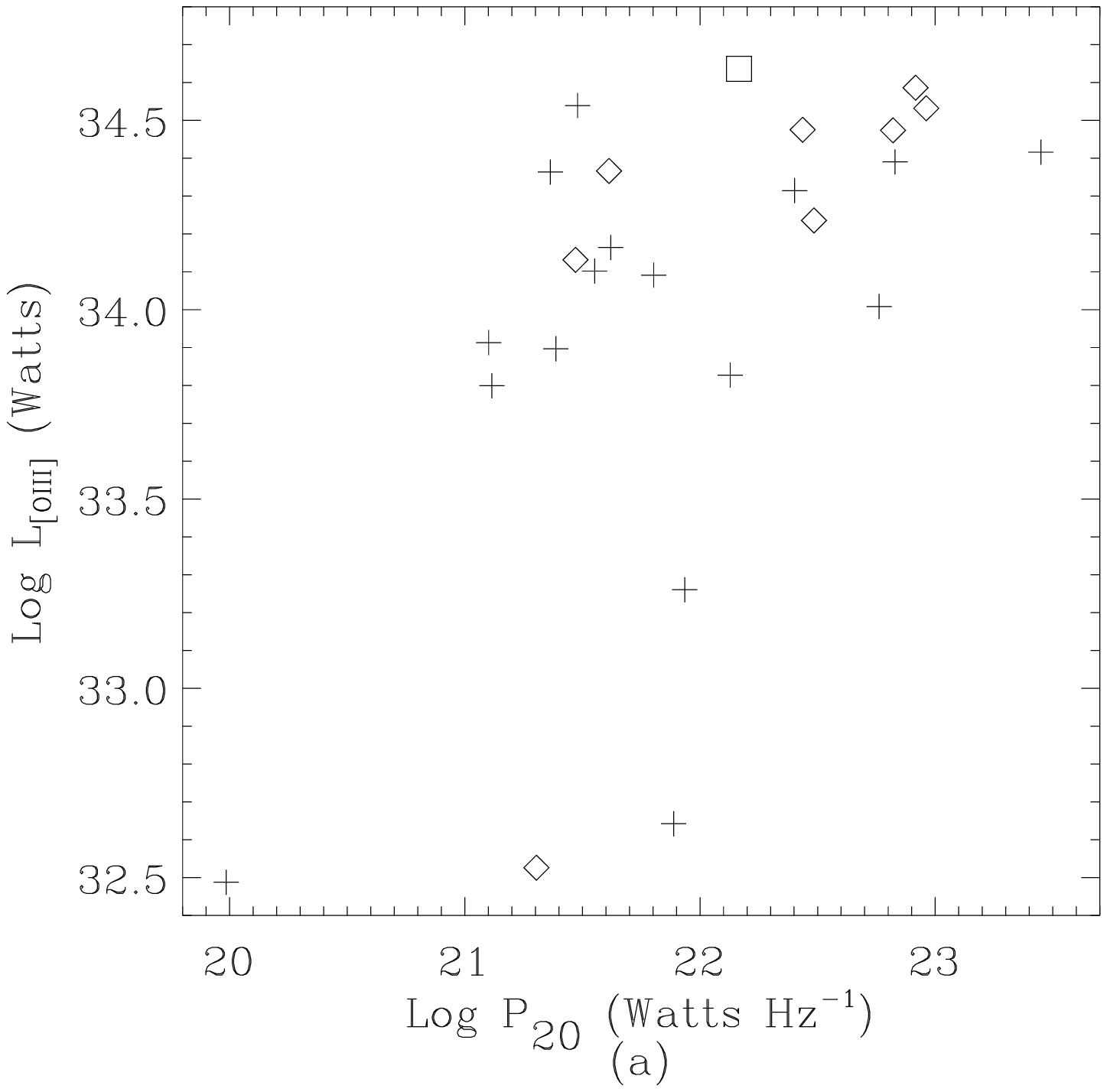}{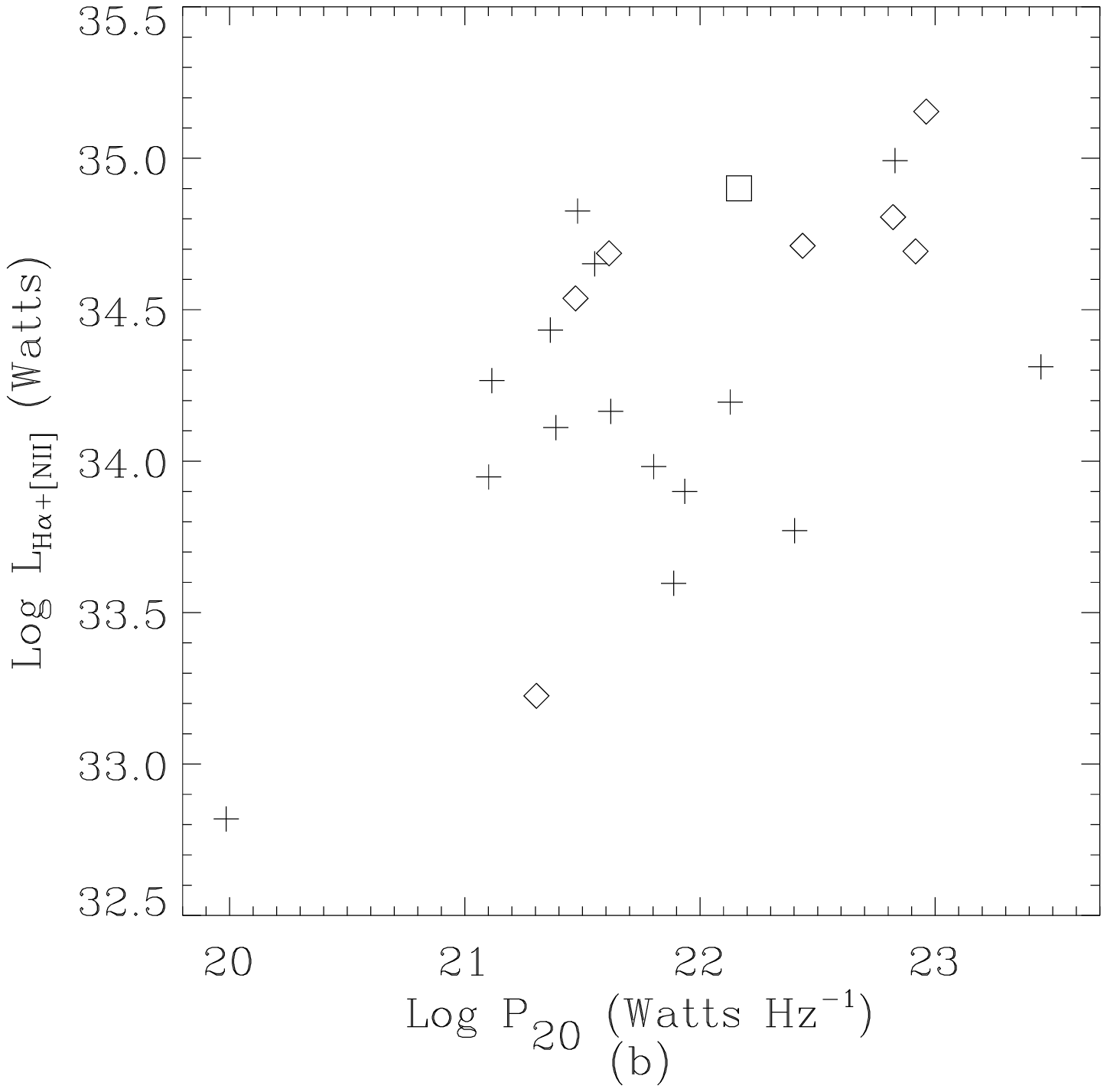}
\caption
{\small{Relationships between the luminosities of the [OIII], H$\alpha$+[NII]
and radio emissions for the early-type Seyfert sample. 
Seyfert 1's (Seyfert 1.0's through Seyfert 1.5's)
are plotted as diamonds, Seyfert 1.8's and Seyfert 1.9's as squares,
and Seyfert 2.0's as crosses.
(a)~Relationship between [OIII] luminosity and 20~cm radio luminosity;
(b)~relationship between H$\alpha$+[NII] luminosity and 
	20~cm radio luminosity.
}}
\end{figure}

\subsection{The alignment between emission-line structure and host galaxy}

The relative orientations of the nuclear [OIII] and 
nuclear green emission with respect
to the major axis of the host galaxy for the 
early-type Seyfert sample of MWZ have been studied by
Mulchaey \& Wilson (1995). They measured the ``nuclear'' orientation of the 
[OIII] and green emission  (on scales of 3{\arcsec}--4{\arcsec}) and
found a close alignment between the 
[OIII] and green emission, and a weaker alignment between the 
nuclear [OIII] emission and the host galaxy major axis as listed in
the UGC and ESO catalogs.

\begin{figure}[!htp]
\figurenum{20}
\plotfiddle{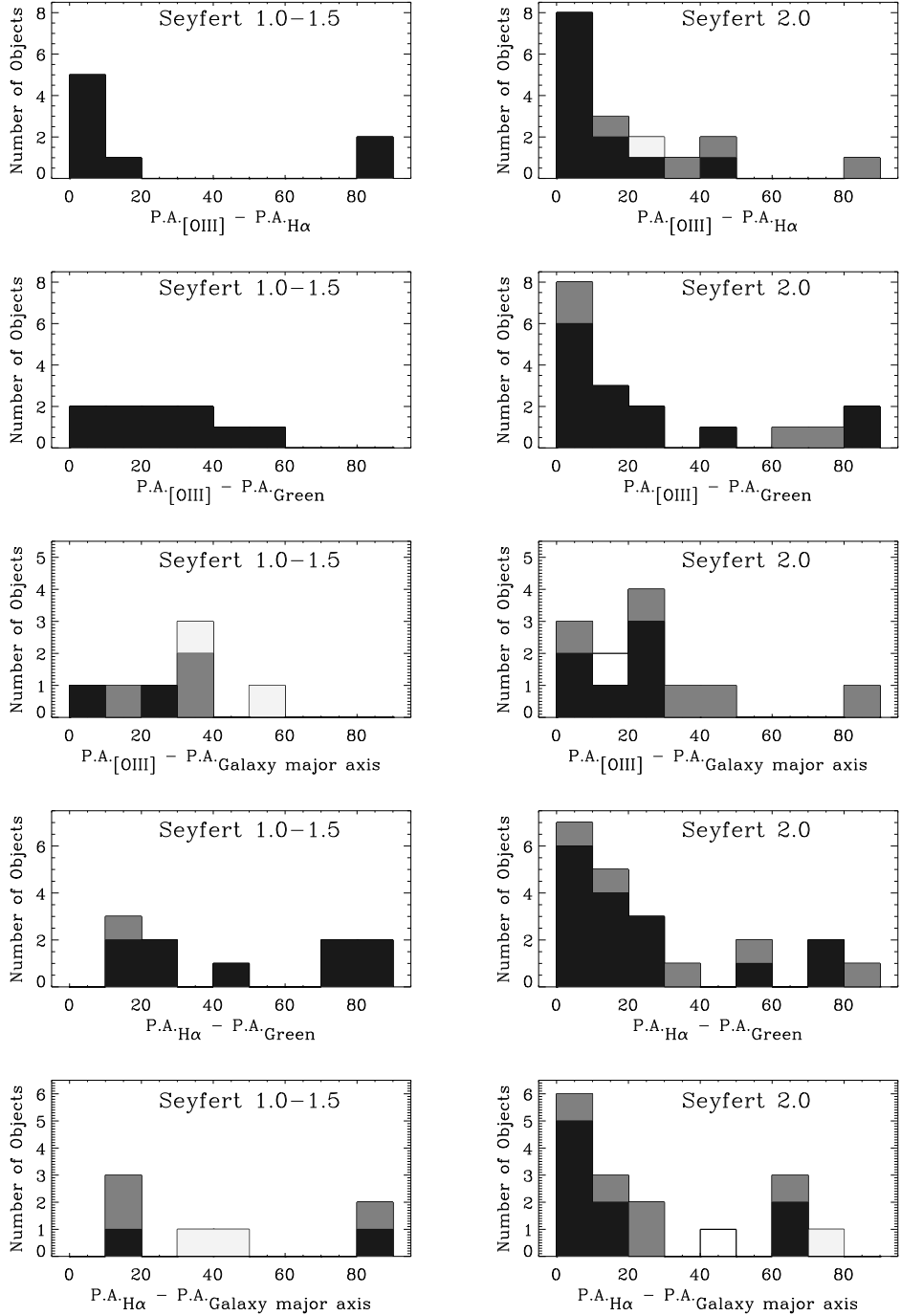}{7.0in}{0}{82}{82}{-250}{-50}
\caption
{\small{Histograms of the differences between the P.A.'s of the [OIII] and 
H$\alpha$ structures, the green 
continuum emission, and the host galaxy major axis as listed in Table~3 
and MWZ's Table~4. 
The green continuum is measured at a surface brightness of 
2~$\times$~10$^{-18}$
ergs s$^{-1}$ cm$^{-2}$ {\AA}$^{-1}$ (arcsec)$^{-2}$, which
is the lowest contour of the images of MWZ (see Section~3 and Table~3).
Quality flags for P.A.$_{Radio}$ and 
P.A.$_{Galaxy~major~axis}$ are explained in Table~4. Quality flags
for P.A.$_{H\alpha}$, P.A.$_{[OIII]}$ and P.A.$_{Green}$ are `a', except for 
values in brackets in Table~3 and MWZ's Table~4 which are assigned `b'. 
Objects with quality flag `a'
for both P.A. values are shaded black. 
Objects with one  flag `a' and the other `b'  for the two
P.A. values are shaded in dark grey. Objects with flag `b' for both P.A.
values are shaded light grey. All others are in white.
}}
\end{figure}

Here we discuss the relative orientations of the larger scale [OIII] and 
H$\alpha$ emission, the larger scale green continuum emission (all from MWZ),
and the host galaxy major axis (determined as described in Tables 3 and 4).
The P.A.'s of the [OIII], H$\alpha$ and green emission are listed in Table~3
(for the early-type Seyferts which are extended in the radio) and in Table~4 
of MWZ (for the entire early-type Seyfert sample).
These P.A.'s have been measured at typical extents of 5{\arcsec}--10{\arcsec},
8{\arcsec}--15{\arcsec} and 20{\arcsec}--40{\arcsec} for the [OIII], 
H$\alpha$ and green emission, respectively (see Table~4 of MWZ for
the exact extents). When determined photometrically, the P.A.'s
of the host galaxy major axis were measured at a larger
extent (typically $\geq$~2) than the green emission of MWZ.
Histograms of the relative orientations are shown in Figure~20.
The [OIII] emission is the better tracer of high excitation gas associated
with nuclear activity, because H$\alpha$ is often contaminated by
HII regions in the disk of the galaxy. Here, however, the [OIII] and
H$\alpha$ emission tend to be aligned.
What is striking about Figure~20 is that, for Seyfert 2.0's,  
the [OIII], H$\alpha$, green  and host galaxy major axes are all 
significantly aligned with each other (see Table~6).
The relationship between these axes is not
as clear for Seyfert 1's, perhaps because of the smaller number of objects.
For the Seyfert 2.0's, the ionized gas tends to align with both the smaller
scale radio emission (Figure~18) and the larger scale galaxy disk (Figure~20).
This is consistent with a scenario in which ambient gas lying preferentially
in the galaxy disk is ionized by an ionization cone from the central engine,
and/or compressed by a gaseous outflow, both of which tend to be aligned
with the axis of the radio source.

\subsection{Seyfert radio extents - at odds with the unified scheme~?}

Seyfert 2's show a higher fraction of resolved radio sources 
as compared to Seyfert 1's (see Section~5.1 and, e.g., Paper VII).
Ulvestad \& Wilson (Paper VII) also found a weak trend for Seyfert 2's
to have more extended nuclear radio structures, though we find (Section~5.1)
that the existence of a few highly extended Seyfert 1's in the early-type
radio sample makes the mean Seyfert
1 radio extent greater than the mean Seyfert 2.0 radio extent.
Within the unified scheme of AGN, which is supported by the fact that
Seyfert 1's and 2's have similar radio luminosities, we would expect the
radio extent to be correlated with Seyfert type, with
Seyfert 1.0's having the lowest and Seyfert 2.0's having the
highest mean projected radio extent.

\begin{figure}[!ht]
\figurenum{21}
\plottwo{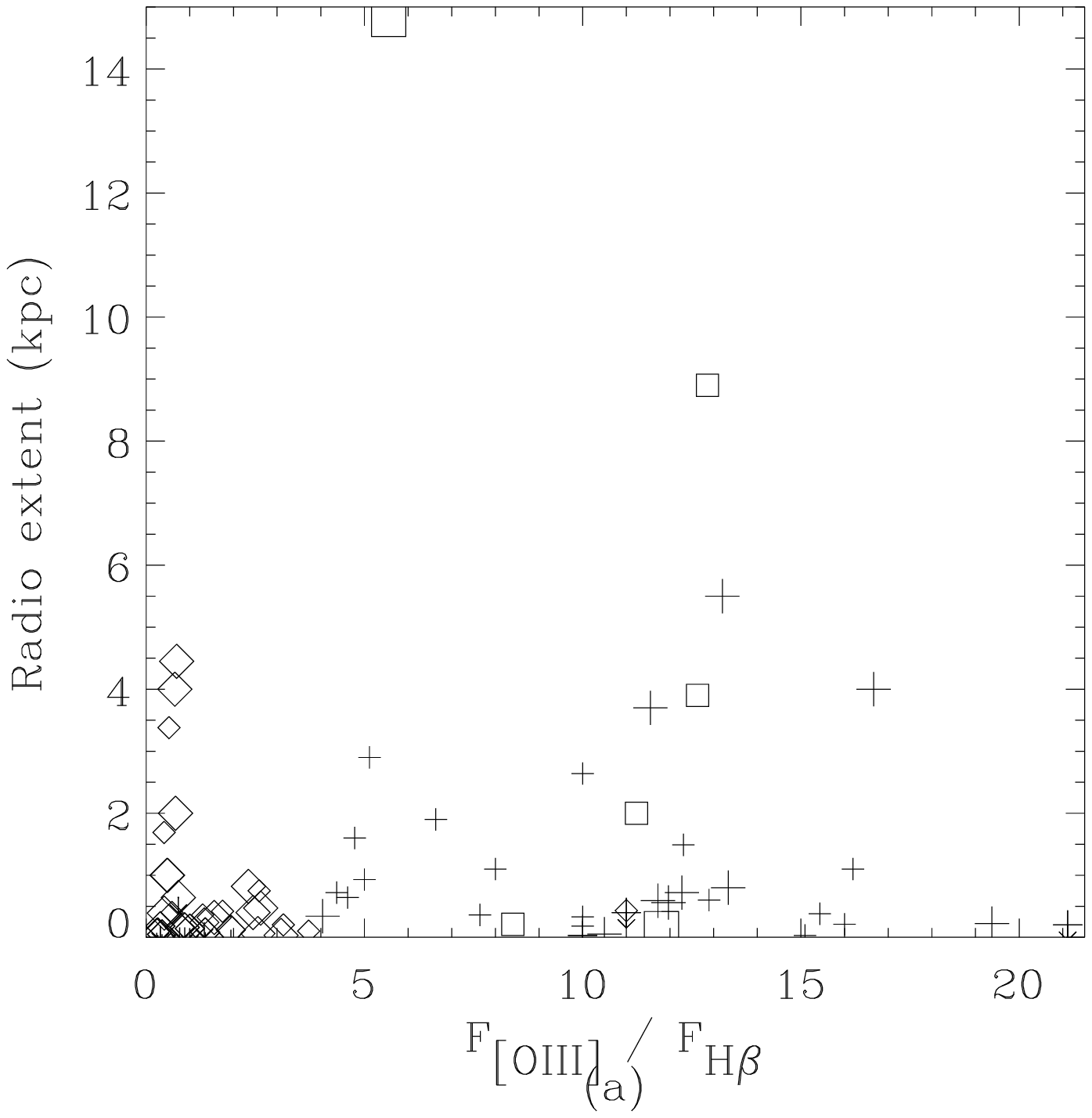}{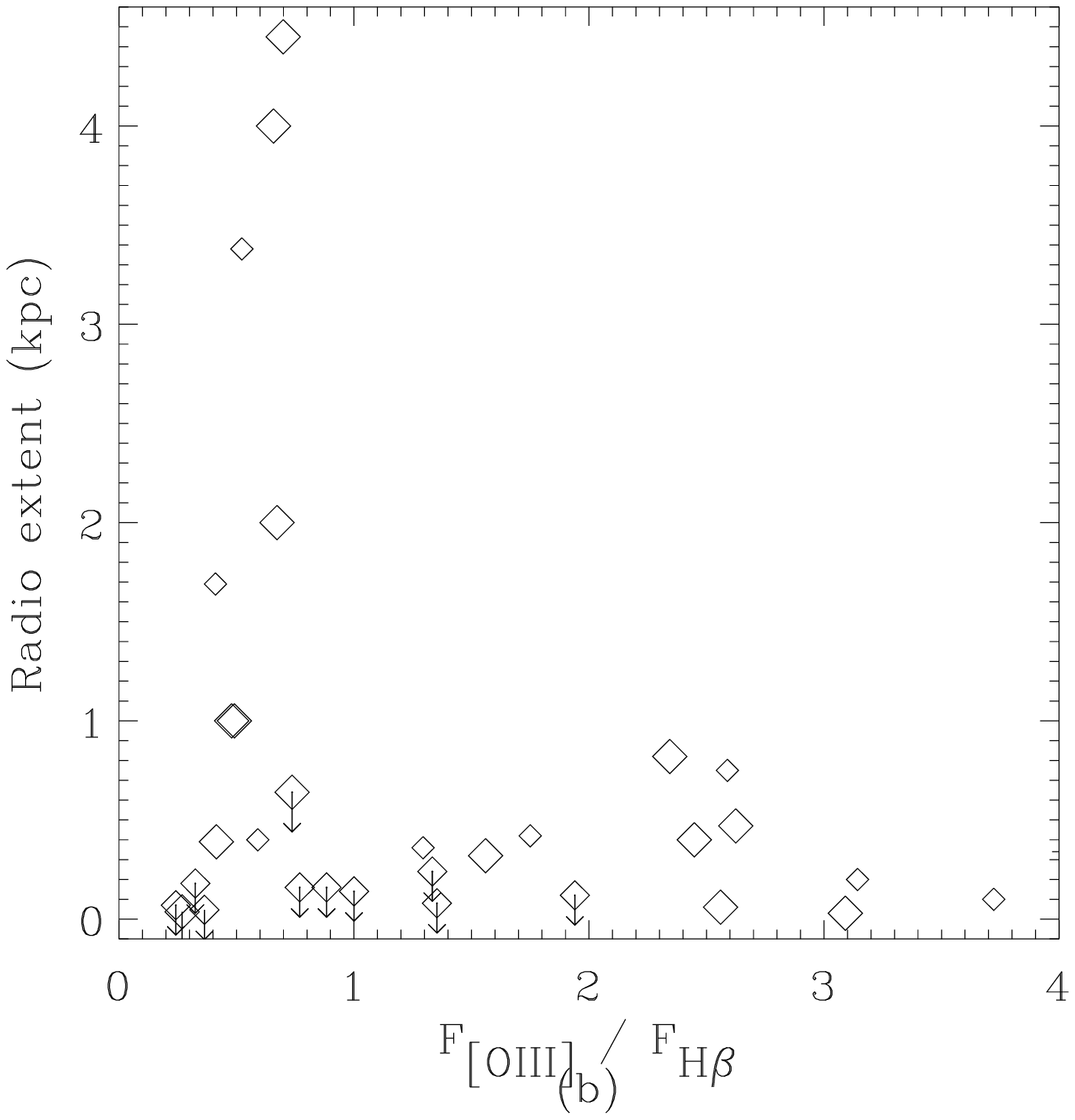}
\caption
{\small{The relationship between Seyfert radio extent and 
R=F$_{[OIII]}$/F$_{H\beta}$ for an enlarged sample of Seyfert galaxies
(see Section~5.4).
Seyfert 1's (Seyfert 1.0's through Seyfert 1.5's)
are plotted as diamonds; Seyfert 2.0's are plotted as
crosses; Seyfert 1.8's and Seyfert 1.9's are plotted as squares.
Large symbols denote higher quality data, 
for which the radio extent, [OIII] flux and H$\beta$
flux have at least one quality flag `a', and the others `b' or higher.
Fluxes from Winkler (1992) have been assigned quality flag `b'.
Smaller symbols are used for the lower quality data.
(a) Relationship between radio extent and R;
(b) same as (a), but with a smaller x and y axis range.
}}
\end{figure}

We therefore investigate the relationship between radio extent and 
R=F$_{[OIII]}$/F$_{H\beta}$.
The parameter R is used (see e.g. \cite{vo87}) both to distinguish
Seyfert 2's from other emission-line nuclei (such as LINERs and HII regions),
and to distinguish the sub-classes 1.0, 1.2, 1.5 and 1.8 among Seyfert 1's
(see Section~3). 
Since the early-type Seyfert sample is too small for this investigation,
we have added to it all Seyferts from the literature which show extended radio
structure (see Nagar \& Wilson 1998 for a list of these objects),
and all radio-unresolved Seyferts from Papers V, VI, VII, 
and Kukula et al. (1995), for which values of [OIII] flux
and H$\beta$ flux are listed in Whittle (1992a) and Winkler (1992).
The final sample includes 21 radio-extended
Seyfert 1's, 11 radio-unresolved Seyfert 1's, 6 radio-extended
Seyfert 1.8's and 1.9's, 27 radio-extended Seyfert 2.0's and
4 radio-unresolved Seyfert 2.0's.
The results are shown in Figure~21. 

It is remarkable that all 5 highly extended
Seyfert 1's (extent $\geq$ 1.5 kpc) have R values in a relatively
narrow range - 0.4$<$R$<$0.7 (Figure~21, right panel). 
The [OIII] luminosities of these highly extended Seyfert 1's are in
the lower $\sim$50 percentile of the log P$_{[OIII]}$ distribution of
the sample,
but their H$\beta$ luminosities are in the upper $\sim$50 percentile of
the log P$_{H\beta}$  distribution. Their R values  
are therefore low and they are all classified as Seyfert 1.2's.
It is clear that some factor other than the relative orientation of 
the nuclear obscuring torus to the line of sight affects the R value,
and that this factor is correlated with the radio extent.

If we ignore the 5--7 highly extended Seyfert 1.2's, the distribution of the
other Seyferts in Figure~21 appears consistent with the expectations
of the unified scheme.
Only two confirmed Seyfert 1.0's (Mrk~335 and NGC~4593) have been imaged
at high resolution in the radio and both of these are unresolved
(two of the Seyfert 1's, NGC~235A and Mrk~231, are of unknown intermediate
type).
The small number of Seyfert 1.8's and 1.9's limits any conclusions, though
it is interesting that all of these are radio-extended 
and that the 6 Seyfert 1.9's tend to be more radio-extended
than the 3 Seyfert 1.8's. In fact, the Seyfert 1.9's
appear to be among the most radio extended objects in the sample, even
though their [OIII] and H$\beta$ luminosities are not significantly different
from those of Seyfert 2.0's.
Among Seyfert 2.0's, the radio extent and R are uncorrelated.

\section{Summary}

We have carried out a high resolution radio imaging survey of a
sample of early-type Seyfert galaxies. 
We have presented radio maps and derived properties
for all newly observed galaxies. The data forms part of
a homogeneous database of a magnitude and distance-limited sample;
most galaxies in the sample have been
previously imaged in the [OIII] and H$\alpha$+[NII]
emission-lines and in continuum (green and red) emission.

The radio luminosities of Seyfert 1's (i.e. Seyfert 1.0's, 1.2's and 1.5's)
and Seyfert 2.0's in the early-type sample are similar, and are
independent of morphological type within the limited range of 
morphological type in the sample. 
However, the fraction of resolved radio sources is larger in the 
Seyfert 2.0's (93\%), than in the Seyfert 1's (64\%).
The statistical results on Seyfert 1's are limited by the small size of the
sample, but their radio and emission-line luminosities are correlated, as found
in earlier studies. For Seyfert 2.0's, we find that the emission-line
axis on the $\simeq$10{\arcsec} scale tends to align with both the nuclear
radio axis measured on a smaller scale and the galaxy disk axis 
measured on a larger scale. These alignments  are consistent with a
picture in which the 
ionized gas lies preferentially in the plane of the galaxy disk.
This ambient gas is ionized by nuclear radiation that escapes along and
around the radio axis and is compressed in shocks driven by the radio ejecta
or by larger-scale winds coaxial with the radio ejecta.
We have used the radio - emission-line alignment to place a constraint on
the product of the velocity of
the radio ejecta and the period, P, of any large angle precession of
the inner accretion disk and jet : V$_{ejecta}~{\times}~P~\geq$~2~kpc.

We have also investigated the relationship between the radio extent and
R=F$_{[OIII]}$/F$_{H\beta}$ (F$_{H\beta}$ is the total [broad plus narrow]
H$_\beta$ flux) for a larger sample of Seyferts.
Remarkably, we find that the 5 highly extended ($\geq$~1.5~kpc) Seyfert 1's 
all have R values in the range 0.4$<$R$<$0.7, leading to a 
classification of Seyfert 1.2. This result implies
that the value of R,  
and hence the Seyfert intermediate type classification, depends 
on some factor other than the angle between the axis of the
obscuring torus and the line of sight.
The distribution of the remainder of the Seyferts is consistent with
the expectations of the unified scheme.

\acknowledgements
NN would like to acknowledge Steven White's guidance with all
things AIPS, and Pierre Ferruit for useful comments on the manuscript.
The National Radio Astronomy Observatory is a facility of the National Science
Foundation operated under cooperative agreement by Associated Universities, 
Inc.
This research has made use of the NASA/IPAC extragalactic database (NED) which
is operated by the Jet Propulsion Laboratory, Caltech, under contract
with the National Aeronautics and Space Administration.
We have made use of the Lyon-Meudon Extragalactic Database                   
(LEDA) supplied by the LEDA team at the CRAL-Observatoire de                   
Lyon (France).           
We have used the Digital Sky Surveys (DSS) which 
were produced at the Space Telescope Science Institute under U.S. 
Government grant NAG W-2166.
The images of these surveys are based on photographic data obtained
using the Oschin Schmidt Telescope on Palomar Mountain and the UK Schmidt
Telescope. The plates were processed into the present
compressed digital form with the permission of these institutions. 
This research has made use of the statistical tests in ASURV 1.2
(\cite{lif92}).
This work was supported by grant AST~9527289 from NSF and grant NAG~81027
from NASA.

\pagestyle{empty}

\newpage
\begin{figure}
\plotfiddle{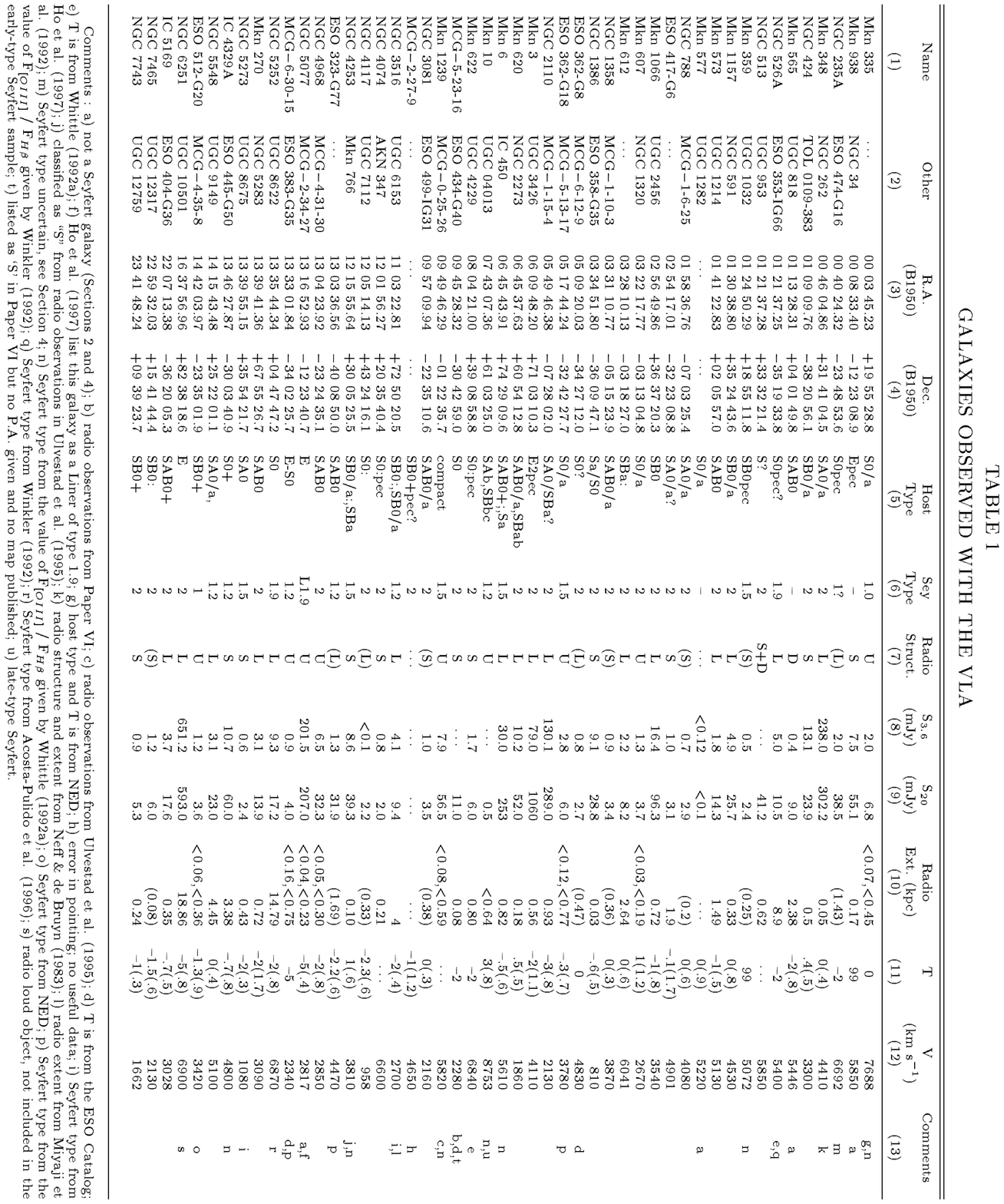}{9in}{0}{100}{100}{-300}{0}
\end{figure}

\newpage
\begin{figure}
\plotfiddle{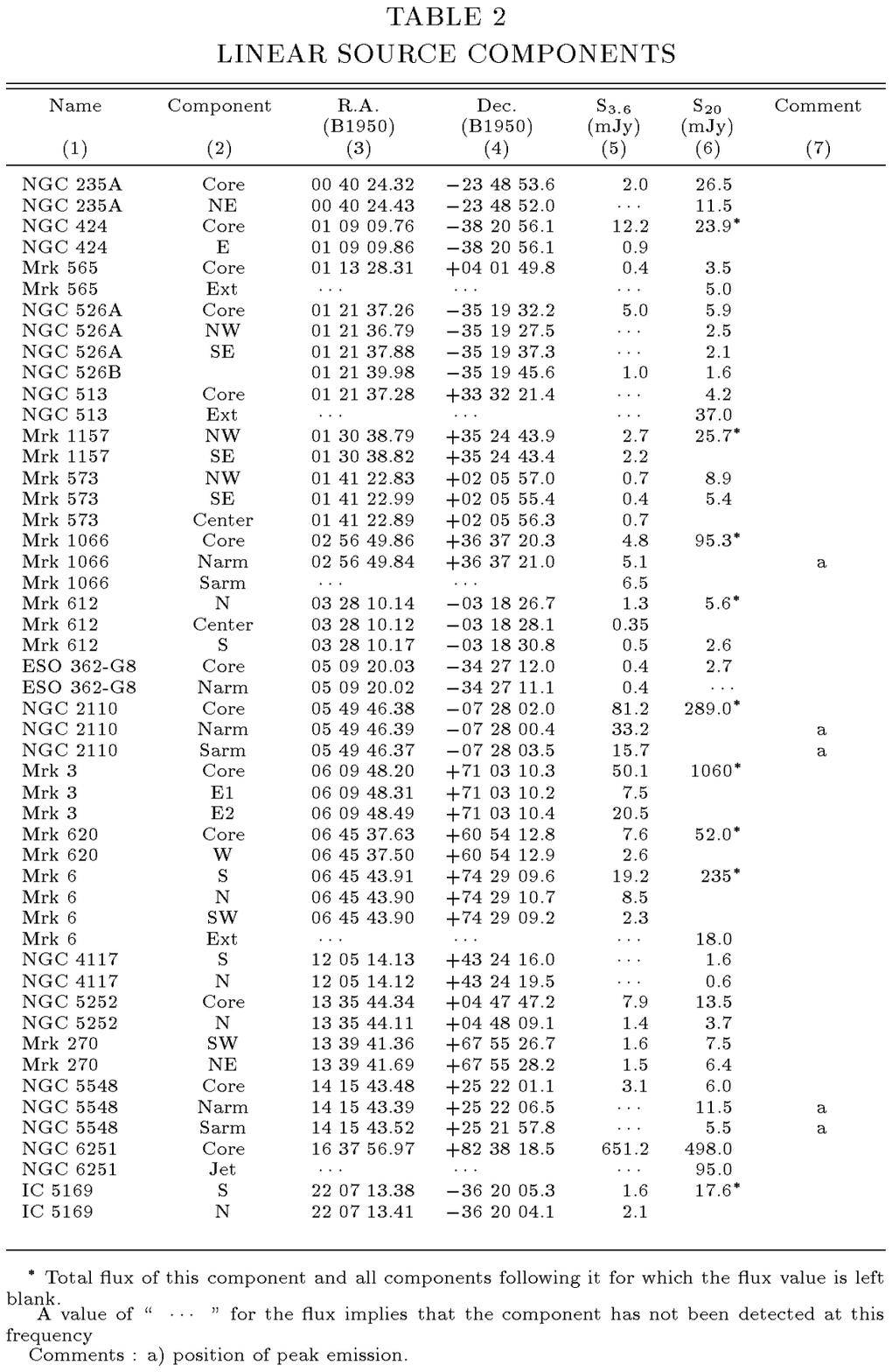}{9in}{0}{100}{100}{-300}{0}
\end{figure}

\newpage
\begin{figure}
\plotfiddle{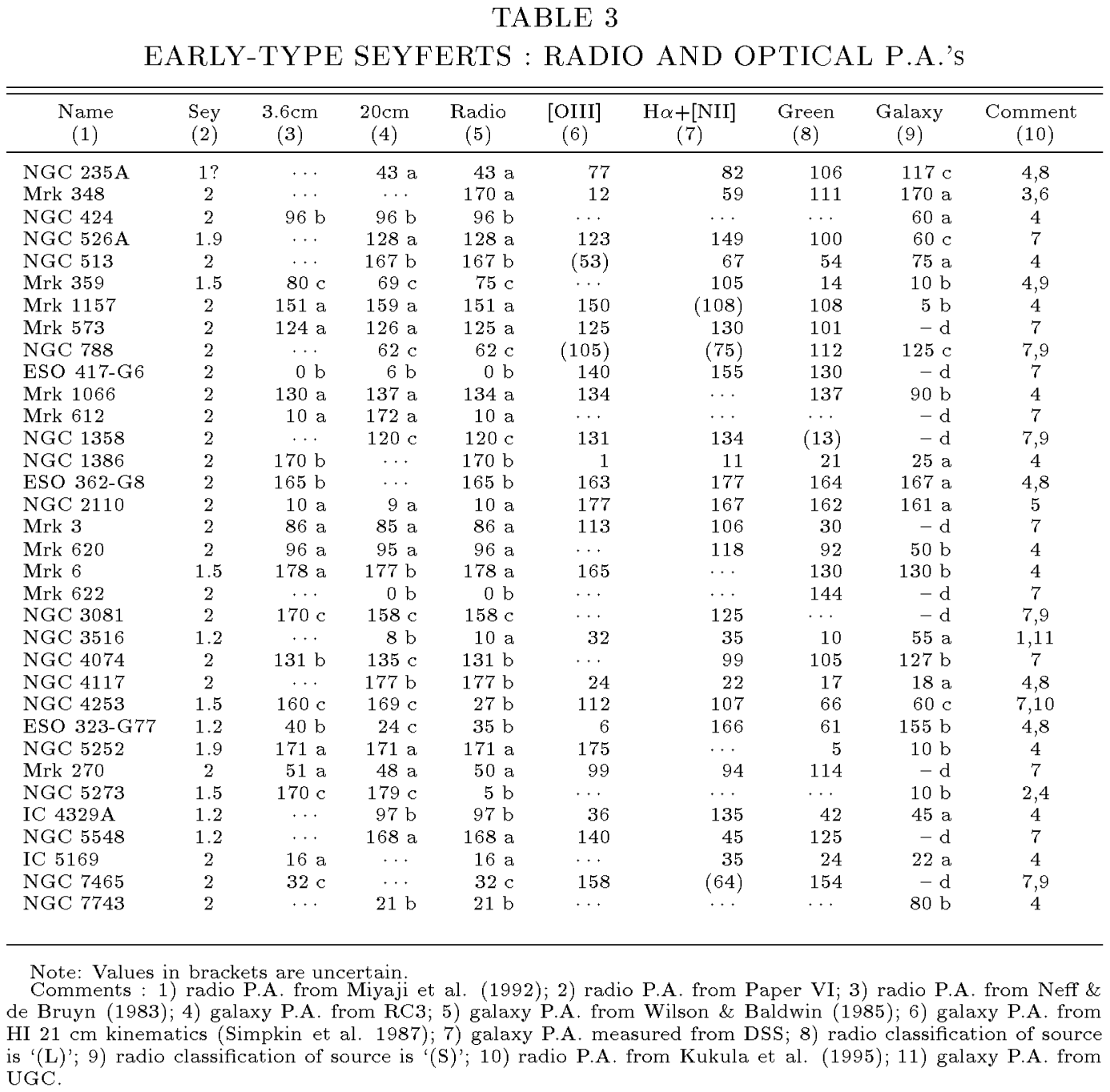}{9in}{0}{100}{100}{-300}{0}
\end{figure}

\newpage
\begin{figure}
\plotfiddle{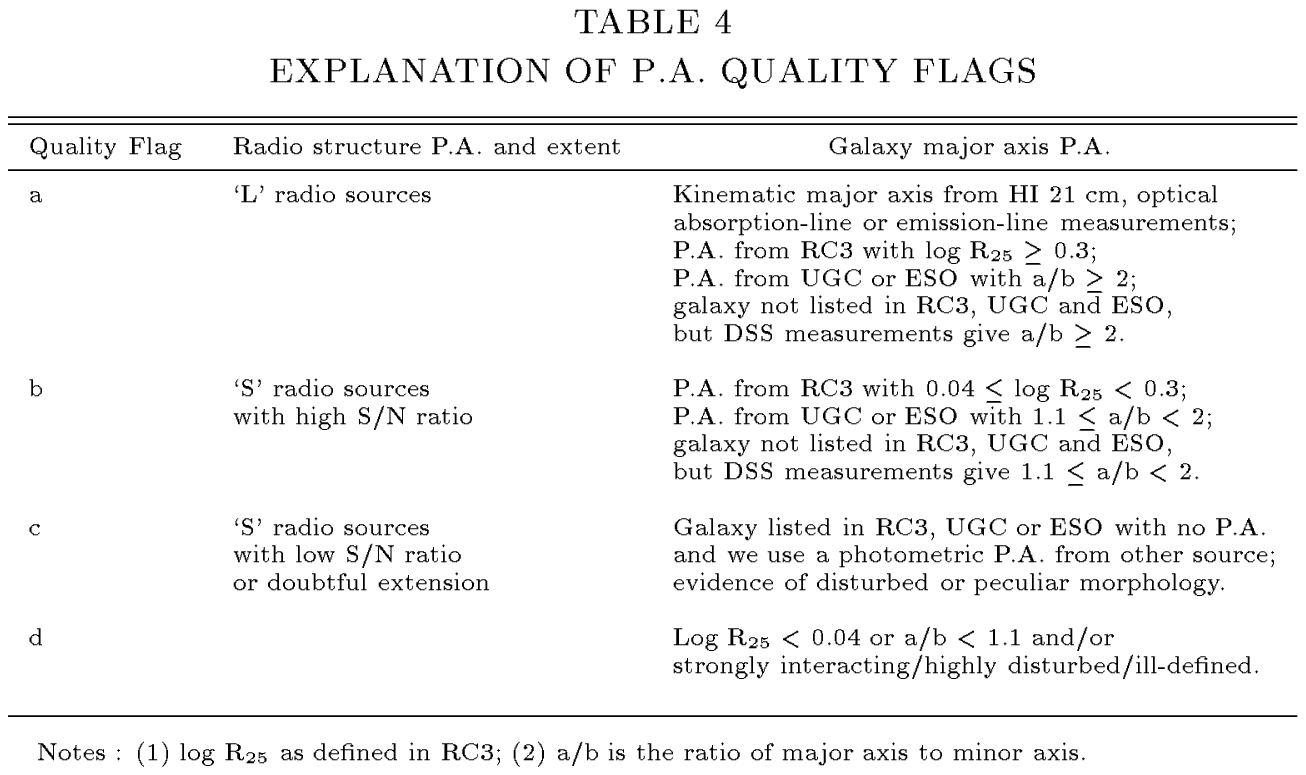}{9in}{0}{100}{100}{-300}{0}
\end{figure}

\newpage
\begin{figure}
\plotfiddle{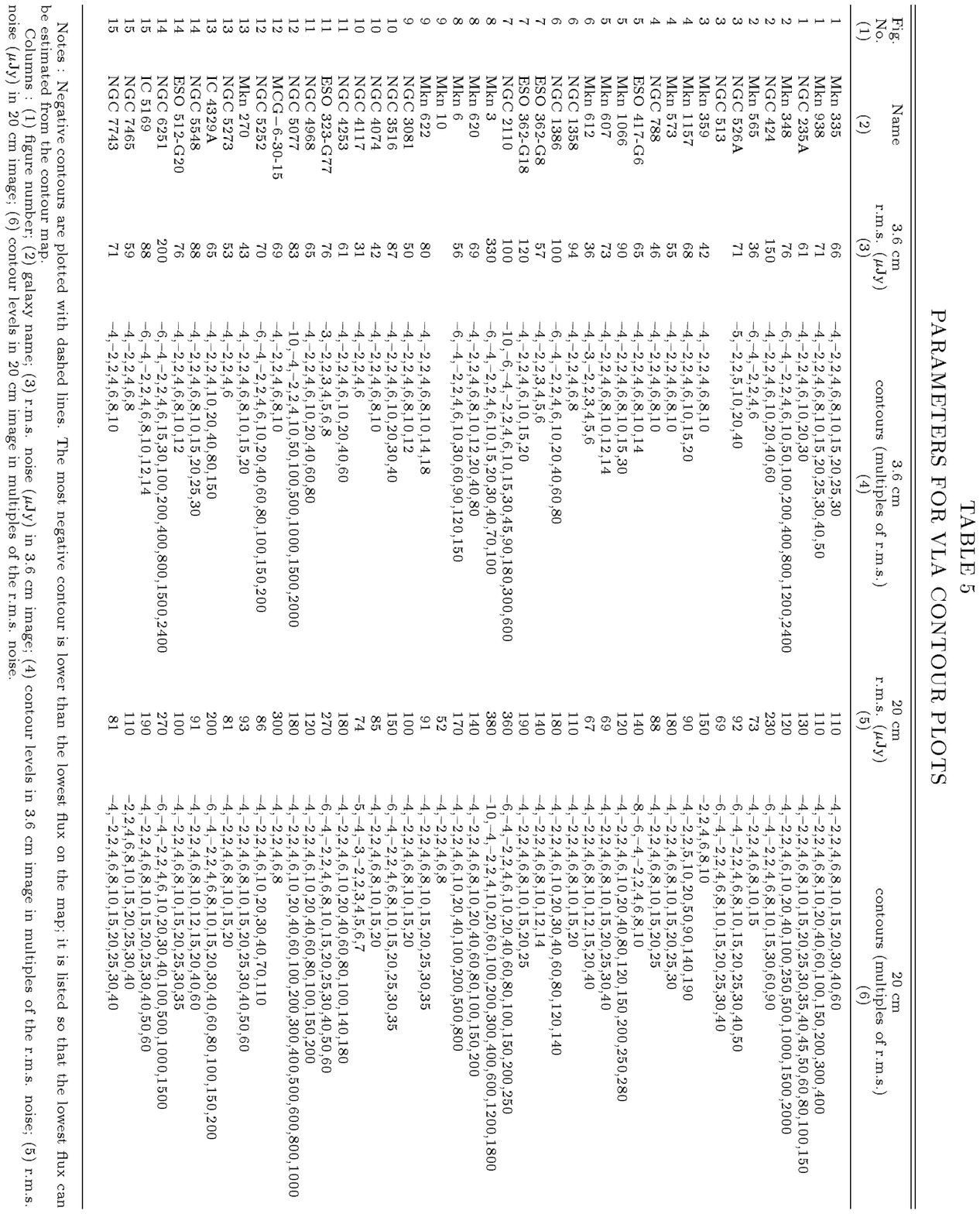}{9in}{0}{100}{100}{-300}{0}
\end{figure}

\newpage
\begin{figure}
\plotfiddle{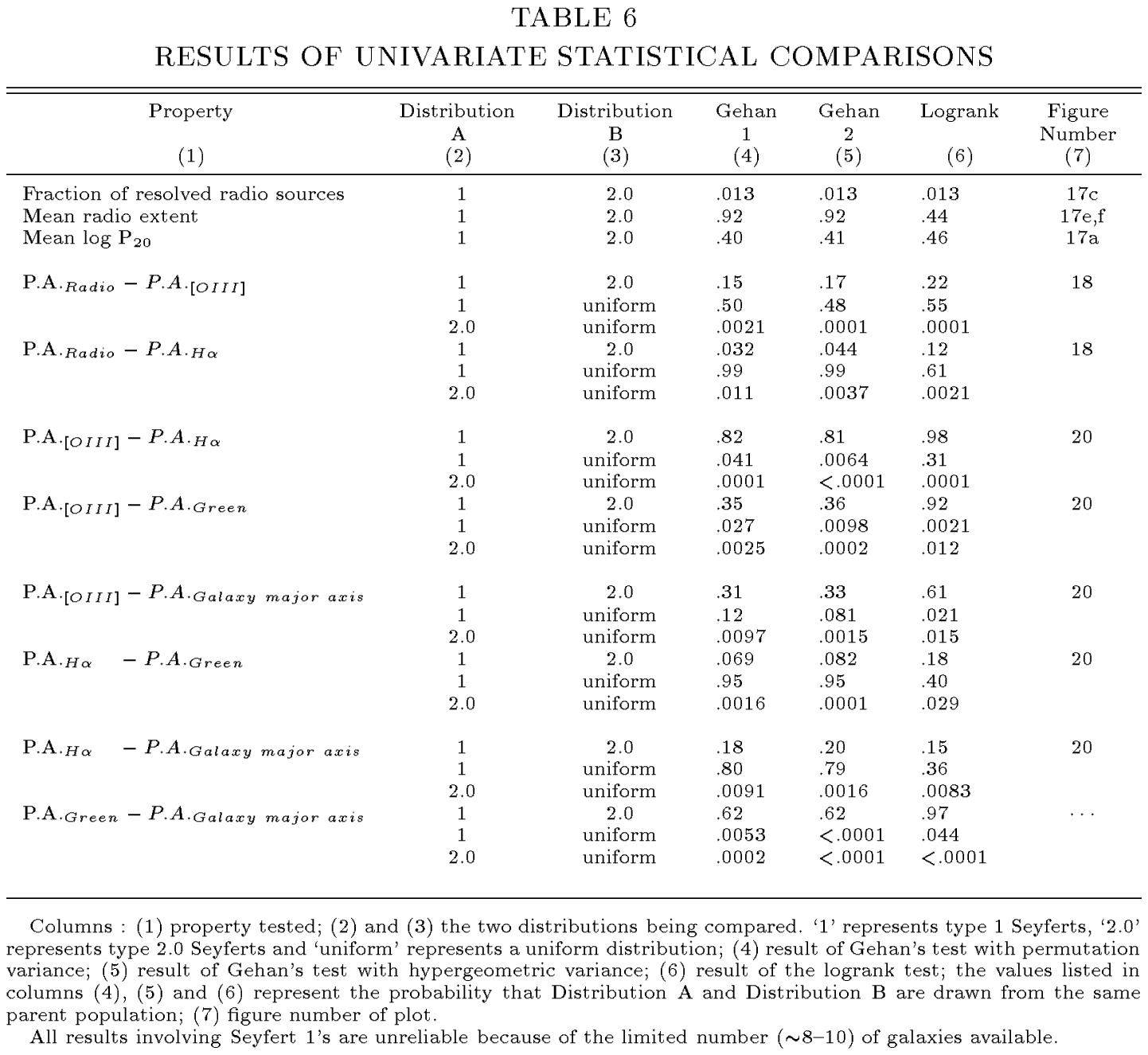}{9in}{0}{100}{100}{-300}{0}
\end{figure}

\newpage
\begin{figure}
\plotfiddle{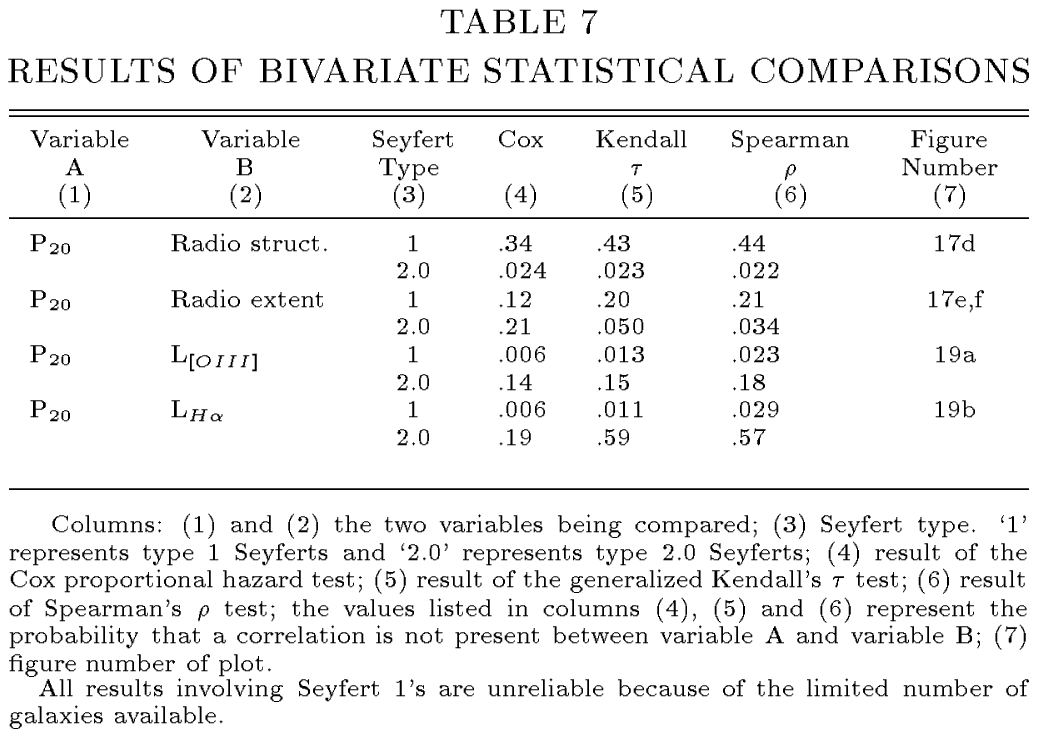}{9in}{0}{100}{100}{-300}{0}
\end{figure}

\end{document}